\documentclass[a4paper,11pt]{book}
\usepackage[english]{babel}
\addto\captionsenglish{}
\usepackage{amsfonts}
\usepackage{amsmath,amssymb}
\usepackage{bbold}
\usepackage{dsfont}
\usepackage{graphicx}
\usepackage{tikz}
\usepackage{hyperref}
\usepackage[utf8]{inputenc}
\usepackage{verbatim}
\usepackage{cite}
\usepackage{siunitx}
\usepackage{color}
\definecolor{tango}{rgb}{0.204,0.396,0.643}

\newcommand{\nablagras}{\mbox{\boldmath$\nabla$}}
\newcommand{\bc}{\begin{center}}
\newcommand{\ec}{\end{center}}
\newcommand{\beq}{\begin{equation}}
\newcommand{\eeq}{\end{equation}}
\newcommand{\bea}{\begin{eqnarray}}
\newcommand{\eea}{\end{eqnarray}}

\newcommand{\ds}{\displaystyle}

\newcommand{\ket}[1]{\ensuremath{| #1 \rangle}} 
 
\newcommand{\braket}[2]{\ensuremath{\langle #1 | #2 \rangle}} 
\newcommand{\ketbra}[2]{\ensuremath{\left| #1 \right\rangle\left\langle #2 \right|}}

\newcommand{\dd}{{\rm d}}

\begin{document}
\thispagestyle{empty}
\noindent 
IV School on Light and Cold Atoms \hfill Hélène Perrin\\
Course on 2D quantum gases \hfill helene.perrin[at]univ-paris13.fr\\
S\~ao Paulo, Oct 20--31, 2025

\vspace{1.5cm}
\begin{center}
{\LARGE\bf Course on two-dimensional quantum gases}\\[1cm]
{\Large IV School on Light and Cold Atoms}\\[2mm]
{\large S\~ao Paulo, October 20--31, 2025}\\[1cm]
{\Large\bf Hélène Perrin}\\[5mm]
{\large Laboratoire de physique des lasers\\
CNRS and Université Sorbonne Paris Nord, France}
\end{center}

\vspace{1cm}
This course on two-dimensional quantum gases, prepared for the S\~ao Paulo ICTP-SAIFR IVth School on Light and Cold Atoms, is organized into three lectures. The first introductory lecture addresses Bose-Einstein condensation in an ideal gas, the effect of interactions in the weakly interacting limit, Gross-Pitaevskii equation and its hydrodynamic formulation, as well as superfluidity. The second lecture is devoted to the main features of two-dimensional quantum gases: importance of the interactions, enhanced phase fluctuations and the emergence of a quasi long-range order, superfluidity through the Berezinskii-Kosterlitz-Thouless mechanism, scaling symmetry. The third lecture has taken the form of a seminar on fast rotating quasi two-dimensional Bose gases and thermal melting of the vortex lattice. The slides accompanying these notes can be found on the school website: \url{https://www.ictp-saifr.org/slca2025/}.

\subsection*{Acknowledgements}
I thank the organizers for the invitation, the efficient organization and the excellent atmosphere, and for encouraging me to write these lecture notes: Romain Bachelard, Patricia Marques Castilho, Raul Celistrino Teixeira, and Mathilde Hugbart. Special thanks to Patricia Castilho who carefully read all these lecture notes. I also thank ICTP SAIFR for their efficient help during the school and the students of the school for their lively participation to the lectures, in particular Amaru Gael Moya Ragal for sharing his handwritten notes with me. Finally, I thank Jérôme Beugnon and Jean Dalibard for enlightening discussions on the correlation function during the preparation of these lecture notes, Maxim Olshanii for a careful reading of the section on the monopole mode, and once again Jean Dalibard for reading all the notes and providing comments and suggestions that helped improve the manuscript.

\clearpage
\tableofcontents
\clearpage

\chapter{Bose-Einstein condensation and superfluidity}

The first lecture is devoted to Bose-Einstein condensation and superfluidity. To prepare the lecture, I have mostly used my lecture notes prepared for Master program ICFP at ENS and the lectures given by Jean Dalibard at Coll\`ege de France during the academic year 2015-2016 \cite{DalibardCF2016en}, which are available online, in French and in English. In addition, the following general references may be useful:
\begin{enumerate}
\item F. Dalfovo, S. Giorgini, L. Pitaevskii and S. Stringari, \textit{Theory of Bose-Einstein condensation in trapped gases}, Rev. Mod. Phys. \textbf{71}, 463 (1999) \cite{Dalfovo1999};
\item Y. Castin, \textit{Bose--Einstein condensates in atomic gases: simple theoretical results}, Proceedings of Les Houches LXXII Summer School \cite{Castin2001};
\item C. J. Pethick and H. Smith, \textit{Bose–Einstein Condensation in Dilute Gases}, Second edition, Cambridge (2008) \cite{Pethick2008};
\item Lev Pitaevskii and Sandro Stringari, \textit{Bose-Einstein condensation}, Oxford (2003) \cite{Pitaevskii2003};
\item Lev Pitaevskii and Sandro Stringari, \textit{Bose-Einstein Condensation and Superfluidity}, Oxford (2016)  \cite{Pitaevskii2016}.
\end{enumerate}

\section{Reminder: BEC in an ideal gas}
Let us first examine the possibility for Bose-Einstein condensation (BEC) for a gas of non interacting bosons. BEC will occur if the number of particles in excited states saturates, i.e. if the total number of particles $N$ exceeds $N_{\rm exc}^{\rm max}(T)$, the maximum number of particles in the excited states at a given temperature $T$.

\subsection{BEC: a saturation of the excited states}
We describe the bosonic gas in the grand canonical ensemble and assume that the ground state of the single-particle Hamiltonian is non degenerate. The average energy is fixed by the temperature $T$ and the average atom number $N$ is fixed by the chemical potential $\mu$. The average occupation of each state $j$ of energy $E_j$, with $E_0$ the energy of the ground state, is given by
\beq
\boxed{
n_j = \ds\frac{1}{e^{(E_j-\mu)/kT}-1},
}
\eeq
where $k$ is Boltzmann's constant. $n_j\geq 0$, which imposes $\mu<E_j$ for all $j$, which is fulfilled as soon as $\mu<E_0$. The occupation in the ground state is
\beq
n_0 = \frac{1}{e^{\beta(E_0-\mu)}-1} = \frac{z}{1-z}
\eeq
where $\beta=1/(kT)$ and we have introduced the \textit{fugacity}
\beq
\boxed{
z = e^{\beta(\mu-E_0)} < 1.
}
\eeq
Using $z$, we can also write $n_j$ as
\beq
n_j = \ds\frac{1}{z^{-1}e^{\beta(E_j-E_0)}-1},
\eeq
such that the number of particles in excited states writes
\beq
N_{\rm exc}(z,T) = \sum_{j>0}\ds\frac{1}{z^{-1}e^{\beta(E_j-E_0)}-1},
\label{eq:Nexc}
\eeq
which is an increasing function of $z$. BEC will occur if $N_{\rm exc}(z,T)$ stays finite when $z$ approaches 1, while $n_0$ is not bounded and diverges as $z\to 1$. This depends on how $E_j$ depends on $j$.

Using the formula
\beq
\sum_{n=1}^{\infty} x^n = \frac{x}{1-x} \qquad \mbox{if } |x|<1,
\eeq
we can write Eq.~\eqref{eq:Nexc} as
\beq
N_{\rm exc}(z,T) = \sum_{j>0}\sum_{n=1}^{\infty}\ds z^n e^{n\beta(E_0-E_j)} = \sum_{n=1}^{\infty} z^n \sum_{j>0} e^{n\beta(E_0-E_j)}.
\eeq

\subsection{Semi-classical approximation: using the density of states}
If $kT$ is much larger than the spacing $E_{j+1}-E_j$ between consecutive energies in the spectrum, we can make a semi-classical approximation and replace the discrete sum by an integral
\beq
\sum_{j>0} \to \int_{E_0}^{+\infty} \rho(\varepsilon) d\varepsilon,
\eeq
where $\rho(\varepsilon)$ is the density of states at energy $\varepsilon$, defined such that $\rho(\varepsilon)d\varepsilon$ is the number of states of the Hamiltonian $\mathcal{H}(\mathbf{r},\mathbf{p})$ with an energy between $\varepsilon$ and $\varepsilon+d\varepsilon$, or more formally
\beq
\boxed{
\rho(\varepsilon) = \frac{1}{h^D}\int d\mathbf{p}\,d\mathbf{r}\,\delta\left(\mathcal{H}(\mathbf{r},\mathbf{p})-\varepsilon\right),
}
\eeq
with $h$ the Planck constant and $D$ the dimension of the system. We thus have
\beq
N_{\rm exc}(z,T) = \int_{E_0}^{+\infty} d\varepsilon\, \rho(\varepsilon)n_\varepsilon(\varepsilon) = \int_{E_0}^{+\infty} d\varepsilon \,\frac{\rho(\varepsilon)}{z^{-1}e^{\beta(\varepsilon-E_0)}-1}.
\label{eq:Nexc_sc}
\eeq
The convergence of the integral at $+\infty$ is ensured as soon as $\rho(\varepsilon)e^{-\beta\varepsilon}$ has a converging integral, which is easy with the exponential: $\rho(\varepsilon)$ should not increase exponentially with the energy. When $z\to 1$, the limit at $\varepsilon\to E_0$ is more tricky: the integrand in Eq.~\eqref{eq:Nexc_sc} is approximately $kT \rho(\varepsilon)/(\varepsilon - E_0)$ such that $\rho(\varepsilon)$ should converge to 0 when $\varepsilon\to E_0$ to ensure the convergence of the integral. For example, $N_{\rm exc}$ diverges if $\rho$ does not depend on $\varepsilon$.

Introducing an infinite sum over $n$, we can recast Eq.~\eqref{eq:Nexc_sc} into:
\beq
N_{\rm exc}(z,T) = \sum_{n=1}^{\infty} z^n \int_{E_0}^{+\infty} d\varepsilon\, \rho(\varepsilon)e^{-n\beta(\varepsilon-E_0)} = \sum_{n=1}^{\infty} z^n \int_0^{+\infty} d\varepsilon \,\rho(\varepsilon-E_0)e^{-n\beta\varepsilon}.
\eeq
Using the change in integration variable $u=n\beta\varepsilon$ or $\varepsilon=ukT/n$, we can also recast this integral as
\beq
\boxed{
N_{\rm exc}(z,T) = kT \sum_{n=1}^{+\infty} \frac{z^n}{n} \int_0^{+\infty} du \,\rho\left(\frac{kT}{n}u-E_0\right)e^{-u}.
}
\eeq

\subsection{Simple expressions for a power-law density of states}
If the density of state $\rho$ is a power law, i.e.
\beq
\rho(\varepsilon) = \frac{1}{\varepsilon_0}\left(\frac{E_0+\varepsilon}{\varepsilon_0}\right)^q,
\eeq
then we get for the total number of particles in the excited states
\beq
N_{\rm exc}(z,T) = \frac{kT}{\varepsilon_0} \sum_{n=1}^{+\infty} \frac{z^n}{n} \int_0^{+\infty} du \,\left(\frac{kT}{n\varepsilon_0}\right)^q u^qe^{-u} = \left(\frac{kT}{\varepsilon_0}\right)^{q+1}\sum_{n=1}^{+\infty} \frac{z^n}{n^{q+1}}\int_0^{+\infty} du \,u^qe^{-u},
\eeq
and finally
\beq
\boxed{
N_{\rm exc}(z,T) = \Gamma(q+1)\left(\frac{kT}{\varepsilon_0}\right)^{q+1}{\rm Li}_{q+1}(z).
}
\eeq
$\Gamma$ is the Euler Gamma function, and
\beq
{\rm Li}_{s}(z) = \sum_{n=1}^{+\infty} \frac{z^n}{n^s}
\eeq
is the polylogarithmic function of order $s$. When it converges in the limit $z\to1$, i.e. for $s>1$, its value for $z=1$ is the Riemann zeta function:
\beq
{\rm Li}_{s}(1) = \zeta(s).
\eeq
We thus have a simple criterion for BEC to occur: if $q>0$, $N_{\rm exc}^{\rm max}(T) = N_{\rm exc}(1,T) = \Gamma(q+1)\zeta(q+1)\left(\frac{kT}{\varepsilon_0}\right)^{q+1}$ is finite and the number of atoms in the excited states saturates if $N>N_c(T)=N_{\rm exc}^{\rm max}(T)$,
\beq
\boxed{
N_c(T) = \Gamma(q+1)\zeta(q+1)\left(\frac{kT}{\varepsilon_0}\right)^{q+1},
}
\label{eq:Nc}
\eeq
or equivalently if $T<T_c(N)$ with
\beq
\boxed{
kT_c = \frac{\varepsilon_0 N^{\frac{1}{q+1}}}{\left[\Gamma(q+1)\zeta(q+1)\right]^{\frac{1}{q+1}}}.
}
\label{eq:Tc}
\eeq

Below $T_c$, the number of particles in the excited states saturates to $N_{\rm exc}=N_c(T)$, and fraction of atoms in the ground state is given by
\beq
\frac{N_0}{N} = 1 - \frac{N_{\rm exc}}{N} = 1 -  \frac{N_c(T)}{N}.
\eeq
Using $N_c(T)\propto T^{q+1}$ and $N=N_c(T_c)$ by definition of $T_c$, we get
\beq
\boxed{\frac{N_0}{N} = 1 - \left(\frac{T}{T_c}\right)^{q+1}.}
\eeq

\subsection{From box traps to harmonic traps}
Let us consider the case where the Hamiltonian describes a particle of mass $M$ in dimension $D$ in a generic,  isotropic\footnote{If the trap is anisotropic, it is easy to come back to the isotropic case by a mere dilation of the axes. The conclusion for BEC still holds.} power-law trap
\beq
V(\mathbf{r}) = V_0\left(\frac{r}{R}\right)^{2/\alpha}
\eeq
with $\alpha\geq0$ some exponent. This expression describes the case of an harmonic trap, with $\alpha=1$, but also a box trap for $\alpha=0$, where $V=0$ for $r<R$ and $V=+\infty$ for $r>R$. The Hamiltonian is then
\beq
\mathcal{H}(\mathbf{r},\mathbf{p}) = \frac{p^2}{2M} + V_0\left(\frac{r}{R}\right)^{2/\alpha}.
\eeq
Let us compute the density of states for this case, taking advantage of the symmetry of the trap and the kinetic energy:
\beq
\rho(\varepsilon) = \frac{1}{h^D}\int_0^{+\infty} C(D) r^{D-1} dr \int_0^{+\infty} C(D) p^{D-1} dp \, \delta\left(\frac{p^2}{2M} + V_0\left(\frac{r}{R}\right)^{2/\alpha}-\varepsilon\right).
\eeq
Here, $C(D)$ is a constant that depends on the dimension and gives the result of angular integration: $C(1)=1$, $C(2)=2\pi$, $C(3)=4\pi$.

We change the integration variable for $K=p^2/2M$, such that $dK = p\,dp/M$ and $p=\sqrt{2MK}$. We get
\beq
\rho(\varepsilon)=\frac{C(D)^2}{2}\left(\frac{2M}{h^2}\right)^{D/2}\int_0^{+\infty}  r^{D-1} dr \int_0^{+\infty} K^{D/2-1} dK \, \delta\left(K + V_0\left(\frac{r}{R}\right)^{2/\alpha}-\varepsilon\right).
\eeq
The delta function integrates directly with $K=\varepsilon - V_0\left(\frac{r}{R}\right)^{2/\alpha}$ with the condition $K\geq0$, i.e. $r\leq r_{\rm max}= R\left(\varepsilon/V_0\right)^{\alpha/2}$:
\beq
\rho(\varepsilon)=\frac{C(D)^2}{2}\left(\frac{2M}{h^2}\right)^{D/2}\int_0^{r_{\rm max}} dr \, r^{D-1} \left(\varepsilon - V_0\left(\frac{r}{R}\right)^{2/\alpha}\right)^{D/2-1}.
\eeq
We now use the variable $u=r/r_{\rm max}$:
\beq
\rho(\varepsilon)=\frac{C(D)^2}{2}\left(\frac{2MR^2}{h^2}\right)^{D/2}\left(\frac{\varepsilon}{V_0}\right)^{D\alpha/2}\varepsilon^{D/2-1}\int_0^{1} du \, u^{D-1} \left(1 - u^{2/\alpha}\right)^{D/2-1}.
\eeq
The integral is now just a number $\eta(\alpha,D)=D^{-1}\Gamma(D/2)\Gamma(1+\alpha D/2)/\Gamma[(1+\alpha)D/2]$. We get:
\beq
\rho(\varepsilon) = \frac{\eta(\alpha,D)C(D)^2}{2}\left(\frac{2MR^2}{h^2}\right)^{D/2}V_0^{-D\alpha/2}\varepsilon^{D(\alpha+1)/2-1} = \frac{1}{\varepsilon_0}\left(\frac{\varepsilon}{\varepsilon_0}\right)^q
\label{eq:DOS}
\eeq
with
\beq
\varepsilon_0 = \left[\frac{2}{\eta(\alpha,D)C(D)^2}\left(\frac{h^2}{2MR^2}\right)^{D/2}V_0^{D\alpha/2}\right]^{\frac{2}{D(\alpha+1)}}
\label{eq:eps0}
\eeq
and
\beq
\boxed{
q = \frac{D(\alpha+1)}{2}-1.
}
\label{eq:q_exponent}
\eeq

We can now conclude: BEC can occur only if $q>0$, i.e. only if
\beq
\boxed{
D(\alpha+1)>2.
}
\eeq
\begin{itemize}
\item In dimension $D=1$, we need $\alpha>1$, i.e. a trap steeper than a harmonic trap, a linear trap for example. BEC doesn't occur in a box, nor in a harmonic trap.
\item In dimension $D=2$, the condition is $\alpha>0$. Any trapping potential with a power-law with positive exponent is sufficient to have a finite critical temperature for BEC. However, BEC doesn't occur in a box trap, which corresponds to the critical case $\alpha=0$.
\item Finally, in dimension $D=3$, the condition writes $\alpha>-1/3$, which is always fulfilled as $\alpha\geq0$. BEC occurs in any trap, even in a box (uniform density).
\end{itemize}

\noindent\hrulefill\\
\textbf{Exercise 1}: give the expressions for $N_c(T)$ and $T_c(N)$ for the cases discussed above: harmonic trap of frequency $\omega_0$ in 2D and 3D, and box trap with an atomic density $n$ in 3D.\\
\noindent\hrule

\subsection{Experimental results}
\begin{figure}[ht]
\centering
\raisebox{8mm}{\includegraphics[width=0.45\linewidth]{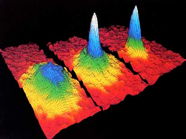}}
\begin{tikzpicture}
\node at (0,0) {\includegraphics[width=0.5\linewidth]{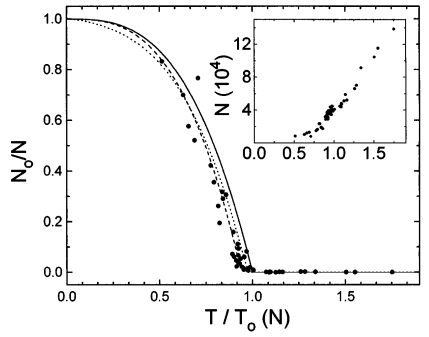}};
\node at (2,-1) {\scriptsize $\ds \frac{N_0}{N} = 1 - \left( \frac{T}{T_c}\right)^3$};
\end{tikzpicture}
\caption{\textbf{Left}: First observation of Bose-Einstein condensation in atomic gases. Rubidium atoms initially confined in a harmonic trap are released in a time-of-flight experiment. The gas expands, and the density distribution (integrated along the imaging axis) after expansion is linked to the initial momentum distribution. Figure from Cornell's group. \textbf{Right}: Condensed fraction as a function of temperature in units of the predicted critical temperature $T_0$ for a non interacting gas, compared with the prediction for a non interacting gas. BEC occurs at a temperature slightly lower than predicted by the non interacting model. Inset: remaining atoms as a function of $T/T_0$. Figure adapted from \cite{Ensher1996}.\label{fig:BEC_Cornell}}
\end{figure}

The first experimental demonstration of atomic Bose-Einstein condensation has been performed in the group of Eric Cornell and Carl Wieman \cite{Anderson1995}. They observed BEC of rubidium 87 atoms confined in a three-dimensional harmonic magnetic trap ($q=2$), by releasing the atoms to access the momentum distribution,\footnote{It was understood soon after that the shape of the central peak is linked to the initial interaction energy rather than to kinetic energy.} see Fig.~\ref{fig:BEC_Cornell}. As temperature is lowered, a central peak appears in the distribution, corresponding to the condensate. In this case, condensation happens both in real and momentum space. They verified approximately the expected law for the condensate fraction \cite{Ensher1996}.

\begin{figure}[ht]
\centering
\includegraphics[width=0.45\linewidth]{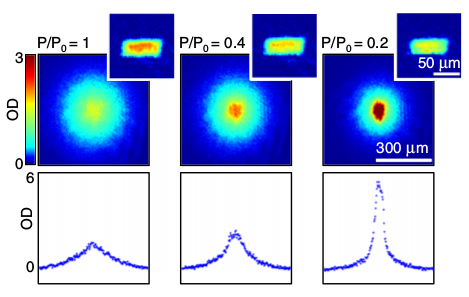}
\includegraphics[width=0.5\linewidth]{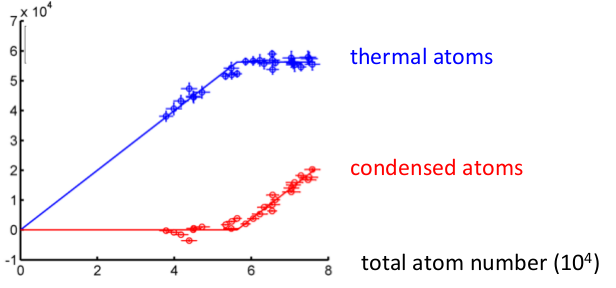}
\caption{\textbf{Left}: Observation of BEC for a 3D Bose gas confined in a box trap. Figure from Ref. \cite{Gaunt2013}. Upper panel: momentum distribution as observed after a 50\,ms time-of-flight expansion. Lower panel: cut in the momentum distribution. Inset: in situ distribution before expansion. \textbf{Right}: Saturation of the number of atoms in the excited states in this situation. Figure from Ref. \cite{DalibardCF2016en}, adapted from a figure provided by Zoran Hadzibabi\'c.\label{fig:BEC_3Dbox}}
\end{figure}

It took almost twenty more years to be able to realize a three-dimensional box potential for cold atoms ($q=1/2$). In this case, condensation occurs in momentum state only. Interactions are kept extremely small, the density being independent of temperature in a box and much smaller than in the previous case. This experiment was first performed in the group of Zoran Hadzibabic in Cambridge \cite{Gaunt2013}, see Fig.~\ref{fig:BEC_3Dbox}. The absence of interactions makes the agreement with theory easier, and they were able to observe the saturation of the excited states \cite{Schmidutz2014}.

\section{Weakly interacting degenerate Bose gas}
From the early results presented in Fig.~\ref{fig:BEC_Cornell}, it appears already that there is a shift in the critical temperature with respect to the prediction for the non interacting case. The expansion of the condensate is also not only due to its very small kinetic energy, but mainly to the effect of the repulsive interactions \cite{Castin1996}. Even if the interactions remain weak, they affect qualitatively the shape of the condensate and its dynamics. In this section, we will derive the equation that describes the condensate in its ground state from a mean-field approach, using a variational procedure.

\subsection{S-wave scattering in a nutshell}
I will not recall the scattering theory here, nor derive explicitly the expression of the effective interaction potential and its link to the scattering length. I refer the interested reader to Ref. \cite{WalravenQuantumGases}. The main idea is that at low energies (for linear momentum less than $\hbar/r_{\rm pot}$ where $r_{\rm pot}$ is the range of the molecular potential, i.e. typically below a few mK), collisions occur only in s-wave and are described by a single parameter, the scattering length $a$.

In brief, we consider a spherical interaction potential $V_{\rm int}(|\mathbf{r}_1-\mathbf{r}_2|)$ depending only on the relative distance $r=|\mathbf{r}_1-\mathbf{r}_2|$ between the two identical colliding particles of mass $M$. We assume that $V_{\rm int}(r)$ is short-range, with an attractive part faster than $1/r^3$ and a sharp potential barrier near $r=0$ due to the repulsion between the electronic clouds of the two atoms. We introduce the center-of-mass frame:
\bea
\hat{\mathbf{P}} &=& \hat{\mathbf{p}}_1+\hat{\mathbf{p}}_2 \qquad \mbox{and} \qquad \hat{\mathbf{p}} = \frac{1}{2} (\hat{\mathbf{p}}_1-\hat{\mathbf{p}}_2 ),\\
\hat{\mathbf{R}} &=& \frac{1}{2} ( \hat{\mathbf{r}}_1+\hat{\mathbf{r}}_2 ) \qquad \mbox{and} \qquad\hat{\mathbf{r}} =  \hat{\mathbf{r}}_1-\hat{\mathbf{r}}_2.
\eea
The Hamiltonian $\hat{H}=\frac{\hat p_1^2}{2M}+\frac{\hat p_2^2}{2M}+ V_{\rm int}(|\hat{\mathbf{r}}_1 - \hat{\mathbf{r}}_2|)$ writes with these new coordinates:
\beq
H = \frac{1}{4M} \left( (\hat{\mathbf{p}}_1+\hat{\mathbf{p}}_2)^2 + (\hat{\mathbf{p}}_1-\hat{\mathbf{p}}_2)^2 \right) + V(\hat{\mathbf{r}}) = \frac{\hat{\mathbf{P}}^2}{4M} + \frac{\hat{\mathbf{p}}^2}{M} + V_{\rm int}(\hat{\mathbf{r}}).
\eeq
$\hat{\mathbf{P}}$ and $\hat{\mathbf{R}}$ describe the motion of the center of mass (mass $2M$) and $\hat{\mathbf{p}}$ and $\hat{\mathbf{r}}$ describe the relative motion, equivalent to the motion of a particle of mass $M/2$. The two motions are independent and we concentrate on the relative motion to describe the collision. The problem corresponds to the scattering of a  fictitious particle of mass $M/2$ on the potential $V_{\rm int}(r)$.

It is solved by looking at the long-distance steady state of the wave function, with $r\to\infty$, under the form:
\beq
\psi_{\mathbf{k}}(\mathbf{r}) \sim e^{i \mathbf{k}.\mathbf{r}} + f(k,\mathbf{n},\mathbf{n}') \frac{e^{i k r}}{r},
\eeq
with $\mathbf{k}=k\mathbf{n}$ corresponding to the incoming wave of energy $\hbar^2k^2/M$, and $\mathbf{r}=r\mathbf{n}'$ to the outgoing wave. $V_{\rm int}(r)$ being isotropic, the scattering amplitude $f(k,\theta)$ only depends on $k$ and on the angle $\theta = (\mathbf{n},\mathbf{n}')$ between the incoming and the outgoing waves. The scattering amplitude is linked to the interaction potential through
\beq
f(k,\mathbf{n},\mathbf{n}') = -\frac{M}{4\pi\hbar^2}\int d\mathbf{r}'\, e^{-ik\mathbf{n}'\cdot\mathbf{r}'} V_{\rm int} (r')\psi_{\mathbf{k}}(\mathbf{r}').
\label{eq:scatt_amplitude}
\eeq

The wave function can be written as a product of a function of $r$ and a function of $\theta$, and the angular part can be decomposed on the spherical harmonics. For partial waves with a non zero angular momentum $\ell>0$, the radial potential is modified by a centrifugal term $\hbar^2\ell(\ell+1)/Mr^2$, which constitutes a potential barrier and prevent atoms with a very small energy to reach the region where $V_{\rm int}(r)$ is non zero. As a result, at low energy only partial waves with $\ell=0$, also called s-waves by analogy with the atomic structure, can contribute to the scattering amplitude. As s-waves are isotropic, the scattering amplitude is independent of $\theta$. The \textit{scattering length} is given by the limit at low energy of $f(k)$:
\beq
a = -\lim_{k\to0} f(k).
\eeq

The scattering properties at low energy are entirely contained in this single parameter. We can then replace the true interaction potential $V_{\rm int}(\mathbf{r})$ by an effective potential that will give the same scattering length $a$. The simplest choice\footnote{In reality, the potential should be regularized in zero and take a more sophisticated form, see Ref. \cite{DalibardCF2021en}.} is
\beq
\boxed{
V_{\rm eff}(\mathbf{r}) = g\delta(\mathbf{r}),
\label{eq:int_pot}
}
\eeq
where we need to choose
\beq
\boxed{
g = \frac{4\pi\hbar^2 a}{M}
\label{eq:int_const}
}
\eeq
to get the correct scattering amplitude (to lowest order, $\psi_{\mathbf{k}}(\mathbf{r}')\sim e^{i \mathbf{k}.\mathbf{r}'}$ in Eq.~\eqref{eq:scatt_amplitude}).

\subsection{Gross-Pitaevskii equation}
To describe the effect of weak interactions on the condensate, we will take a mean-field approach. We will assume that the interactions are weak enough to neglect the correlations between particles in the many-body wave function \ket{\Psi}, and write it as a product state of $N$ identical single particle states \ket{\phi}, where $N$ is the number of atoms in the condensate:
\beq
\ket{\Psi} = \ket{\phi}_1\otimes\ket{\phi}_2\otimes\ket{\phi}_3\otimes\dots\otimes\ket{\phi}_N.
\eeq
However, we now allow that this single-particle state \ket{\phi} differs from the ground state of the single-particle Hamiltonian $h^{(1)}=p^2/2M + V(\mathbf{r})$, that we take with the kinetic term and an optional potential term. We then apply a variational method and look for the single particle wave function \ket{\phi} which would minimize the energy $E[\phi]$ for the state \ket{\Psi}, under the constraint $\braket{\phi}{\phi}=1$.

Using the effective potential introduced in the previous section, the many-body Hamiltonian for $N$ interacting atoms writes
\beq
H = \sum_{i=1}^N h_i^{(1)} + \frac{g}{2}\sum_{i\neq j}\delta(\mathbf{r}_i - \mathbf{r}_j)
\eeq
where $h_i^{(1)} = \mathbb{1}\otimes\mathbb{1}\otimes\dots\otimes h^{(1)}\otimes\mathbb{1}\otimes\dots\otimes\mathbb{1}$ with $h^{(1)}=p^2/2M + V(\mathbf{r})$ at the $i^{\rm th}$ position is the single-particle Hamiltonian acting on atom $i$. The second term, with $N(N-1)$ terms in the sum, describes the contact interactions.

The energy $E[\phi]$ writes, given the state \ket{\psi} of wave function $\phi(\mathbf{r})=\braket{\mathbf{r}}{\phi}$:
\bea
E[\phi]&=&N\int d\mathbf{r}\left\{\phi^*(\mathbf{r})\left[-\frac{\hbar^2}{2M}\nabla^2\phi\right]+\phi^*(\mathbf{r})V(\mathbf{r})\phi(\mathbf{r})\right\}\nonumber\\
&& + \frac{g}{2}N(N-1)\int d\mathbf{r} \, d\mathbf{r}' \, \phi^*(\mathbf{r}')\phi^*(\mathbf{r})\delta(\mathbf{r}-\mathbf{r}')\phi(\mathbf{r})\phi(\mathbf{r}')\\
&\simeq&N\int d\mathbf{r}\left\{\phi^*(\mathbf{r})\left[-\frac{\hbar^2}{2M}\nabla^2\phi\right]+\phi^*(\mathbf{r})V(\mathbf{r})\phi(\mathbf{r})\right\}+ g\frac{N^2}{2}\int d\mathbf{r} \, \phi^*(\mathbf{r})^2\phi(\mathbf{r})^2.\nonumber
\eea
In the last line, we have replaced $N(N-1)$ by $N^2$, which is valid if the condensate contains many atoms. Instead of minimizing $E[\phi]$ with the constraint
\beq
\int d\mathbf{r}\phi^*(\mathbf{r})\phi(\mathbf{r})=1,
\eeq
we use the Lagrange multiplier approach and minimize
\beq
F(\phi,\phi^*) = E[\phi] - \lambda N \int d\mathbf{r}\phi^*(\mathbf{r})\phi(\mathbf{r}),
\eeq
where $\phi$ and $\phi^*$ act as independent variables. The change in $F$ when $\phi^*$ is modified by an infinitesimal amount $\delta\phi^*$ writes
\beq
\frac{\delta F}{N} = \int d\mathbf{r}\delta\phi^*(\mathbf{r})\left[-\frac{\hbar^2}{2M}\nabla^2\phi+V(\mathbf{r})\phi(\mathbf{r})-\lambda\phi(\mathbf{r})+gN\phi^*(\mathbf{r})\phi(\mathbf{r})^2\right].
\eeq
$F$ is minimum if $\phi$ obeys the following equation:
\beq
-\frac{\hbar^2}{2M}\nabla^2\phi+V(\mathbf{r})\phi(\mathbf{r})+gN|\phi(\mathbf{r})|^2\phi(\mathbf{r})=\lambda\phi(\mathbf{r}).
\label{eq:GPE1}
\eeq

There is a simple interpretation of the Lagrange multiplier $\lambda$. The energy of the ground state with $N$ particles writes
\beq
E(N) = NE_1 + \frac{N(N-1)}{2}E_2
\eeq
with $E_1$ the single-particle energy corresponding to \ket{\phi} and $E_2$ the two-particle energy for two atoms in state \ket{\phi}. The energy needed to add a particle in the condensate is given by
\beq
E(N+1)-E(N) = E_1 + N\frac{N+1 - (N-1)}{2}E_2 = E_1 + N E_2 = \lambda.
\eeq
The last equality comes directly from the integration of Eq.~\eqref{eq:GPE1} after multiplication by $\psi^*(\mathbf{r})$. In other words, $\lambda$ is the \textit{chemical potential} of the gas.

Introducing the wave function normalized to $N$ atoms $\psi=\sqrt{N}\phi$ and using $\mu$ as a notation for the \textit{chemical potential}, we arrive at the usual form of the time-independent \textit{Gross-Pitaevskii equation}:
\beq
\boxed{
-\frac{\hbar^2}{2M}\nabla^2\psi+V(\mathbf{r})\psi(\mathbf{r})+g|\psi(\mathbf{r})|^2\psi(\mathbf{r})=\mu\psi(\mathbf{r}).
\label{eq:GPE_st}
}
\eeq

Under this form, the square modulus of the wave function is simply the atomic density of the condensate:
\beq
\boxed{
n(\mathbf{r})=|\psi(\mathbf{r})|^2.
\label{eq:density}
}
\eeq
\subsection{Thomas-Fermi limit}
It is easy to check using a Gaussian ansatz for $\psi$ that the condensate would be unstable with attractive interactions, i.e. $g<0$, unless the atom number is below some maximum atom number \cite{Castin2001}. This maximum atom number being quite small (typically of the order of 100 atoms), we will only consider the case of repulsive interactions in this lecture, i.e. $g>0$.

Let us estimate the three contributions to the energy per particle in the Gross-Pitaevskii equation (GPE), assuming that the condensate with $N$ atoms has a size of order $R$. We get (in 3D):
\bea
E_k&\simeq&\frac{\hbar^2}{2MR^2}\label{eq:Ek}\\
E_{\rm int}&\simeq& \frac{gN}{R^3} = 8\pi\frac{\hbar^2}{2M}\frac{Na}{R^3}\label{eq:Eint}\\
E_{\rm pot}&\simeq& V(R) = \frac{1}{2}M\omega_0^2R^2 = \frac{\hbar^2}{2M}\frac{R^2}{a_0^4}
\eea
if we assume a harmonic potential of frequency $\omega_0$ and $a_0=\sqrt{\hbar/M\omega_0}$ is the size of its ground state. These energies can be recast in units of $\hbar\omega_0 = \hbar^2/Ma_0^2$:
\bea
E_k&\simeq&\frac{\hbar\omega_0}{2}\frac{a_0^2}{R^2}\\
E_{\rm int}&\simeq& \frac{\hbar\omega_0}{2}\times 8\pi\frac{Na}{a_0}\frac{a_0^3}{R^3}\\
E_{\rm pot}&\simeq& \frac{\hbar\omega_0}{2}\frac{R^2}{a_0^2}.
\eea
If $N$ is sufficiently large, the interaction term will be large, and the size of the condensate $R$ will be significantly larger than the harmonic ground state $a_0$ due to the repulsive interactions. In this limit, both the interaction term and the potential term (which scales as $R^2/a_0^2$) will be much larger than the kinetic term, which instead is reduced by a factor $a_0^2/R^2$ with respect to the non interacting situation. For a harmonic trap, the potential energy will be much larger than $\hbar\omega_0$, indicating that interactions induce a population of the excited states of the harmonic oscillator.

The Thomas-Fermi approximation consists in neglecting the kinetic term in the GPE \eqref{eq:GPE_st}, which becomes a simple equation for $\psi$ (no derivative). This justified if $8\pi Na \gg R$, see Eqs.~\eqref{eq:Ek} and \eqref{eq:Eint}. We can simplify by $\psi$, which leads to the solution for the density:
\beq
\boxed{
n_{\rm TF}(\mathbf{r}) = \frac{1}{g}\left[\mu - V(\mathbf{r})\right], \qquad \mbox{for } V(\mathbf{r})\leq\mu.
\label{eq:density_TF_regime}
}
\eeq

\begin{figure}[ht]
\centering
\includegraphics[width=0.45\linewidth]{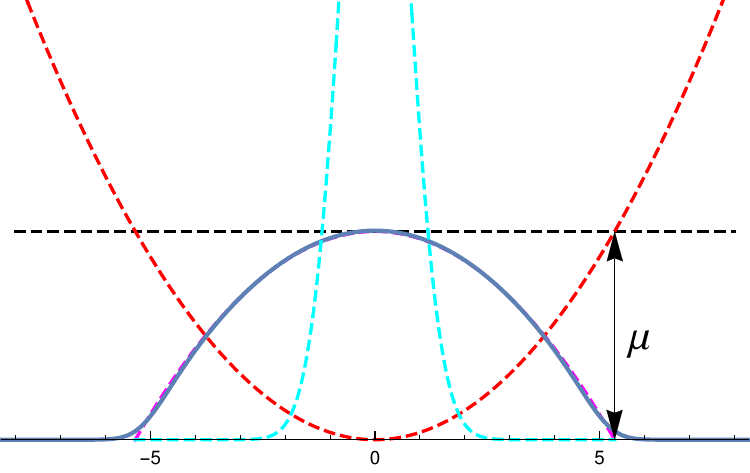}\includegraphics[width=0.45\linewidth]{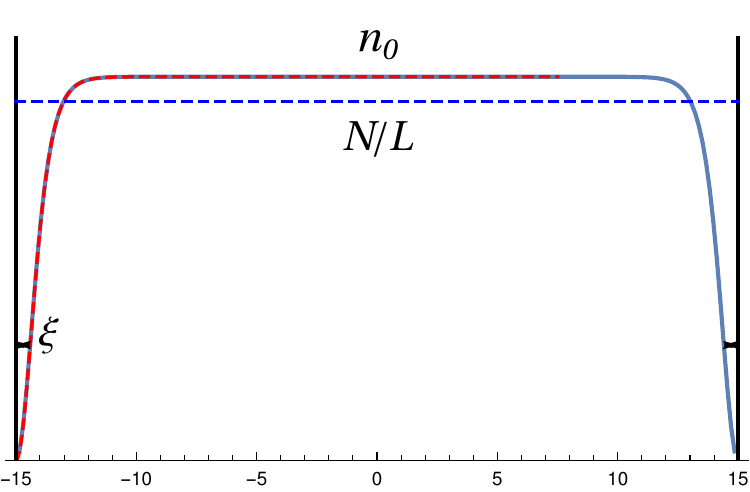}
\caption{\textbf{Left}: Solution of Gross-Pitaevskii equation for a condensante in a harmonic trap (full blue line), compared to the Thomas-Fermi profile (dashed magenta, very well superposed except on the edges) and the non interacting solution (dashed cyan). Also shown are the trapping potential (dashed red) and the chemical potential (dashed black). \textbf{Right}: Same thing in a box potential. The density is uniform, except in a small region of size $\xi$ near the edge.\label{fig:TF_profile}}
\end{figure}

The condensate thus takes a shape that is the opposite of the potential ---a parabola with a maximum in the trap center if the trap is harmonic, see Fig.~\ref{fig:TF_profile}. The condition $V(\mathbf{r})\leq\mu$ sets the limits of the condensate, where the density vanish. For an isotropic trap for instance, the radius is given by $V(R)=\mu$. Around this region, of course, the term $gn(\mathbf{r})\simeq \mu - V(\mathbf{r})$ in the GPE becomes very small, and the kinetic term must play a role.

The size of this region can be estimated easily in the case of a cubic box trap. The size $\xi$ over which the density drops is found by equating the kinetic energy due to the wave function bending, $\hbar^2/2M\xi^2$, and the interaction energy $gn_0=\mu$. We find
\beq
\boxed{
\xi = \frac{\hbar}{\sqrt{2M\mu}},
}
\label{eq:xi}
\eeq
which is called the \textit{healing length}. Near one of the edges, for example in $(x,L/2,L/2)$ with $x\ll L$ if the box is defined by $x,y,z\in (0,L)$, the wave function is given by
\beq
\psi(x,0,0)\simeq\sqrt{n_0} \tanh\left(\frac{x}{\xi\sqrt{2}}\right).
\eeq
Its square (the density) is represented in Fig.~\ref{fig:TF_profile}, right. The central density is a little higher due to this depletion at the edges, on the order of $n_0\simeq N/(L-2\xi)^3$.

\noindent\hrulefill\\
\textbf{Exercise 2}: using the Thomas-Fermi approximation, give the expression for the chemical potential $\mu$ in the case of atoms confined in a harmonic trap in dimension 2 and 3, with trapping frequencies $\omega_x$ and $\omega_y$ (and $\omega_z$). Deduce the Thomas-Fermi radius $R_i$ in each direction, i.e. the radius at which the density vanishes.\\
\noindent\hrule

\section{Time-dependent GPE and hydrodynamic equation}
\subsection{Time-dependent GPE}
The time-dependent version of GPE is very similar to its time-independent counterpart Eq.~\eqref{eq:GPE_st} \cite{Pitaevskii2003,Castin2001}:
\beq
\boxed{
i\hbar\partial_t\psi = -\frac{\hbar^2}{2M}\nabla^2\psi+V(\mathbf{r})\psi(\mathbf{r})+g|\psi(\mathbf{r})|^2\psi(\mathbf{r}).
\label{eq:GPE}
}
\eeq
If a wave function $\psi_0$ satisfies the time-independent GPE, then $\psi(\mathbf{r},t) = \psi_0(\mathbf{r})\exp(-i\mu t/\hbar)$ is a solution of the time-dependent GPE, Eq.~\eqref{eq:GPE}. It corresponds to the ground state of the condensate at rest. Beyond this case, the time-dependent GPE will allow us to explore the excitations and the dynamics of the condensate.

\subsection{Hydrodynamic formulation}
We can give an equivalent of Eq.~\eqref{eq:GPE}, which implies a single complex classical field, in terms of coupled hydrodynamic equations on two real classical fields, the density and the velocity field.

Let us write the wave function in terms of density $n=|\psi|^2$ and phase:
\beq
\psi(\mathbf{r},t) = \sqrt{n(\mathbf{r},t)}e^{i \theta(\mathbf{r},t)}.
\eeq
We will inject this expression into \eqref{eq:GPE}, multiply by $\psi^*$ and get two equations for the imaginary part and the real part.

We start from the easy terms that do not involve derivatives. We get a real term:
\beq
\psi^*(\mathbf{r})\left[V(\mathbf{r})\psi(\mathbf{r})+g|\psi(\mathbf{r})|^2\psi(\mathbf{r})\right] = n(\mathbf{r})\left[V(\mathbf{r})+gn(\mathbf{r})\right].
\eeq

Then we examine the Laplacian term:
\bea
\psi^*\nabla^2\psi &=& \sqrt{n}e^{-i\theta}\nabla\cdot\left[\nabla\left(\sqrt{n}e^{i\theta}\right)\right]\nonumber\\
&=&\sqrt{n}e^{-i\theta}\nabla\cdot\left[\left(\nabla\sqrt{n}\right)e^{i\theta}+i\sqrt{n} e^{i\theta}\nabla \theta\right]\nonumber\\
&=&\sqrt{n}e^{-i\theta}\left[\left(\nabla^2\sqrt{n}\right)e^{i\theta}+2i e^{i\theta}\nabla\left(\sqrt{n}\right)\cdot\nabla \theta+i\sqrt{n} e^{i\theta}\nabla\cdot\nabla \theta - \sqrt{n}e^{i\theta}\nabla \theta^2\right]\nonumber\\
&=&n\left[\frac{\nabla^2\sqrt{n}}{\sqrt{n}}-\nabla \theta^2\right]+i\left(\nabla n\cdot\nabla\theta + n\nabla\cdot\nabla\theta \right).
\eea
By analogy with the probability current in quantum mechanics $\mathbf{J}=i\hbar(\psi\nabla\psi^*-\psi^*\nabla\psi)/2M$, we introduce the fluid velocity
\beq
\boxed{
\mathbf{v}=\frac{\hbar}{M}\nabla\theta.
}
\label{eq:velocity}
\eeq
Using this notation, the Laplacian writes finally
\beq
-\frac{\hbar^2}{2M}\psi^*\nabla^2\psi = n\left[-\frac{\hbar^2}{2M}\frac{\nabla^2\sqrt{n}}{\sqrt{n}}+\frac{1}{2}Mv^2\right]-i\frac{\hbar}{2}\left(\nabla n\cdot\mathbf{v} + n\nabla\cdot\mathbf{v} \right).
\eeq
The time derivative of the left-hand-side writes
\bea
i\hbar\psi^*\partial_t\psi&=&i\hbar\sqrt{n}e^{-i\theta}\left[\frac{1}{2\sqrt{n}}\partial_t n e^{i\theta} + i\partial_t\theta\sqrt{n}e^{i\theta}\right]\nonumber\\
&=&i\frac{\hbar}{2}\partial_t n - n\hbar \partial_t\theta.
\eea

Identifying the imaginary terms, we get the \textit{continuity equation}:
\beq
\boxed{
\partial_t n + \nablagras\cdot(n\mathbf{v}) = 0.
}
\label{eq:continuity}
\eeq
This equation describes the conservation of the flow.

Simplifying the real parts by $n$ and taking the gradient, we arrive at a \textit{Euler-type equation}:
\beq
\boxed{
M\partial_t\mathbf{v} = -\nablagras\left[ -\frac{\hbar^2}{2M}\frac{\nablagras^2\sqrt{n}}{\sqrt{n}}+\frac{1}{2}Mv^2 + V + gn \right].
}
\label{eq:Euler}
\eeq
Eqs.~\eqref{eq:continuity} and \eqref{eq:Euler} together are equivalent to Eq.~\eqref{eq:GPE}. Eq.~\eqref{eq:Euler} can also be recast into
\beq
M\frac{D\mathbf{v}}{dt} = -\nablagras\left[ -\frac{\hbar^2}{2M}\frac{\nablagras^2\sqrt{n}}{\sqrt{n}} + V + gn \right]
\eeq
where
\beq
\frac{D\mathbf{v}}{dt} = \partial_t\mathbf{v} + \mathbf{v}\cdot\nablagras \mathbf{v}
\eeq
is the particle derivative in the flow. The right-hand side has a potential energy, the interaction term $gn$ that plays the role of the pressure and a quantum pressure term involving $\hbar$ which has no classical counterpart. This term plays a role on smaller scales (the healing length $\xi$ typically) and is responsible for the edges of the wave function in a trap.

Neglecting this term, which is less restrictive than the Thomas-Fermi approximation that would also neglect the $v^2$ term, leads to a description of the condensate dynamics at larger scales. This is the \textit{hydrodynamic approximation}, useful for example to derive the collective modes of a trapped Bose gas.

\subsection{Wave-like excitations}
The time-dependent GPE, or the hydrodynamics equations, gives us access to the elementary excitations of the condensate. Let us first consider the easy case of a homogeneous gas. At equilibrium, the density is $n_0=\mu/g$, independent of position, and the velocity field is zero. We look for small amplitude excitations. We write the density as $n(\mathbf{r},t) = n_0 + \delta n(\mathbf{r},t)$ and we will keep only first order terms in $\delta n$ and $\mathbf{v}$.

Under these assumptions, the continuity equation at first order writes
\beq
\partial_t\delta n + n_0 \nablagras\cdot\mathbf{v} = 0.
\label{eq:cont_1st_order}
\eeq
Using the expansion $\sqrt{n} = \sqrt{n_0} + \delta n/2\sqrt{n_0}$, we get for the Euler equation at first order:
\beq
M\partial_t \mathbf{v} = -\nablagras\left[-\frac{\hbar^2}{2M}\frac{\nablagras^2 \delta n}{2n_0}+g\delta n\right].
\label{eq:Euler_1st_order}
\eeq
Taking the time derivative of Eq.~\eqref{eq:cont_1st_order}, we get
\beq
\partial_t^2 n =\frac{n_0}{M}\nablagras^2\left[-\frac{\hbar^2}{2M}\frac{\nablagras^2 \delta n}{2n_0}+g\delta n\right]=\nablagras^2\left[-\frac{\hbar^2}{4M^2}\nablagras^2 \delta n+\frac{gn_0}{M}\delta n\right].
\eeq

We look for solutions of the form $A\cos(\omega t-\mathbf{k}\cdot\mathbf{r}+\varphi)$ for both the density $\delta n$ and the velocity, which leads to the following dispersion relation for the small amplitude excitations:
\beq
\boxed{
\omega(k) = \sqrt{\frac{\hbar^2k^4}{4M^2}+\frac{gn_0k^2}{M}} = \sqrt{\frac{\hbar^2k^4}{4M^2}+c^2k^2},
}
\label{eq:Bogo}
\eeq
where we have introduced the quantity
\beq
\boxed{
c=\sqrt{\frac{gn_0}{M}} = \sqrt{\frac{\mu}{M}}
\label{eq:speed_sound}
}
\eeq
which has the dimension of a velocity.

\begin{figure}
\centering
\includegraphics[width=0.5\linewidth]{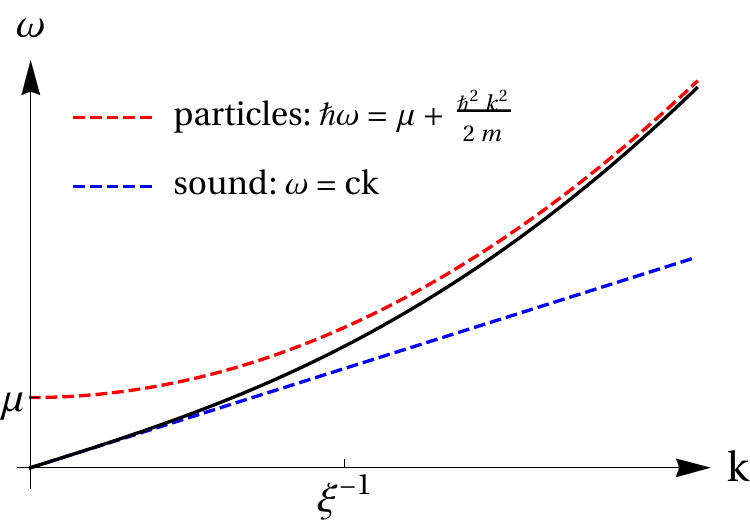}
\caption{Bogolubov spectrum in a homogeneous gas.\label{fig:Bogo_spectrum}}
\end{figure}

The relation $\omega(k)$ in Eq.~\eqref{eq:Bogo} is known as the \textit{Bogolubov spectrum}. It is represented in Fig.~\ref{fig:Bogo_spectrum}. It behaves differently at low or high momenta:\\
\quad$\bullet$ In the low momentum limit, the term in $k^2$ dominates and the frequency writes approximately $\omega(k)=ck$. The dispersion relation is thus linear. This corresponds to \textit{sound waves}, with a speed of sound given by $c$. This is very different from the ideal gas, which has a dispersion relation in $k^2$.\\
\quad$\bullet$ In the high momentum limit, the term in $k^4$ is larger. We can make an expansion to first order in $c^2k^2$, and get
\beq
\hbar\omega(k)= \frac{\hbar^2k^2}{2M}\sqrt{1+\frac{4M}{\hbar^2k^2}\mu}\simeq\frac{\hbar^2k^2}{2M}+\mu.
\eeq
The spectrum corresponds to free particles with an energy shifted by $\mu = 2\mu-\mu$ due to the exchange energy: the interaction of the excited (discernable) particle with the majority atoms in the condensate is $2gn_0=2\mu$ while removing a particle from the condensate decreases the energy by $\mu$.

The boundary between these two regimes occurs for
\beq
\frac{\hbar^2k^2}{2M} = \mu \qquad \mbox{or} \qquad k = \frac{\sqrt{2M\mu}}{\hbar}=\xi^{-1}.
\eeq

\subsection{A first criterion for superfluidity: existence of a critical velocity}
The linear behavior of the dispersion relation has an important consequence. Indeed, let us consider a condensate at rest, in which we launch a very small object of mass $m$ at speed $\mathbf{v}$. At which condition will the motion of this object be damped through the creation of some excitation in the fluid? Note that the situation is equivalent to the one of the fluid moving at $-\mathbf{v}$, if we look in the frame moving at $-\mathbf{v}$, such that it also describes the possible damping of the flow by excitation of the fluid due to small imperfections in a pipeline.

The problem is translationally invariant (at least along $\mathbf{v}$), and energy and total momentum are conserved. Let us write energy and momentum conservation (we take $T=0$):
\bc
\begin{tabular}{l|c|c|}
&Before excitation&After excitation\\
\hline
&&\\[-3mm]
Energy& $\ds 0 + \frac{1}{2}mv^2$ & $\ds c\hbar k + \frac{1}{2}mv'^2$\\[3mm]
\hline
&&\\[-4mm]
Momentum& $\mathbf{0}+m\mathbf{v}$& $\hbar\mathbf{k} + m\mathbf{v'}$\\[1mm]
\hline
\end{tabular}
\ec
Equating the total momentum before and after excitation gives the new velocity $\mathbf{v}'=\mathbf{v}-\hbar\mathbf{k}/m$. Reporting this expression into the condition for energy conservation, we get
\beq
\frac{1}{2}mv^2=c\hbar k + \frac{1}{2}mv^2 - \hbar\mathbf{k}\cdot\mathbf{v}+\frac{\hbar^2k^2}{2m} \Leftrightarrow \mathbf{v}\cdot\mathbf{k} = ck+\frac{\hbar^2k^2}{2m}.
\eeq
As $\hbar^2k^2/2m$ is always positive, the last equality implies $\mathbf{v}\cdot\mathbf{k} \geq ck$ and thus $v\geq c$. The speed of sound $c$ appears to be a \textit{critical velocity} for the creation of excitations. If the object has a velocity smaller than $c$, its motion is not damped ---or if the fluid flows at a speed smaller than $c$, it is not damped. This corresponds to the \textit{Landau criterion for superfluidity}, a first hint of superfluidity for quantum gases. As the speed of sound is proportional to $\sqrt{g}$, it vanishes in the absence of interactions. Interactions are required for the quantum gas to be a superfluid.

\begin{figure}
\centering
\includegraphics[width=0.25\linewidth]{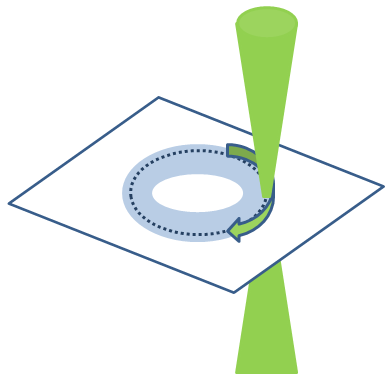}
\begin{tikzpicture}[font=\footnotesize]
\node at (0,0) {\includegraphics[width=0.4\linewidth]{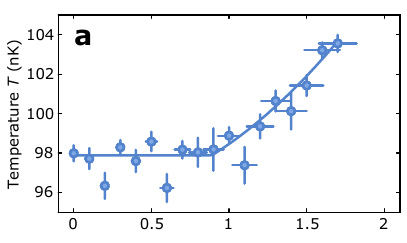}};
\node at (0.5,-1.9) {\sf Excitation velocity \textit{v} (mm/s)};
\end{tikzpicture}
\caption{Demonstration of a critical velocity in a 2D gas. \textbf{Left}: A focused laser is stirred around the gas, at some fixed distance from the center, and at a fixed linear speed $v$. \textbf{Right}: The energy transferred to the gas is measured in a time-of-flight experiment. At low $v$, no energy is transferred. Above a critical value $v_c$, the energy increases quadratically with $v-v_c$, due to the creation of excited particle out of the BEC. Figure from Ref.~\cite{Desbuquois2012}.\label{fig:vc}}
\end{figure}

The existence of a critical velocity in quantum gases has been demonstrated by the group of Wolfgang Ketterle in 1999. A nice illustration is given in Fig.~\ref{fig:vc}, from an experiment performed in the group of Jean Dalibard in Paris. The measured critical velocity $v_c$ is below $c$ as the focused laser beam used to create the excitation also depletes the density locally, resulting in a smaller local speed of sound.

\subsection{Collective modes in a trap}
We consider now the case where the gas is confined in a trap $V(\mathbf{r})$, and we look again for the elementary excitations away from the density at equilibrium given by $n_0(\mathbf{r})\simeq[\mu-V(\mathbf{r})]/g$, and write $n(\mathbf{r},t)=n_0(\mathbf{r})+\delta n(\mathbf{r},t)$. In the trap, with a large enough atom number, we can neglect the quantum pressure and use the hydrodynamic approximation, such that the hydrodynamic equations read
\bea
&&\partial_t n + \nablagras(n\mathbf{v}) = 0,\\
&&M\partial_t\mathbf{v} = -\nablagras\left[\frac{1}{2}Mv^2+V(\mathbf{r})+gn\right].
\eea
We keep only the first order terms in $\delta n$ and $\mathbf{v}$:
\bea
&&\partial_t \delta n + \nablagras(n_0\mathbf{v}) = 0,\\
&&M\partial_t\mathbf{v} = -\nablagras\left[V(\mathbf{r})+gn_0 + g\delta n\right].\label{eq:Euler_hydro_trap}
\eea
We use $V(\mathbf{r})+gn_0 \simeq \mu$ to neglect its gradient in Eq.~\eqref{eq:Euler_hydro_trap}. We take again the time derivative of the continuity equation and inject it in Euler's equation. We get \cite{Stringari1996}
\beq
\partial_t^2\delta n = \nablagras\cdot\left[ \frac{gn_0(\mathbf{r})}{M}\nablagras\delta n\right] = \nablagras\cdot\left[ c^2(\mathbf{r})\nablagras\delta n\right].
\eeq
We have introduced the local speed of sound $c(\mathbf{r})=\sqrt{gn_0(\mathbf{r})/M}$.

In the case of an isotropic 3D harmonic trap of frequency $\omega_0$, the rotational invariance allows us to give the frequencies of the collective modes as a function of three quantum numbers $(n_r,\ell,m)$ for the number radial nodes, total angular momentum $\ell$ and its projection $m=-\ell\dots\ell$ along $z$. The frequency depends only on $n_r$ and $\ell$ and reads \cite{Stringari1996}
\beq
\omega_{n_r,\ell}=\omega_0\sqrt{2n_r^2+2n_r\ell+3n_r+\ell}.
\label{eq:3D_modes}
\eeq
These frequencies get a small correction if we take into account the terms beyond the Thomas-Fermi approximation \cite{Pitaevskii1998}.

In two dimensions, the relevant quantum numbers are $n_r$ and $m$, and we get almost the same formula (notice however the factor 2 instead of 3) \cite{Stringari1998}
\beq
\omega_{n_r,m}=\omega_0\sqrt{2n_r^2+2n_r |m|+2n_r+|m|}.
\label{eq:2D_modes}
\eeq

Let us discuss a few important modes:
\begin{itemize}
\item The lowest frequency modes are the center-of-mass or dipole modes, at the trap frequencies.
\item The quadrupole modes, corresponding to a quadrupole deformation of the trap with $n_r=0$ and $m=\pm 2$ ($\ell=2$ in the 3D case), oscillate at frequency $\sqrt{2}\omega_0$. They are specific of a superfluid.
\item The first mode with $n_r=1$ (and $m=0$ of $\ell=0$) is the monopole or breathing mode. Its frequency is $\sqrt{5}\omega_0$ in 3D and $2\omega_0$ in 2D. In this latter case, it is the same as the monopole frequency of a thermal gas. We will see in Lecture 2 that this mode is not damped in 2D, due to the underlying scaling symmetry \cite{Pitaevskii1997,Chevy2002}.
\end{itemize}

In anisotropic traps, another mode can also exist: an oscillation along a trap axis, called the scissors mode \cite{GueryOdelin1999}. This mode is also a signature of superfluidity and can be used to probe the superfluid state \cite{Marago2000,DeRossi2016}.

\subsection{Another kind of excitation: vortices}
\begin{figure}
\centering
\begin{tikzpicture}
\node at (0,0) {\includegraphics[width=0.25\linewidth]{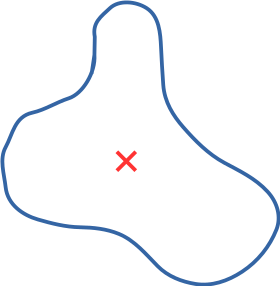}};
\node at(-1,1) {\textcolor{tango}{$\mathcal{C}$}};
\node[red] at (0,-0.6) {$O$};
\node at (-0.4,0) {$q$};
\draw[->,thick] (1.75,-0.7) -- (1.25,-0.2);
\node at (1.3,-0.8) {$\mathbf{d\ell}$};
\draw[->,thick] (1.75,-0.7) -- (1.75,0.2);
\node at (1.95,-0.2) {$\mathbf{v}$};
\end{tikzpicture}
\hspace{0.15\linewidth}\includegraphics[width=0.25\linewidth]{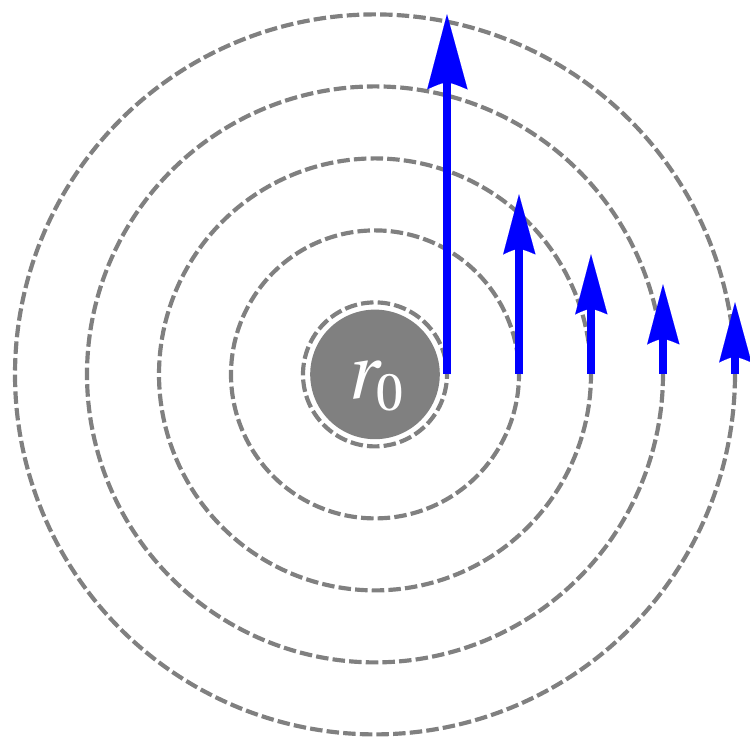}
\caption{\textbf{Left}: Circulation of the velocity field around a contour $\mathcal{C}$. The integration is done on a path where the velocity is well-defined, while there may be points inside the contour, here point $O$, where the density vanishes and the velocity is not defined, which may be a vortex of charge $q$. The circulation is an integer number of $h/M$, $qh/M$ in this latter case. \textbf{Right}: Typical velocity field around a vortex. The central region of radius $r_0\sim\xi$ is depleted and the density vanishes in the center of the core. \label{fig:circulation}}
\end{figure}
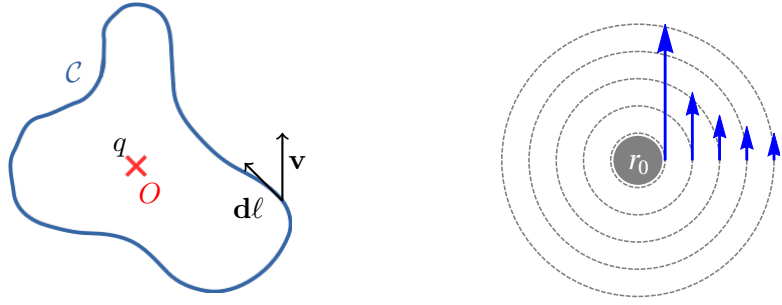

The velocity of the fluid, given by Eq.~\eqref{eq:velocity}, is proportional to the gradient of the phase, which is defined everywhere except at the points where $\psi$ vanishes. Outside these singularities, $\mathbf{v}$ is well-defined, and its rotational is equal to zero:
\beq
\nablagras\times\mathbf{v}=0.
\eeq
As a consequence of the expression of $\mathbf{v}$, we will see that the circulation $\Gamma_\mathcal{C}$ of the velocity along a close contour $\mathcal{C}$ that does not cross a singularity is quantized. Consider first the 2D case (see Fig.~\ref{fig:circulation}, left):
\beq
\Gamma_\mathcal{C} = \oint_\mathcal{C} \mathbf{v}\cdot\mathbf{d\ell} = \frac{\hbar}{M}\oint_\mathcal{C} \nablagras\theta\cdot\mathbf{d\ell}=\frac{\hbar}{M}\Delta\theta,
\eeq
where $\Delta\theta$ is the phase difference of the wave function $\psi$ after a close loop. $\psi$ is singly-valued in a given point while its phase $\theta$ is defined modulo $2\pi$, which means that $\Delta\theta$ should be a multiple of $2\pi$, say $q2\pi$ with $q\in\mathbb{Z}$. We find finally
\beq
\Gamma_\mathcal{C} = q\frac{h}{M}, \qquad q\in\mathbb{Z}.
\eeq
The circulation of the velocity is quantized in units of $h/M$, the \textit{quantum of circulation}. This is a striking difference with respect to classical fluids, where the circulation can take any value.

A singularity around which the circulation has a value $qh/M$, such as the point $O$ in Fig.~\ref{fig:circulation}, is called a \textit{vortex of charge $q$}.  The density vanishes on this point and recovers its background value on a characteristic size $R_v$, the vortex radius. Around a vortex, the velocity field is rotating. If we assume that, at least locally around the vortex, the velocity is tangential with a modulus that only depends to the distance to the vortex core $\mathbf{v}(\mathbf{r})=v(r)\mathbf{e}_\varphi$ in polar coordinates $(r,\varphi)$, see Fig.~\ref{fig:circulation}, right, we can deduce the value of $v(r)$ for a given circulation. Let us compute the circulation over a circle of radius $r$:
\bea
\Gamma_\mathcal{C} &=& \int_0^{2\pi} d\varphi v(r) r = 2\pi r v(r) = q\frac{h}{M}\nonumber\\
\Rightarrow \quad v(r)&=&q\frac{\hbar}{Mr}.\label{eq:vortex_velocity}
\eea

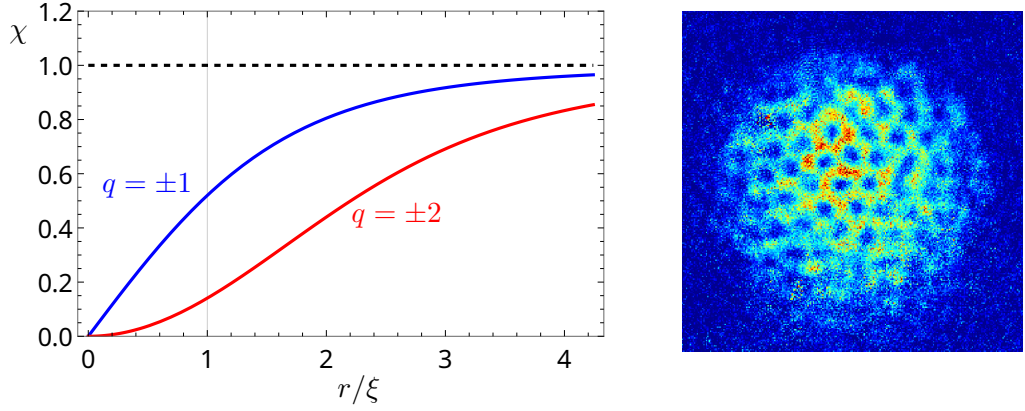
\begin{figure}
\centering
\begin{tikzpicture}
\node at (0,0) {\includegraphics[width=0.5\linewidth]{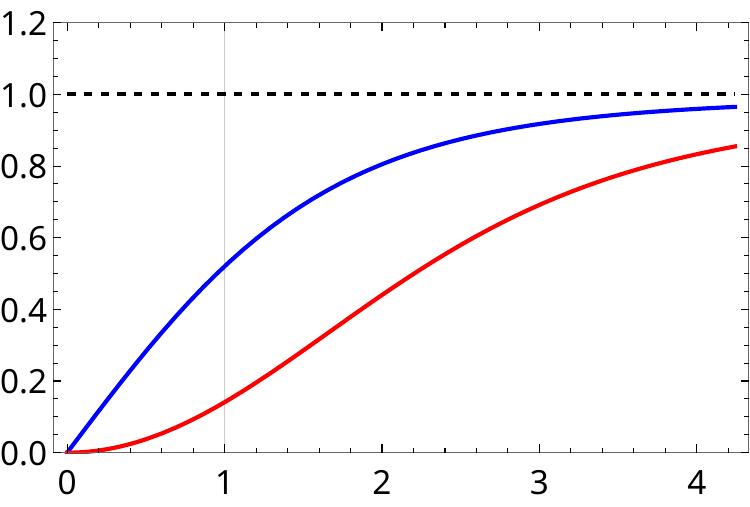}};
\node at (-4,2) {$\chi$};
\node at (0.5,-2.7) {$r/\xi$};
\node[blue] at (-2.3,0) {$q=\pm 1$};
\node[red] at (1,-0.4) {$q=\pm 2$};
\end{tikzpicture}
\qquad\raisebox{0.055\linewidth}{\includegraphics[width=0.3\linewidth]{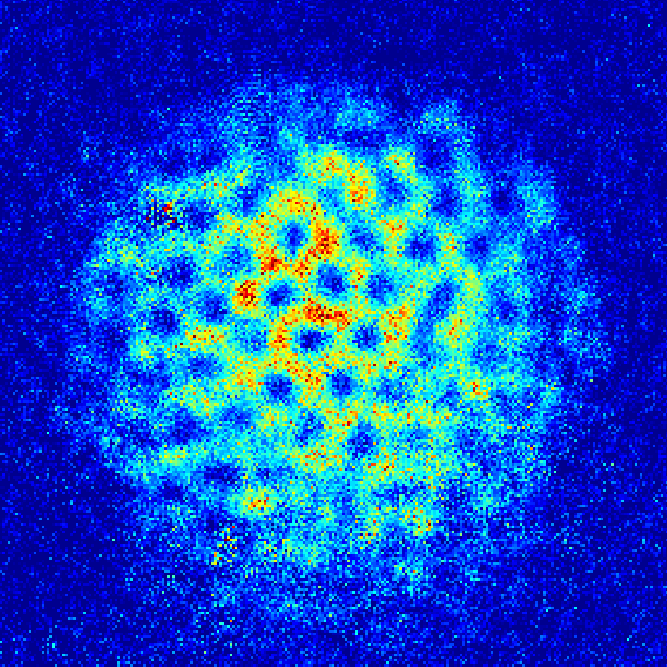}}
\caption{\textbf{Left}: Modulus of the wave function $\chi(r)$ of a vortex of charge $q=\pm 1$ (blue line) and of charge $q=\pm 2$ (red line), as a function of the distance from the core in units of $\xi$. \textbf{Right}: Vortex lattice with approximately 50 singly-charged vortices obtained by rotating a quasi two-dimensional gas in a harmonic trap. Figure from LPL, see also Ref.~\cite{Sharma2024}.\label{fig:vortex_wf}}
\end{figure}

If the fluid is in 3D and not in 2D, singular vortex points are replaced by vortex lines along which the density vanishes and around which the fluid rotates. To estimate the size of the vortex core, we assume that the density doesn't depend on $z$ and write the wave function $\psi$ as a modulus that depends only on the distance $r$ to the vortex core, and a phase winding $q\varphi$ that leads to the correct velocity given at Eq.~\eqref{eq:vortex_velocity}:
\beq
\psi = \sqrt{\frac{\mu}{g}}\chi(r)e^{iq\varphi}
\eeq
where we have introduced the density $\mu/g$ far from the vortex core, linked to the healing length through $\mu=\hbar^2/2M\xi^2$, see Eq.~\eqref{eq:xi}. We write the GPE in cylindrical coordinates\footnote{The Laplacian of a function $f(r,\varphi,z)$ in cylindrical coordinates writes $\nablagras^2 f = \partial_r^2 F + \frac{1}{r}\partial_r f + \frac{1}{r^2}\partial_\varphi^2 f + \partial_z^2 f$.}
\beq
\mu\sqrt{\frac{\mu}{g}}\chi e^{iq\varphi} = -\frac{\hbar^2}{2M}\left(\chi''+\frac{1}{r}\chi' - \frac{q^2}{r^2}\chi\right)\sqrt{\frac{\mu}{g}} e^{iq\varphi}+g\frac{\mu}{g}\sqrt{\frac{\mu}{g}}\chi^3 e^{iq\varphi}
\eeq
where the term in $q^2$ comes from the second derivative along $\varphi$, and $\chi'$ and $\chi''$ are the derivatives of $\chi(r)$. We simplify by $\sqrt{\mu/g}e^{iq\varphi}$ and replace $\mu$ by $\hbar^2/2M\xi^2$ to get
\beq
\chi  = -\xi^2\left(\chi''+\frac{1}{r}\chi' - \frac{q^2}{r^2}\chi\right)+\chi^3.
\eeq
We see that the natural length appearing to describe the shape of a vortex is $\xi$, the healing length. The solution is shown in Fig.~\ref{fig:vortex_wf}, left, for two values of $|q|$.

Let us finally estimate the energy of a cylindrical condensate of radius $R$ with a vortex of charge $q$ in its center. The density is homogeneous almost everywhere, $n_0\simeq N/(\pi R^2 L)$, except in the core where it drops to zero. We will model the density profile as a step function, with zero density in a central cylinder of radius $|q|\xi$ and $n_0$ elsewhere.

Let us estimate the energy cost of the nucleation of a single, centered vortex. There is an additional interaction energy $E_{\rm int}$, required to `drill a hole' in the density because the density has to increase everywhere else by $\Delta n$, of order $N g \Delta n/2$. This estimation gives
\beq
E_{\rm int}\simeq N\frac{g}{2}\left[\frac{N}{\pi L(R^2-q^2\xi^2)} - \frac{N}{\pi L R^2}\right]\simeq N\frac{g n_0}{2}\frac{q^2\xi^2}{R^2}=q^2\frac{\pi}{4}\frac{\hbar^2}{M}\frac{N}{\pi R^2}\sim q^2\frac{\hbar^2n_{2D}}{M}
\eeq
where we have introduced the 2D, transversally integrated density $n_{2D}=n_0L=N/(\pi R^2)$.

The dominant energy is however the total kinetic energy
\beq
E_q = n_0 L \int_{|q|\xi}^R 2\pi r \, dr \, \frac{1}{2}Mv(r)^2 = \pi n_0 L \int_{|q|\xi}^R \, dr \, \frac{q^2\hbar^2}{Mr} = q^2\frac{\pi\hbar^2 n_{2D}}{M} \log\left(\frac{R}{|q|\xi}\right) \simeq q^2 E_1
\eeq
where $E_1=\pi\hbar^2 n_{2D} \log(R/\xi)/M$ is the energy of the condensate with a vortex of charge 1. Here, we have neglected the density increase $n_0+\Delta n$ and we have used $R\gg\xi$ to neglect the correction due to $|q|$ in the logarithm. In contrast with the interaction energy, it diverges logarithmically with the system size.

We can compare this energy with the one of a system with $|q|$ independent vortices, which has the same total circulation. Its energy is now $|q|E_1$ instead of $q^2E_1$. We see here that it is energetically more favorable to break a multiply-charged vortex into several singly-charged vortices. A system placed in fast rotation, which tends to accommodate vortices, will indeed result in a vortex lattice with many vortices of charge unity, as can be seen in Fig.~\ref{fig:vortex_wf}, right.

Note that we simply added here the energy of the $|q|$ vortices independently. This is valid if they are very far from each other. If this is not the case, an additional term appears which can be interpreted as a vortex-vortex interaction. This will be discussed in Lecture 2, Section~\ref{sec:vortex_pairs}.

\clearpage
\section{Solution of the exercises}
\subsection*{Exercise 1. Critical atom number and temperature}
\subsubsection*{BEC in a 3D box}
In 3D, $C(D)=4\pi$. In a box, $\alpha=0$. The parameter $\eta(\alpha,D)$ is equal to $\eta(0,3)=1/3$, and $q=1/2$, see Eq.~\eqref{eq:q_exponent}.
We get for $\varepsilon_0$ from Eq.~\eqref{eq:eps0}:
\beq
\varepsilon_0 = \left[\frac{2\times 3}{16\pi^2}\left(\frac{h^2}{2MR^2}\right)^{3/2} \right]^{2/3} =
\left[\frac{1}{2\pi \mathcal{V} }\left(\frac{h^2}{2M}\right)^{3/2} \right]^{2/3}
\eeq
where $\mathcal{V}=4\pi R^3/3$ is the volume of the box. From Eq.~\eqref{eq:Nc} and $\Gamma(3/2)=\sqrt{\pi}/2$ we get
\beq
N_c(T) = \zeta\left(\frac{3}{2}\right)\frac{\sqrt{\pi}}{2} 2\pi \mathcal{V} \left(\frac{2MkT}{h^2} \right)^{3/2} = \zeta\left(\frac{3}{2}\right) \left(\frac{2\pi MkT}{h^2} \right)^{3/2} = \zeta\left(\frac{3}{2}\right) \mathcal{V} \lambda_{\rm dB}^{3/2}.
\eeq
We get the famous formula for the critical phase space density for BEC in a uniform gas of density $n_0$:
\beq
\boxed{
n_0\lambda_{\rm dB}^{3/2} = \zeta\left(\frac{3}{2}\right) = 2.612\dots
}
\eeq
which also gives $T_c$ for a given atom number $N$.

\subsubsection*{BEC in a 3D harmonic trap}
In an harmonic trap, $\alpha=1$. We have $C(3)=4\pi$, $\eta(1,3)=\pi/16$ and $q=2$. The potential is $V(r)=V_0 r^2/R^2$ or with more usual notations $V(r) = M\omega_0^2r^2/2$, which is consistent if we take $V_0=\hbar\omega_0/2$ and $R=\sqrt{\hbar/M\omega_0}$. We then have $h^2/2MR^2 = 2\pi^2\hbar\omega_0$. We get for $\varepsilon_0$:
\beq
\varepsilon_0 = \left[\frac{2\times 16}{\pi\times 16\pi^2}\left(2\pi^2\hbar\omega_0\times\frac{\hbar\omega_0}{2}\right)^{3/2}\right]^{1/3} =2^{1/3}\hbar\omega_0.
\eeq
The critical atom number is
\beq
N_c(T) = 2\zeta(3)\left(\frac{kT}{2^{1/3}\hbar\omega_0} = \right)^3 = \zeta(3)\left(\frac{kT}{\hbar\omega_0}\right)^3 \simeq 1.202\left(\frac{kT}{\hbar\omega_0}\right)^3.
\eeq
Inverting this relation gives the critical temperature at fixed atom number $N$:
\beq
\boxed{
kT_c = \zeta(3)^{-1/3}\hbar\omega_0 N^{1/3} \simeq 0.940\,\hbar\omega_0 N^{1/3}.
}
\eeq

\subsubsection*{BEC in a 2D trap}
The only difference between the previous case is $D=2$, which affects $C(2)=2\pi$, $q=1$ and $\eta(1,2)=1/2$. We get for $\varepsilon_0$:
\beq
\varepsilon_0 = \left[\frac{2\times 2}{4\pi^2}\pi^2(\hbar\omega_0)^2\right]^{1/2} = \hbar\omega_0.
\eeq
The critical atom number is
\beq
N_c(T) = \zeta(2)\left(\frac{kT}{\hbar\omega_0}\right)^2 =\frac{\pi^2}{6}\left(\frac{kT}{\hbar\omega_0}\right)^2.
\eeq
Inverting this relation gives the critical temperature at fixed atom number $N$:
\beq
\boxed{
kT_c = \frac{\sqrt{6}}{\pi}\,\hbar\omega_0 \sqrt{N}.
}
\eeq

\subsection*{Exercise 2: chemical potential and TF radius in a harmonic trap}
\subsubsection*{BEC in a 3D hamonic trap}
We consider a condensate with $N$ atoms in a harmonic trap with frequencies $\omega_x$, $\omega_y$, $\omega_z$, of geometric average $\bar\omega = (\omega_x\omega_y\omega_z)^{1/3}$. The atomic density in the Thomas-Fermi regime,see Eq.~\eqref{eq:density_TF_regime}, writes
\beq
n(\mathbf{r})=\frac{1}{g}\left[\mu - \left(\frac{1}{2}M\omega_x^2x^2 + \frac{1}{2}M\omega_y^2y^2 + \frac{1}{2}M\omega_z^2z^2\right)\right] = n_0\left(1-\frac{x^2}{R_x^2}-\frac{y^2}{R_y^2}-\frac{z^2}{R_z^2}\right)
\eeq
where we have introduced the peak density in the center $n_0=\mu/g$ and the Thomas-Fermi radius in each direction $i=x,y,z$:
\beq
\boxed{
R_i=\frac{1}{\omega_i}\sqrt{\frac{2\mu}{M}}.
\label{eq:R_TF}
}
\eeq
This represents the distance from the trap center at which the density vanishes in the direction $i$. The surface in 3D space where the density vanishes is an ellipsoid of radii $R_x$, $R_y$ and $R_z$, defined by the equation
\beq
\frac{x^2}{R_x^2}+\frac{y^2}{R_y^2}+\frac{z^2}{R_z^2} =1.
\eeq
In order to determine $\mu$, we integrate the density inside this ellipsoid of volume $\mathcal{V}$, which is equa to the number of atoms in the condensate:
\beq
N = \int_\mathcal{V}d\mathbf{r}\,n(\mathbf{r}) = n_0\int_\mathcal{V}dx\,dy\,dz\,\left(1-\frac{x^2}{R_x^2}-\frac{y^2}{R_y^2}-\frac{z^2}{R_z^2}\right).
\eeq
We change the integration variables for $u_i=x_i/R_i$, such that the integration now runs over the volume of a sphere of radius 1, $\mathcal{V}_1$, which allows us to use spherical coordinates $(u,\theta,\varphi)$:
\bea
N &=& n_0R_xR_yR_z\int_{\mathcal{V}_1} du_x \, du_y \, du_z\, (1-u_x^2-u_y^2-u_z^2)\nonumber\\
& =& \frac{\mu}{g}\frac{1}{\bar\omega^3}\left(\frac{2\mu}{M}\right)^{3/2}\int u^2 du\, \sin\theta d\theta\,d\varphi\,(1-u^2) \nonumber\\
&=& \left(\frac{2\mu}{\hbar\bar\omega}\right)^{5/2}\frac{M\times\hbar^2}{4\pi\hbar^2a\times 2M^{3/2}}\sqrt{\frac{\hbar}{\bar\omega}}\times 4\pi\int_0^1 du\, u^2(1-u^2)\nonumber\\
&=&\left(\frac{2\mu}{\hbar\bar\omega}\right)^{5/2}\frac{1}{a}\sqrt{\frac{\hbar}{M\bar\omega}}\times\frac{1}{15}.
\eea
We introduce the size of the mean harmonic oscillator
\beq
a_{\rm osc}=\sqrt{\frac{\hbar}{M\bar\omega}}
\label{eq:a_osc}
\eeq
and we get for the chemical potential:
\beq
\boxed{
\mu = \frac{\hbar\bar\omega}{2}\left(\frac{15Na}{a_{\rm osc}}\right)^{2/5}\quad\mbox{valid if } Na\gg a_{\rm osc}.
}
\eeq
The Thomas-Fermi radius in the direction $i$ then writes
\bea
R_i &=& \frac{1}{\omega_i}\sqrt{\frac{\hbar\bar\omega}{M}}\left(\frac{15Na}{a_{\rm osc}}\right)^{1/5} =\sqrt{ \frac{\hbar}{M\omega_i}}\sqrt{\frac{\bar\omega}{\omega_i}}\left(\frac{15Na}{a_{\rm osc}}\right)^{1/5}\nonumber\\
R_i &=& a_i\sqrt{\frac{\bar\omega}{\omega_i}}\left(\frac{15Na}{a_{\rm osc}}\right)^{1/5}
\eea
where $a_i=\sqrt{\hbar/M\omega_i}$ is the size of the quantum harmonic oscillator in the direction $i$.

\subsubsection*{BEC in a 2D harmonic trap}
We apply the same procedure in 2D, and Eq.~\eqref{eq:R_TF} still holds, with now $\bar\omega=\sqrt{\omega_x\omega_y}$. The integration now runs over the area $\mathcal{A}$ of the ellipse defined by
\beq
\frac{x^2}{R_x^2}+\frac{y^2}{R_y^2} =1.
\eeq
We also use rescaled integration coordinates $u_i=x_i/R_i$ and the rescaled ellipse is the unit disk $\mathcal{D}_1$:
\bea
N &=& n_0\int_\mathcal{A}dx\,dy\,\left(1-\frac{x^2}{R_x^2}-\frac{y^2}{R_y^2}\right)=n_0R_xR_y\int_{\mathcal{D}_1} du_x \, du_y\, (1-u_x^2-u_y^2)\nonumber\\
&=& \frac{\mu}{g_{2D}}\frac{1}{\bar\omega^2}\frac{2\mu}{M}\int_0^1 2\pi u(1-u^2)\, du  \nonumber\\
&=& \frac{2\mu^2}{\tilde{g}\hbar^2\bar\omega^2}\times\frac{\pi}{2} = \frac{\pi}{\tilde{g}}\left(\frac{\mu}{\hbar\bar\omega}\right)^2
\eea
where we have written $g_{2D}=\tilde{g}\hbar^2/M$, see Lecture 2.
We finally get
\beq
\boxed{
\mu_{2D}=\sqrt{\frac{N\tilde{g}}{\pi}}\hbar\bar{\omega}\quad\mbox{valid if }N\tilde{g}\gg 1.
}
\eeq
The Thomas-Fermi radii then write
\beq
R_i=\frac{1}{\omega_i}\sqrt{\frac{2\hbar\bar\omega}{M}}\left(\frac{N\tilde{g}}{\pi}\right)^{1/4} = a_i \sqrt{\frac{\bar\omega}{\omega_i}}\left(\frac{4N\tilde{g}}{\pi}\right)^{1/4}.
\eeq


\chapter[Two-dimensional quantum Bose gases]{Two-dimensional quantum Bose gases: phase fluctuations, BKT mechanism and scaling symmetry}

Restricting a physical system to lower dimensions not only reduces the space in which a particle can move. It also affects qualitatively the behavior of a many-body system \cite{Bloch2008RMP}. As underlined by the Hohenberg--Mermin--Wagner theorem \cite{Mermin1966,Hohenberg1967}, fluctuations are enhanced for a many-body system with short-range interactions in dimensions two and one, preventing long-range order to establish. In one dimension, particles can not avoid each other, which strongly restricts collisional properties and eventually leads to the integrability of motion for a uniform one-dimensional (1D) many-body system of particles with contact interactions \cite{Lieb1963a,Lieb1963b}.

In two dimensions, the interaction parameter for contact interaction at low temperatures is dimensionless in units of $\hbar^2/M$ where $M$ is the atomic mass, in contrast to the three-dimensional (3D) case where it is described by a unique parameter, the scattering length $a$ \cite{WalravenQuantumGases}. This leads to a specific scaling symmetry, resulting in an equation of state depending only on the ratio between the chemical potential $\mu$ and the temperature $T$, and in an undamped breathing mode. For a quasi two-dimensional gas obtained by confining strongly a 3D gas in a transverse direction, the interaction parameter can depend explicitly on the atomic density \cite{Petrov2000a}.

Two-dimensional (2D) systems are particularly interesting as the dimension two is a critical dimension from many aspects ---which gives rise to many logarithmic divergences. Long-range order is prevented by the Hohenberg--Mermin--Wagner theorem. Bose-Einstein condensation (BEC) doesn't occur at any finite temperature in a uniform ideal two-dimensional gas, but still exists at $T=0$, and an arbitrarily weak trapping potential $V(r)\propto r^\gamma$ with $\gamma>0$ arbitrarily small allows us to recover BEC. In the case where weak interactions are present, no strict long-range order exists but a quasi long-range order subsists, with an algebraic decay of the one-body correlation function $G_1(r)$ that can be very slow. In practice, for a finite-size system, the condensate fraction can be rather large. The system sustains a normal to superfluid transition, with a mechanism involving the unbinding of topological defects ---vortex pairs--- described by Vadim Berezinskii, Michael Kosterlitz and David Thouless \cite{Kosterlitz1973,Berezinskii1971,Berezinskii1972}. Kosterlitz and Thouless were awarded the 2016 Nobel prize together with Duncan Haldane \cite{KosterlitzNobel,HaldaneNobel}.

The second lecture is devoted to 2D quantum Bose gases. To prepare the lecture, I have mostly used the lectures given by Jean Dalibard at Coll\`ege de France during the academic year 2016-2017 \cite{DalibardCF2017en}, which are available online, in French (and hopefully soon in English). In addition, the following general references may be useful:
\begin{enumerate}
\item \textit{Quantum Gases in Low Dimensions}, edited by L. Pricoupenko, H. Perrin and M. Olshanii, J. Phys IV \textbf{116} (2004) \cite{QGLD2003}

Les Houches lectures by Shlyapnikov \cite{Petrov2004}, Cirac \cite{Paredes2004b} and Douçot \cite{Doucot2004}.
\item \textit{Many body physics with ultra cold gases}, I. Bloch, J. Dalibard and W. Zwerger, Rev. Mod. Phys. \textbf{80}, 885 (2008) \cite{Bloch2008RMP}
\item Nobel lectures by J. Michael Kosterlitz and F. Duncan M. Haldane, Rev. Mod. Phys. \textbf{89}, 040501 \& 040502 (2017) \cite{KosterlitzNobel,HaldaneNobel}
\end{enumerate}

\section{A first glance at two-dimensional quantum gases}
\subsection{Non interacting 2D gas: BEC and log divergences}
Let us first examine the possibility for Bose-Einstein condensation (BEC) in two dimensions, for a gas of non interacting bosons. BEC will occur if the number of particles in excited states saturates, i.e. if the total number of particles $N$ exceeds $N_{\rm exc}^{\rm max}(T)$, the maximum number of particles in the excited states at a given temperature $T$.

We have seen in Lecture 1 that BEC doesn't occur in a box in two dimensions, but does occur in any trap, including a harmonic trap. With the notations of Lecture 1, we have $E_0=0$, $C(2)=2\pi$, $\eta(\alpha,2)=1/2$ independent of $\alpha$, and $q=\alpha$. We deduce for $\varepsilon_0$
\beq
\varepsilon_0 = \left(2\frac{\hbar^2}{MR^2}V_0^\alpha\right)^{1/(\alpha+1)}
\eeq
such that
\beq
\rho(\varepsilon) = \frac{MR^2}{2\hbar^2}\left(\frac{\varepsilon}{V_0}\right)^\alpha.
\eeq
The number of particles in the excited states is
\beq
N_{\rm exc}(z,T)=\Gamma(\alpha+1)\frac{Mk_BTR^2}{2\hbar^2}\left(\frac{k_BT}{V_0}\right)^\alpha {\rm Li}_{\alpha+1}(z),
\eeq
or using $\lambda_{\rm dB}= h/\sqrt{2\pi Mk_BT}$:
\beq
\frac{N_{\rm exc}(z,T)}{\pi R^2}\lambda_{\rm dB}^2=\Gamma(\alpha+1)\left(\frac{k_BT}{V_0}\right)^\alpha {\rm Li}_{\alpha+1}(z).
\eeq
\begin{itemize}
\item In a \textbf{box trap}, we recognize in the left term the atomic density in the excited states $n_{\rm exc}=N_{\rm exc}/\pi R^2$, and $\mathcal{D}_{\rm exc}(z,T) = n_{\rm exc}(z,T)\lambda_{\rm dB}^2$ is the phase space density in the excited states. Using $\alpha=0$, $\Gamma(1)=1$ and ${\rm Li}_{1}(z) = -\log(1-z)$, we get
\beq
\mathcal{D}_{\rm exc}(z,T) = -\log(1-z) = -\log\left(1-e^{\beta\mu}\right).
\label{eq:psd_exc}
\eeq
where $\beta=1/k_BT$.
$\mathcal{D}_{\rm exc}$ diverges logarithmically as $\mu\to 0$ or $z\to 1$. We will meet many more logarithmic divergences in 2D quantum gases. This is the sign that the number of atoms in the excited states doesn't saturate and BEC doesn't occur (we have $\mathcal{D}\simeq \mathcal{D}_{\rm exc}$), but that the behavior of $\mathcal{D}_{\rm exc}$ with $T$ is very slow, emphasizing the critical character of dimension 2.
\item In a \textbf{harmonic trap} $V(r) = M\omega_0^2r^2/2$, which corresponds to $V(r)=V_0 r^2/R^2$ when we take $\alpha=1$, $R=\sqrt{\hbar/M\omega_0}$ and thus $V_0=\hbar\omega_0/2$. We get $\varepsilon_0=\hbar\omega_0$, and using $\Gamma(2)=1$:
\beq
N_{\rm exc}(z,T)=\left(\frac{k_BT}{\hbar\omega_0}\right)^2 {\rm Li}_{2}(z).
\eeq
${\rm Li}_{2}(1)=\zeta(2) = \pi^2/6$ is finite, and BEC occurs for $N>N_c(T)$ with
\beq
\boxed{
N_c(T)=\frac{\pi^2}{6}\left(\frac{k_BT}{\hbar\omega_0}\right)^2,
}
\eeq
or equivalently $T<T_c(N)$, with
\beq
\boxed{
k_BT_c(N) = \frac{\sqrt{6N}}{\pi}\hbar\omega_0.
}
\label{eq:Tc_BEC_2D_harmonic}
\eeq
The condensate fraction depends on $T/T_c$ as
\beq
\boxed{\frac{N_0}{N} = 1 - \left(\frac{T}{T_c}\right)^2.}
\eeq
Let us however estimate the phase space density at the center of the harmonic trap. The density is obtained by integrating the Bose-Einstein distribution $f(\mathbf{r},\mathbf{p})$ on $\mathbf{p}$ only:
\beq
n(\mathbf{r}) = 
\frac{1}{h^2}\int_0^{+\infty}\frac{2\pi p\,dp\,e^{-\beta(p^2/2M+V(\mathbf{r})-\mu)}}{1-e^{-\beta(p^2/2M+V(\mathbf{r})-\mu)}} = -\frac{2\pi Mk_BT}{h^2}\log\left(1-e^{-\beta(V(\mathbf{r})-\mu)}\right).
\eeq
Using $z=e^{\beta\mu}$ and $\lambda_{\rm dB}$, we get
\beq
\boxed{
\mathcal{D}(\mathbf{r}) = n(\mathbf{r})\lambda_{\rm dB}^2 = -\log\left(1-ze^{-\beta V(\mathbf{r})}\right).
}
\eeq
Below $T_c$, $z=1$ and we find that the density diverges in the trap center for which $V(\mathbf{0})=0$. Near the center we get $\mathcal{D}(\mathbf{r})\simeq -\log[\beta V(\mathbf{r})]$, once again a logarithmic divergence. Of course, beyond a certain density our assumption of non interacting gas will break down. This indicates that interactions will play a crucial role in 2D.
\end{itemize}

\subsection{Gross-Pitaevskii equation and 2D scale invariance}
\label{sec:2D_GPE}
We have seen that we need to take into account interactions between particles. Let us write the Gross-Pitaevskii equation we expect in 2D in the presence of weak interactions:
\beq
-i\hbar\partial_t \psi(\mathbf{r},t) = -\frac{\hbar^2}{2M}\nablagras^2 \psi(\mathbf{r},t) + V(\mathbf{r})\psi(\mathbf{r},t) + g_{2D}|\psi(\mathbf{r},t)|^2\psi(\mathbf{r},t).
\eeq
We have introduced the interaction constant $g_{2D}$ for our two-dimensional problem. However, it differs from its 3D counterpart. The dimension of the term $g_{2D}|\psi|^2$ in front of $\psi$ must correspond to an energy, just as $V$ for example. This means that $g_{2D}$ has the dimension of an energy multiplied by a surface. In 3D, $g_{3D}=4\pi\hbar^2 a/M$ has the dimension of an energy multiplied by a volume. In other words, the dimension of $g_{2D}$ is simply the one of $\hbar^2/M$, without the need to include any length in the problem.

We generally introduce the dimensionless interaction constant $\tilde{g}$, such that
\beq
\boxed{
g_{2D} = \frac{\hbar^2}{M}\tilde{g}.
}
\eeq

In 3D, the weak interaction regime is defined by $na^3\ll 1$, the density being compared to the characteristic length for the interactions. In 2D, instead, the weak interaction regime is simply defined by $\tilde{g}\ll 1$.

The absence of characteristic length for the interactions is at the heart of the scaling symmetry present in a two-dimensional quantum gas with contact interactions. The consequences of this scaling symmetry include the absence of damping of the monopole (or breathing) mode, and the universality of the equation of state, which depends only on $\mu/k_BT$ and not independently on $\mu$ and $T$.

\section{Quasi long-range order in 2D}
\label{sec:QLRO}
\subsection{From 3D to 2D}
\label{sec:3D_2D}
To estimate the interaction constant $\tilde{g}$ for the 2D gas, we should in principle solve the scattering problem in two dimensions. This comes with other logarithmic divergences, and a peculiar behavior where the interaction parameter can depend on the atomic density \cite{Petrov2000a} if the characteristic length of the transverse confinement is smaller than the 3D scattering length $a$.

However, if we consider that the quasi-2D gas results from a transverse confinement in the $z$ direction by a strong harmonic trap with frequency $\omega_z$, but that the corresponding size $a_z=\sqrt{\hbar/M\omega_z}$ is still significantly larger than $a$, we can use an approximate model. We consider that the 3D description of the atom-atom scattering is still correct and obtain the 2D Gross-Pitaevskii equation by transverse averaging over the direction $z$.

Specifically, we write the 3D wave function $\psi_{3D}(x,y,z)$ as a product between the wave function $\psi(\mathbf{r})$ with $\mathbf{r}=(x,y)$ for the in-plane dynamics and the ground state $\chi(z) = \pi^{-1/4}\exp(-z^2/2a_z^2)/\sqrt{a_z}$ of the transverse harmonic oscillator. $\chi$ is solution of the one-dimensional Schrödinger equation
\beq
-\frac{\hbar^2}{2M}\partial_z^2\chi + \frac{1}{2}M\omega_z^2z^2\chi(z) = \frac{\hbar\omega_z}{2}\chi(z),
\eeq
with a zero-point energy $\hbar\omega_z/2$ and a normalization $\int |\chi|^2dz =1$.
We then inject $\psi_{3D}(\mathbf{r},z) = \psi(\mathbf{r})\chi(z)$ in the 3D GPE, Eq.~(49) of Lecture 1:
\bea
\mu_{3D}\psi\chi &=& -\frac{\hbar^2}{2M}\chi\nablagras_{2D}^2\psi-\frac{\hbar^2}{2M}\psi\partial_z^2\chi+V(\mathbf{r})\psi\chi+\frac{1}{2}M\omega_z^2z^2\psi\chi+\frac{4\pi\hbar^2a}{M}|\psi|^2\psi|\chi|^2\chi\nonumber\\
\left(\mu_{3D}-\frac{\hbar\omega_z}{2}\right)\psi\chi&=&-\frac{\hbar^2}{2M}\chi\nablagras_{2D}^2\psi+V(\mathbf{r})\psi\chi+\frac{4\pi\hbar^2a}{M}|\psi|^2\psi|\chi|^2\chi.
\eea

We now take the average over $z$; i.e. we multiply by $\chi^*$ and integrate over $z$. The terms linear in $\chi$ will give 1 because of the normalization of $\chi$. However there is a term in $|\chi|^4$, whose integral is
\beq
\int|\chi(z)|^4dz = \frac{1}{\pi a_z^2}\int e^{-2z^2/a_z^2}dz = \frac{1}{\sqrt{2\pi}a_z}.
\eeq
We then arrive to the 2D GPE
\beq
\mu_{2D}\psi = -\frac{\hbar^2}{2M}\nablagras^2\psi+V(\mathbf{r})\psi+\tilde{g}\frac{\hbar^2}{2M}|\psi|^2\psi
\eeq
where $\nablagras\equiv\nablagras_{2D}$, we have introduced the chemical potential for the quasi-2D gas
\beq
\boxed{
\mu_{2D}=\mu_{3D}-\frac{\hbar\omega_z}{2}
}
\eeq
and the dimensionless interaction constant is
\beq
\boxed{
\tilde{g}=\sqrt{8\pi}\frac{a}{a_z}.
\label{eq:gtilde}
}
\eeq
This approach holds only if $a\ll a_z$, i.e. if $\tilde{g}$ is small. Moreover, the condition for the gas to stay in the 2D regime is to have all its energies much smaller than the transverse energy $\hbar\omega_z$, i.e. if
\beq
k_BT\ll\hbar\omega_z\quad\mbox{and}\quad\mu_{2D}\ll\hbar\omega_z.
\eeq
If the 2D chemical potential is of order $g_{2D}n$ where $n$ is the two-dimensional density, this last condition writes $\frac{\hbar^2}{M}\tilde{g}n\ll\hbar^2/Ma_z^2$ or $\tilde{g}na_z^2\ll 1$.

\subsection{First order correlation function in the 2D degenerate Bose gas}
A striking difference between the 3D and the 2D case is the importance of phase fluctuations in reduced dimensions, which eventually break the long-range order. To characterize the long-range behavior in the gas, we need to estimate the one-body correlation function  \cite{Petrov2004,DalibardCF2017en}, which probes the phase relation between two points $\mathbf{r}$ and $\mathbf{r}'$:
\beq
G_1(\mathbf{r},\mathbf{r}') = \left\langle\mathbf{r}|\rho_1|\mathbf{r}'\right\rangle
\eeq
with $\rho_1$ the one-body density matrix. If the system is in a pure state, $\rho_1 = \ketbra{\psi}{\psi}$ and $G_1$ writes:
\beq
G_1(\mathbf{r},\mathbf{r}') = \psi(\mathbf{r})\psi^*(\mathbf{r}').
\eeq
In a uniform system which is translationally invariant, $G_1$ only depends on the relative vector $\mathbf{r}-\mathbf{r}'$ between the two points
\beq
G_1(\mathbf{r}) = \left\langle\mathbf{r}|\rho_1|\mathbf{0}\right\rangle\qquad\mbox{or}\qquad G_1(\mathbf{r}) = \psi(\mathbf{r})\psi^*(\mathbf{0}).
\eeq
We note that we always have $G_1(0)=n(0)=n_0$, where $n_0$ is the average density.

The one-body correlation function $G_1$ is the Fourier transform of the normalized momentum distribution $\mathcal{N}(\mathbf{p})$ (see Appendix~\ref{sec:G1_TF}):\beq
\boxed{
G_1(\mathbf{r}) = n_0\int d\mathbf{p}\,\mathcal{N}(\mathbf{p})\,e^{i\mathbf{r}\cdot\mathbf{p}/\hbar}.
}
\label{eq:G1_from_FT}
\eeq

\paragraph{Thermal gas}
The first order correlation function in a thermal gas is a Gaussian, with a characteristic length given by the de Broglie wave length:
\beq
\boxed{
G_1(\mathbf{r})=n_0\,e^{-\pi r^2/\lambda_{\rm dB}^2}.
}
\eeq
This decay is thus rather fast and depends directly on temperature. The derivation is given in Appendix~\ref{sec:G1_thermal}.

\paragraph{Ideal degenerate Bose gas}
In the case of an ideal gas in the degenerate regime, the long range behavior of the first order correlation function is
\beq
\boxed{
G_1(r) \propto \sqrt{\frac{\ell}{r}}e^{-r/\ell}\qquad \mbox{for } r\gg \ell.
}
\eeq
It decays on a typical length scale
\beq
\ell = \frac{\lambda_{\rm dB}}{\sqrt{4\pi}}e^{\mathcal{D}/2}.
\eeq
Again, there is no long-range order as $G_1$ decays faster than exponentially, but the decay length can be much larger than $\lambda_{\rm dB}$ and diverges as the phase space density increases. The derivation is given in Appendix~\ref{sec:G1_ideal}.

\paragraph{Weakly interacting degenerate Bose gas}
In this case, $G_1$ decays more slowly, giving rise to a quasi long-range order. The derivation is rather long and is given in Appendix~\ref{sec:G1_degenerate}. In short, the decay of the correlation is dominated by the thermal phase fluctuations due to the population of phononic modes, up to an energy $k_BT$. The cut-off in momentum for the populated mode is obtained by equating the Bogolubov energy with $k_BT$, yielding
\beq
q_T = \xi^{-1}\left(\sqrt{1+\left(\frac{2\pi}{\tilde{g}\mathcal{D}_s}\right)^2}-1\right)^{1/2}
\label{eq:qT}
\eeq
with $\mathcal{D}_s$ the phase-space density of the superfluid fraction and $n_0$ its density. We recall from Eq.~(58) of Lecture 1 the expression of the healing length $\xi$:
\beq
\xi = \frac{\hbar}{\sqrt{2Mgn_0}} = \frac{1}{\sqrt{2\tilde{g}n_0}}.
\eeq

The large $r$ behavior of the phase fluctuations is dominated by
\beq
\langle(\theta(\mathbf{r})-\theta(\mathbf{0}))^2\rangle \simeq \frac{2}{\mathcal{D}_s}\log(rq_T).
\eeq
This yields for the first order correlation function
\beq
\boxed{
G_1(r) \simeq n_0\left(\frac{\ell_T}{r}\right)^{1/\mathcal{D}_s}\quad\mbox{for }r\to\infty
}
\eeq
with $\ell_T=1/q_T$ and $q_T$ given at Eq.~\eqref{eq:qT}. The first order correlation function decays only algebraically. Moreover, $\ell_T$ increases as $\ell_T\propto \sqrt{n_0}/T$ at large $\mathcal{D}$. Finally, the exponent $1/\mathcal{D}_s$ becomes very small at large phase space density. Therefore, in realistic finite size systems the phase coherence subsists even at the edge of the sample. This situation is described as \textit{quasi long-range order}.

\paragraph{Summary} We summarize the situation for the ideal gas and the weakly interacting gas:

\vspace{5mm}
\noindent
\begin{tikzpicture}
\draw[dashed] (-6,0.3) -- (-6,1);
\draw[dashed] (-6,-0.3) -- (-6,-1);
\node at (-6,0) {$T=0$};
\draw[dashed] (-2,-0.3) -- (-2,-1);
\node at (-2,0) {$\mathcal{D}\sim$?}; 
\draw[dashed] (2,0.3) -- (2,1);
\draw[dashed] (2,-0.3) -- (2,-1);
\node at (2,0) {$\mathcal{D}\sim 1$}; 
\node at (-7.5,1) {ideal gas};
\draw[->,>=latex] (-6,1) -- (6,1);
\draw (-6,0.9) -- (-6,1.1);
\node at (6,0.7) {$T$};
\node at (-2,1.3) {$G_1(r)\propto e^{-r/\ell}/\sqrt{r}$}; 
\node at (4,1.3) {$G_1(r)\propto e^{-\pi r^2/\lambda_{\rm dB}^2}$};
\node at (-2,0.7) {faster than exponential}; 
\node at (4,0.7) {Gaussian};
\node at (-7.5,-1) {interacting gas};
\draw[->] (-6,-1) -- (6,-1);
\draw (-6,-0.9) -- (-6,-1.1);
\node at (6,-0.7) {$T$};
\node at (-4,-0.7) {algebraic}; 
\node at (0,-0.7) {exponential};
\node at (4,-0.7) {Gaussian};
\node at (-4,-1.5) {$G_1(r)\propto\left(\frac{\ell_T}{r}\right)^{1/\mathcal{D}}$}; 
\node at (0,-1.5) {$G_1(r)\propto e^{-r/\ell}$};
\node at (4,-1.5) {$G_1(r)\propto e^{-\pi r^2/\lambda_{\rm dB}^2}$};
\end{tikzpicture}

In both cases, the first order correlation function is Gaussian in the thermal regime. At intermediate temperatures for which the gas is weakly degenerate, the correlation function of the weakly interacting gas is exponential. The goal of the next section is to describe the transition between this regime and the strongly degenerate regime where $G_1$ is algebraic and a quasi long-range order is present. We will also give an estimation of the phase space density $\mathcal{D}$ where this transition occurs.

\section{The Berezinskii--Kosterlitz--Thouless mechanism}
\label{sec:BKT}
In the 2D uniform weakly interacting Bose gas, a transition occurs at a critical phase-space density $\mathcal{D}_c>1$ from a normal to a superfluid state \cite{Berezinskii1971,Berezinskii1972,Kosterlitz1973,KosterlitzNobel}, and $G_1$ transitions from an exponential decay to an algebraic decay, leading to quasi long-range order. This transition involves binding / unbinding of vortex pairs of opposite sign.
\subsection{Vortex pairs arising from thermal fluctuations}
\label{sec:vortex_pairs}
We will give here a simple argument for the production of vortex pairs from thermal fluctuations \cite{DalibardCF2017en}. In a degenerate gas described by an order parameter $\psi(\mathbf{r})$, at finite temperature fluctuations lead to variations of $\psi$ in space, and $\psi$ can even vanish if density fluctuations are strong enough. Yet, this doesn't lead to the birth of a single vortex, a topologically protected excitation that would require to modify the circulation along a contour around this zero of density. However, a pair of vortices with charges $q=\pm 1$ can be created at this point, which preserves the total circulation.

Jean Dalibard gives an intuitive picture of this phenomenon in his lectures at Collège de France \cite{DalibardCF2017en}. Let us decompose the wave function $\psi$ into its real part $\mathcal{R}$ and its imaginary part $\mathcal{I}$: $\psi(\mathbf{r}) = \mathcal{R}(\mathbf{r}) + i\mathcal{I}(\mathbf{r})$. Both vary continuously in space, and can vanish along specific lines. If two such lines cross, $\psi$ vanishes at the intersection. A singly charged vortex sits at that point, with a sign that depends on the size of $\mathcal{R}$ and $\mathcal{I}$ on either side of the zero line.

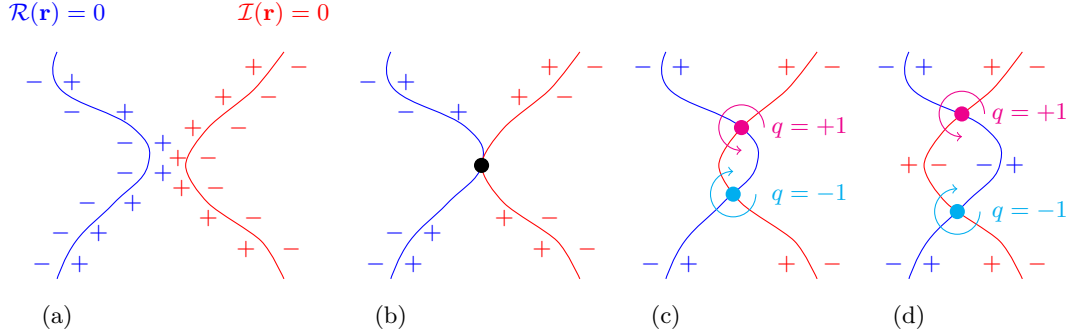
\begin{figure}
\begin{tikzpicture}[font=\footnotesize]
\draw[blue] plot[smooth,tension=0.75] coordinates {(0,-2) (0.25,-1.5) (0.75,-1) (1.2,-0.5) (1,0) (0.05,0.5) (0,1)};
\node[blue] at (-0.2,-1.8) {$-$};
\node[blue] at (0.05,-1.4) {$-$};
\node[blue] at (0.4,-1) {$-$};
\node[blue] at (0.9,-0.6) {$-$};
\node[blue] at (0.9,-0.2) {$-$};
\node[blue] at (0.2,0.2) {$-$};
\node[blue] at (-0.3,0.6) {$-$};
\node[blue] at (0.3,-1.8) {$+$};
\node[blue] at (0.55,-1.4) {$+$};
\node[blue] at (1,-1) {$+$};
\node[blue] at (1.4,-0.6) {$+$};
\node[blue] at (1.4,-0.2) {$+$};
\node[blue] at (0.9,0.2) {$+$};
\node[blue] at (0.2,0.6) {$+$};
\node[blue] at (0,1.5) {$\mathcal{R}(\mathbf{r})=0$};
\draw[red] plot[smooth,tension=0.5] coordinates {(3,-2) (2.7,-1.5) (2,-1) (1.7,-0.5) (2,0) (2.6,0.5) (3,1)};
\node[red] at (3.1,-1.6) {$-$};
\node[red] at (2.6,-1.2) {$-$};
\node[red] at (2.1,-0.8) {$-$};
\node[red] at (2,-0.4) {$-$};
\node[red] at (2.4,0) {$-$};
\node[red] at (2.8,0.4) {$-$};
\node[red] at (3.2,0.8) {$-$};
\node[red] at (2.5,-1.6) {$+$};
\node[red] at (2,-1.2) {$+$};
\node[red] at (1.65,-0.8) {$+$};
\node[red] at (1.6,-0.4) {$+$};
\node[red] at (1.8,0) {$+$};
\node[red] at (2.2,0.4) {$+$};
\node[red] at (2.6,0.8) {$+$};
\node[red] at (3,1.5) {$\mathcal{I}(\mathbf{r})=0$};
\node at (0,-2.5) {(a)};
\end{tikzpicture}
\begin{tikzpicture}[font=\footnotesize]
\draw[blue] plot[smooth,tension=0.75] coordinates {(0.5,-2) (0.75,-1.5) (1.25,-1) (1.7,-0.5) (1.5,0) (0.55,0.5) (0.5,1)};
\node[blue] at (0.3,-1.8) {$-$};
\node[blue] at (0.55,-1.4) {$-$};
\node[blue] at (0.7,0.2) {$-$};
\node[blue] at (0.2,0.6) {$-$};
\node[blue] at (0.8,-1.8) {$+$};
\node[blue] at (1.05,-1.4) {$+$};
\node[blue] at (1.4,0.2) {$+$};
\node[blue] at (0.7,0.6) {$+$};
\draw[red] plot[smooth,tension=0.5] coordinates {(3,-2) (2.7,-1.5) (2,-1) (1.7,-0.5) (2,0) (2.6,0.5) (3,1)};
\node[red] at (3.1,-1.6) {$-$};
\node[red] at (2.6,-1.2) {$-$};
\node[red] at (2.8,0.4) {$-$};
\node[red] at (3.2,0.8) {$-$};
\node[red] at (2.5,-1.6) {$+$};
\node[red] at (2,-1.2) {$+$};
\node[red] at (2.2,0.4) {$+$};
\node[red] at (2.6,0.8) {$+$};
\fill[black] (1.7,-0.5) circle (0.1);
\node at (0.5,-2.5) {(b)};
\end{tikzpicture}
\begin{tikzpicture}[font=\footnotesize]
\draw[blue] plot[smooth,tension=0.75] coordinates {(0.5,-2) (0.75,-1.5) (1.25,-1) (1.7,-0.5) (1.5,0) (0.55,0.5) (0.5,1)};
\node[blue] at (0.3,-1.8) {$-$};
\node[blue] at (0.2,0.8) {$-$};

\node[blue] at (0.8,-1.8) {$+$};
\node[blue] at (0.7,0.8) {$+$};

\draw[red] plot[smooth,tension=0.5] coordinates {(2.5,-2) (2.2,-1.5) (1.5,-1) (1.2,-0.5) (1.5,0) (2.1,0.5) (2.5,1)};
\node[red] at (2.7,-1.8) {$-$};
\node[red] at (2.7,0.8) {$-$};

\node[red] at (2.1,-1.8) {$+$};
\node[red] at (2.1,0.8) {$+$};

\fill[magenta] (1.5,0) circle (0.1);
\fill[cyan] (1.39,-0.88) circle (0.1);
\node[magenta] at (2.4,0) {$q=+1$};
\node[cyan] at (2.4,-0.88) {$q=-1$};
\draw[->,magenta] (1.8,0) arc (0:270:0.3);
\draw[->,cyan] (1.69,-0.88) arc (0:-270:0.3);
\node at (0.5,-2.5) {(c)};
\end{tikzpicture}
\begin{tikzpicture}[font=\footnotesize]
\draw[blue] plot[smooth,tension=0.75] coordinates {(0.5,-2) (0.75,-1.5) (1.25,-1) (1.7,-0.5) (1.5,0) (0.55,0.5) (0.5,1)};
\node[blue] at (0.3,-1.8) {$-$};
\node[blue] at (1.5,-0.5) {$-$};
\node[blue] at (0.2,0.8) {$-$};

\node[blue] at (0.8,-1.8) {$+$};
\node[blue] at (1.9,-0.5) {$+$};
\node[blue] at (0.7,0.8) {$+$};

\draw[red] plot[smooth,tension=0.5] coordinates {(2,-2) (1.7,-1.5) (1,-1) (0.7,-0.5) (1,0) (1.6,0.5) (2,1)};
\node[red] at (2.2,-1.8) {$-$};
\node[red] at (0.9,-0.5) {$-$};
\node[red] at (2.2,0.8) {$-$};

\node[red] at (1.6,-1.8) {$+$};
\node[red] at (0.5,-0.5) {$+$};
\node[red] at (1.6,0.8) {$+$};

\fill[magenta] (1.2,0.18) circle (0.1);
\fill[cyan] (1.14,-1.11) circle (0.1);
\node[magenta] at (2.1,0.18) {$q=+1$};
\node[cyan] at (2.1,-1.11) {$q=-1$};
\draw[->,magenta] (1.5,0.18) arc (0:270:0.3);
\draw[->,cyan] (1.44,-1.11) arc (0:-270:0.3);
\node at (0.5,-2.5) {(d)};
\end{tikzpicture}
\caption{Mechanism for the birth of a vortex pair from the fluctuations of the wave function. (a): two zero lines of $\mathcal{R}$ and $\mathcal{I}$ that do not cross, no vortex pair present. (b): the two lines touch, birth of a vortex pair sitting at the same point. (c,d): vortex pair of opposite charge vortices at increasing distance. Inspired from Ref. \cite{DalibardCF2017en}. \label{fig:vortex_pair_nucleation}}
\end{figure}

With thermal fluctuations, two lines may cross: as they first touch and then cross in two points, a vortex pair is born, and the distance between the vortices of opposite charge increases when the two crossing points fly apart, as illustrated in Fig.~\ref{fig:vortex_pair_nucleation}.

The nucleation of a vortex pair with opposite charges requires an energy that is less than the one of a single vortex, because the phase profile is homogeneous far from this vortex dipole, which means that the velocity field is back to zero far from the dipole. We recall the formula of Eq.~(93) of Lecture~1 that gives the energy of a single vortex of charge $q$ centered in a homogeneous disk of radius $R$:
\beq
E_1(q)=q^2\frac{\pi\hbar^2 n_0}{M} \log\left(\frac{R}{\xi}\right)
\eeq
where $n_0$ is now the two-dimensional density far from the vortex core. It diverges logarithmically with the system size.

On the other hand, the kinetic energy associated to two vortices split by some distance $r\ll R$ and placed close to the center of the disk writes
\bea
E_2(q_1,q_2,r) &=& \frac{\pi\hbar^2n_0}{M}\left[(q_1+q_2)^2\log\left(\frac{R}{\xi}\right) - 2q_1q_2\log\left(\frac{r}{\xi}\right)\right]\quad\mbox{for }r\ll R \label{eq:energy_2v_1}\\
&=& \frac{\pi\hbar^2n_0}{M}\left[(q_1^2+q_2^2)\log\left(\frac{R}{\xi}\right) + 2q_1q_2\log\left(\frac{R}{r}\right)\right].\label{eq:energy_2v_2}
\eea
In the second form Eq.~\eqref{eq:energy_2v_2} of this energy, we recognize the sum of the energy of each vortex $E_1(q_1)+E_1(q_2)$, plus a Coulomb-like interaction term, repulsive for same-sign vortices and attractive for opposite-sign vortices. This interaction is responsible for the arrangement of the vortices into a triangular Abrikosov lattice in fast rotating superfluids, when many same-sign singly charged vortices are present, as shown in Fig.~7 of Lecture 1.

In the case where $q_1=-q_2=1$, the first form of the energy Eq.~\eqref{eq:energy_2v_1} gives the energy of a vortex dipole of size $d$:
\beq
E_d(d) = \frac{2\pi\hbar^2n_0}{M}\log\left(\frac{d}{\xi}\right).
\eeq
At fixed density $n_0$, this energy is now independent of the system size $R$: far from the dipole, the superfluid is not perturbed. In addition, the kinetic energy of a vortex dipole, computed outside its core of radius $\xi$, is vanishingly small as $d\to 0$: the nucleation of a vortex pair doesn't cost any kinetic energy far from the hole, but only an interaction energy $E_{\rm int}$ to `drill a hole' in the density, which increases the density everywhere else, and thus the interaction energy $N\tilde{g}\hbar^2 n/2M$. An estimation gives
\beq
E_{\rm int}\simeq N\frac{\tilde{g}\hbar^2}{2M}\left[\frac{N}{\pi(R^2-\xi^2)} - \frac{N}{\pi R^2}\right]\simeq N\frac{\tilde{g}\hbar^2n_0}{2M}\frac{\xi^2}{R^2}=\frac{\pi}{4}\frac{\hbar^2n_0}{M}\sim \frac{\hbar^2n_0}{M}.
\eeq
It is much smaller than $E_1(q=1)$ in the limit $R\gg\xi$.
\subsection{A transition triggered from unbinding vortex pairs}
We have seen that creating a vortex dipole costs a very small energy, but that this energy increases with the distance $d$. If the thermal energy is large enough, the vortex pairs eventually break into free, independent vortices. This happens if the free energy $F=E-TS$ associated to vortices becomes negative.

In this expression, $E=E_1(1)$ is the energy of a single vortex. The entropy is equal to $S=k_B \log N_s$, where $N_s$ is the number of available states in the system for this vortex, i.e. the number of different positions in the sample where the vortex can appear. As each vortex occupies approximately an area $\pi\xi^2$, the number of states in the system of total area $\pi R^2$ is $N_s=R^2/\xi^2$, yielding
\beq
S = 2k_B\log\frac{R}{\xi}.
\eeq
The free energy thus writes
\beq
F = E-TS = \left(\frac{\pi\hbar^2n_0}{M}-2k_BT\right)\log\frac{R}{\xi} = \frac{k_BT}{2}\left(\mathcal{D}_s-4\right)\log\frac{R}{\xi}
\eeq
where we have used the expression of the de Broglie wave length $\lambda_{\rm dB}^2=2\pi\hbar^2n_0/Mk_BT$ and the superfluid phase space density $\mathcal{D}_s=n_0\lambda_{\rm dB}^2$. We find a critical value for the superfluid phase space density $\mathcal{D}_s<4$ below which free vortices proliferate, eventually breaking superfluidity. Below this critical phase space density, $\mathcal{D}_s$ jumps from 4 to 0. This corresponds to the superfluid transition governed by the BKT mechanism.

Note that the criterion involves the superfluid phase space density ---the one required to describe the vortices as done above. The estimation of the total phase space density at the transition has been done with a Monte Carlo approach, and yields \cite{Prokofiev2001}
\beq
\mathcal{D}_c \simeq \log\frac{C}{\tilde{g}}\quad\mbox{with }C\simeq 380.
\eeq

The expression for $\mathcal{D}_c$ underlines the fact that interactions are required for superfluidity to happen in 2D: the higher $\tilde{g}$, the smaller the criterion on the phase space density. With $\tilde{g}\simeq 0.1$, which is a typical value, we find $\mathcal{D}_c\simeq 8$: this is much larger than the criterion for degeneracy, and at the transition, about half of the gas becomes superfluid.

The critical point also gives the boundary between the algebraic and the exponential behavior for the first order correlation function $G_1(r)$. We can now complete the diagram shown above:

\vspace{5mm}
\noindent
\begin{tikzpicture}
\draw[dashed] (-6,0.3) -- (-6,1);
\draw[dashed] (-6,-0.3) -- (-6,-1);
\node at (-6,0) {$T=0$};
\draw[dashed] (-2,-0.3) -- (-2,-1);
\node at (-2,0) {$\mathcal{D}\sim\log(C/\tilde{g})$}; 
\draw[dashed] (2,0.3) -- (2,1);
\draw[dashed] (2,-0.3) -- (2,-1);
\node at (2,0) {$\mathcal{D}\sim 1$}; 
\node at (-7.5,1) {ideal gas};
\draw[->,>=latex] (-6,1) -- (6,1);
\draw (-6,0.9) -- (-6,1.1);
\node at (6,0.7) {$T$};
\node at (-2,1.3) {$G_1(r)\propto e^{-r/\ell}/\sqrt{r}$}; 
\node at (4,1.3) {$G_1(r)\propto e^{-\pi r^2/\lambda_{\rm dB}^2}$};
\node at (-2,0.7) {faster than exponential}; 
\node at (4,0.7) {Gaussian};
\node at (-7.5,-1) {interacting gas};
\draw[->] (-6,-1) -- (6,-1);
\draw (-6,-0.9) -- (-6,-1.1);
\node at (6,-0.7) {$T$};
\node at (-4,-0.7) {algebraic}; 
\node at (0,-0.7) {exponential};
\node at (4,-0.7) {Gaussian};
\node at (-4,-1.5) {$G_1(r)\propto\left(\frac{\ell_T}{r}\right)^{1/\mathcal{D}_s}$}; 
\node at (0,-1.5) {$G_1(r)\propto e^{-r/\ell}$};
\node at (4,-1.5) {$G_1(r)\propto e^{-\pi r^2/\lambda_{\rm dB}^2}$};
\end{tikzpicture}

Below the critical temperature for the KT transition, the gas becomes superfluid. In principle, there is no condensate in the thermodynamic limit, as $G_1$ decays to 0 with $r$. However, we may wonder if a condensate fraction survives in realistic systems of finite size $R$, or in other words, if $G_1(R)$ is still close to $n_0$. Below the transition, $\mathcal{D}_s>4$, and $G_1(R)/n_0>(\ell_T/R)^{1/4}$.

$\ell_T=1/q_T$ is given by Eq.~\eqref{eq:qT}. At the critical point $\mathcal{D}_s=4$, and in the limit where $\tilde{g}\ll 1$, it reads approximately $\ell_T\simeq \sqrt{2\tilde{g}/n_0}/\pi$, such that $\ell_T/R\simeq \sqrt{2\tilde{g}/\pi N}$. A typical estimation with $\tilde{g}\simeq 0.1$ and $N\simeq 10^5$ atoms yields $G_1(R)/n_0\simeq 0.17$. There is thus still a significant condensate fraction at the transition, which further increases at lower temperature.

\subsection{Experimental observations}
\subsubsection{Initial observation with helium films}
\begin{figure}[h]
\centering
\includegraphics[width=0.7\linewidth]{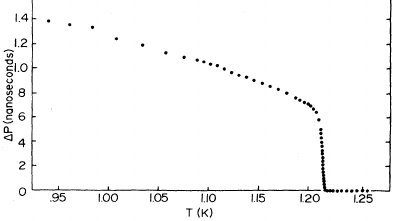}
\caption{First observation of the Kosterlitz-Thouless transition and the superfluid jump in 2D liquid helium, through the frequency shift of a torsion oscillator, see text for details. Figure from Ref. \cite{Bishop1978}.\label{fig:Bishop_Reppy}}
\end{figure}

The first experimental observation of the KT transition was performed in a superfluid helium film on the surface of a Mylar tape substrate, wounded as a spiral to act as a torsion oscillator \cite{Bishop1978}. Bishop and Reppy measured the shit of frequency of this torsion oscillator as temperature is lowered, and observe a sharp increase of the resonant frequency below a threshold, see Fig.~\ref{fig:Bishop_Reppy}, indicating that the 2D helium film becomes superfluid. The jump in frequency is consistent with BKT theory and the predicted superfluid jump $n_s\lambda_{\rm dB}^2=4$.

\subsubsection{First observation in quantum gases}
\begin{figure}[h]
\centering
\begin{minipage}{0.4\linewidth}
\includegraphics[width=\linewidth]{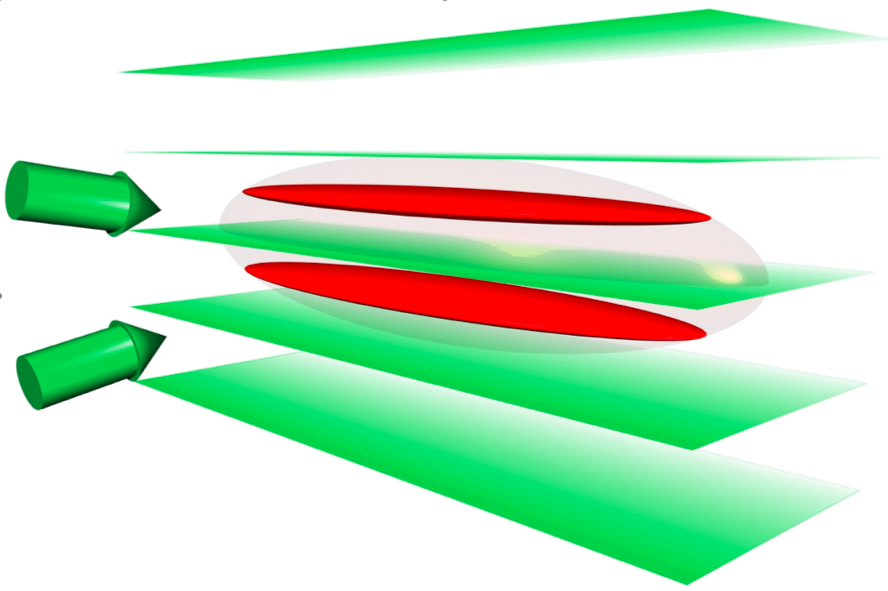}
\end{minipage}
\begin{minipage}{0.15\linewidth}
\centering
\includegraphics[width=\linewidth]{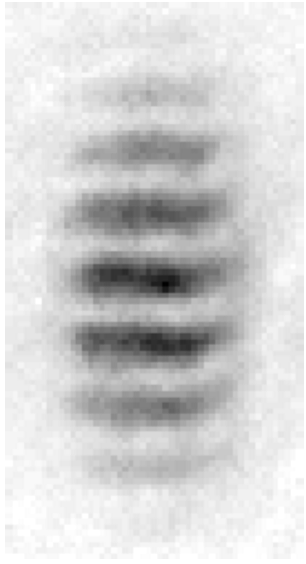}\\
{\small low $T$}
\end{minipage}
\begin{minipage}{0.15\linewidth}
\begin{tikzpicture}[scale=1.5]
\node at (0,0) {\includegraphics[width=\linewidth]{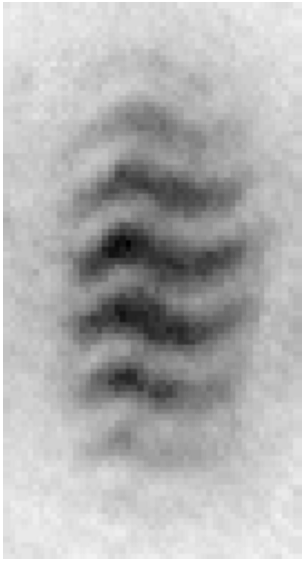}};
\node at (0,-1.7) {\small higher $T$};
\draw[dashed,red] (-0.2,-1.4) -- (-0.2,1.6);
\draw[dashed,red] (0.2,-1.4) -- (0.2,1.6);
\node[red] at (0,1.6) {\small $L_x$};
\end{tikzpicture}
\end{minipage}
\begin{minipage}{0.15\linewidth}
\centering
\begin{tikzpicture}
\node at (0,0) {\includegraphics[width=\linewidth]{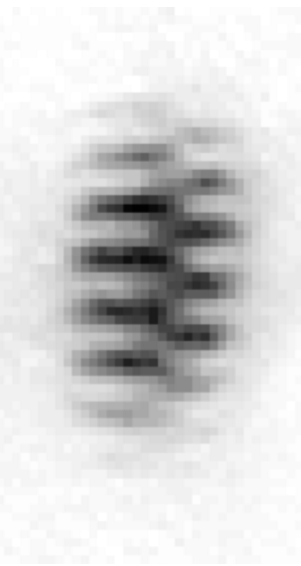}};
\node at (0,-1.5) {\small vortex};
\node at (0,2) {$\Delta\theta=$};
\node at (-0.5,1.5) {$0$};
\node at (0.5,1.5) {$\pi$};
\end{tikzpicture}
\end{minipage}
\caption{Measurement of phase correlations in a 2D gas. (a) Two 2D gases are prepared on top of each other in two successive nodes of an optical lattice with a period of several micrometers. (b) Typical interference fringes obtained after releasing the two gases. At low temperature, the phase is uniform in each sample ($\theta_1$ for one gas, $\theta_2$ for the other) and the interference fringes are straight, with an absolute position that fluctuates from one realization to the other and depend on the phase difference between the two samples $\Delta\theta\theta_2-\theta_1$. AT higher temperature, $\theta_1$ and $\theta_2$ fluctuate and depend on the position in the 2D gas, resulting in a deformation in the interference pattern. This is used to estimate the correlation function $G_1$ by averaging over a width $L$, in red. When a vortex is present in one of the gases, the pattern is phase shifted by $\pi$ from one side to the other of the vortex. Figure adapted from Ref. \cite{Hadzibabic2006}.\label{fig:2Dfringes}}
\end{figure}

The first observation of the KT transition with trapped quantum gases dates back to 2006 in the group of Jean Dalibard in Paris \cite{Hadzibabic2006}. The group prepared two quantum gases on top of each other, confined by a harmonic trap in the horizontal plane and by an optical lattice with a large period $d=\SI{3}{\micro\metre}$ in the vertical direction, obtained from two laser beams at a small angle. They use two 2D gases to enable interferences between these two samples when the gases were released in a time-of-flight experiment. Phase fluctuations in the samples lead to wiggles in the interference fringes, whereas coherent samples produced straight fringes. The presence of a vortex in one of the two samples is evidenced by a $\pi$ phase shift in the fringe pattern, as the phase difference between two opposite sides of a vortex is $\pi$, see Fig.~\ref{fig:2Dfringes}.

The absorption pictures give access to the integrated interference pattern along the imaging laser, say $y$. The integrated density pattern after expansion is of the form
\beq
n_{\rm int}(x,z)=\mathcal{A}(x,z)\left[1+c(x)\cos\left(\frac{2\pi z}{D_t}+\theta(x)\right)\right]
\eeq
where the fringe period $D_t=ht/(Md)$ depends on the time of free expansion $t=\SI{20}{\milli\second}$ and  on the initial splitting $d$, $\mathcal{A}$ results from the global cloud envelope, $c(x)$ is the contrast of the integrated density and $\theta(x)$ is the phase of the interference pattern at position $x$.

The first order correlation function $G_1$ is then related to the average complex contrast $C(x)=c(x)e^{i\theta(x)}$. If we label by $a$ and $b$ the two planes at initial position $z=\pm d/2$, the contrast at point $\mathbf{r}$ results from the interference of $\psi_a(\mathbf{r})$ and $\psi_b(\mathbf{r})$. The complex contrast $C(\mathbf{r})=\psi_a(\mathbf{r})\psi_b^*(\mathbf{r})$ averages to zero because the two planes are independent and their phase is not correlated, while the correlation of the complex contrast is related to $G_1$ \cite{DalibardCF2017en}:
\bea
\langle C(\mathbf{r})C^*(\mathbf{r}') \rangle &=& \langle \psi_a(\mathbf{r})\psi^*_b(\mathbf{r})\psi^*_a(\mathbf{r}')\psi_b(\mathbf{r}')\rangle\nonumber\\
&=& \langle \psi_a(\mathbf{r})\psi^*_a(\mathbf{r}')\rangle\langle\psi^*_b(\mathbf{r})\psi_b(\mathbf{r}')\rangle \quad (a \mbox{ and } b \mbox{ independent)}\nonumber\\
&=&G_1(\mathbf{r},\mathbf{r}')G_1^*(\mathbf{r},\mathbf{r}')=|G_1(\mathbf{r},\mathbf{r}')|^2.
\eea

The Paris group also took advantage of the fact that the square of $G_1$ averaged over $y$ and a given length $L_x$ along $x$ is shown\footnote{In principle, this also requires that the length $L_y$ over which the average is taken satisfies $L_y\ll L_x$ \cite{Polkovnikov2005}.} to scale as a power law with $L_x$ \cite{Polkovnikov2005,Hadzibabic2006}:
\beq
\frac{1}{L_x}\int_{-L_x/2}^{L_x/2}\!\! dx \left|G_1(x,0)\right|^2  \propto \frac{1}{L_x^{2\eta}}.
\eeq
The exponent $\eta$ is expected to be $\eta=1/4$ in the superfluid phase and $\eta=1/2$ in the normal phase. The integration over $y$ comes directly with the absorption along the imaging axis, while an average of the square contrast over a length $L_x$ for various choices of $L_x$, see Fig.~\ref{fig:2Dfringes}, gives access to the exponent $\eta$.

\begin{figure}[h]
\centering
\begin{tikzpicture}
\node at (-3.5,-0.16) {\includegraphics[width=46.5mm,clip=true,trim=15mm 0mm 0mm 0mm]{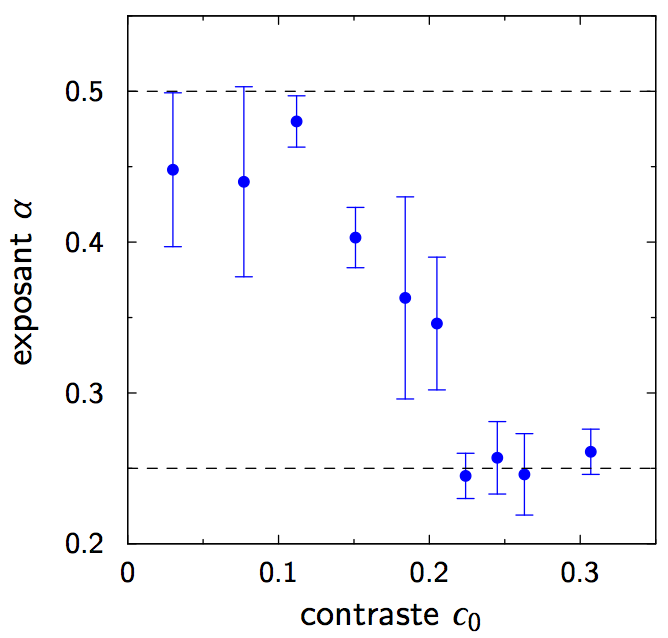}};
\node[rotate=90] at (-6,0) {\scriptsize exponent $\eta$};
\node[fill=white] at (-3.1,-2.4) {\scriptsize contrast $c_0$};
\node at (-2,0.5) {\sf SF};
\node at (-4.2,-0.5) {\sf thermal};
\node at (3,0) {\includegraphics[width=45mm]{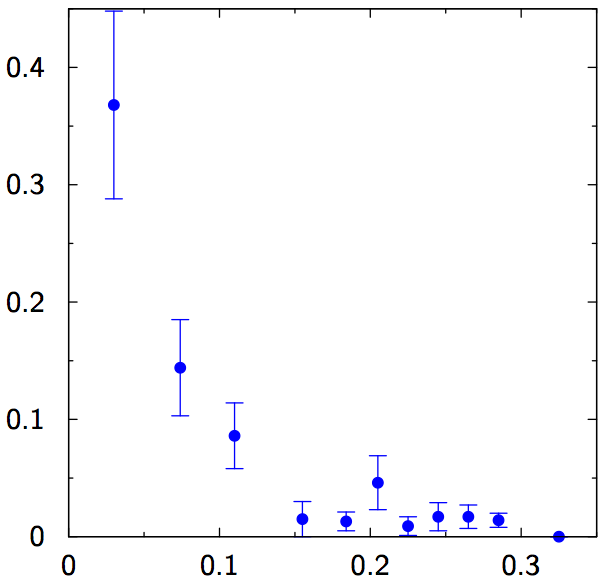}};
\node at (3,-2.4) {\scriptsize contrast $c_0$};
\node[rotate=90] at (0.5,0) {\scriptsize fraction of pictures with vortices};
\node at (3,1) {\includegraphics[width=15mm]{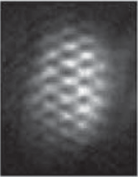}};
\end{tikzpicture}
\caption{Left: Exponent $\eta$ as a function of the contrast $c_0$ in the central region $x=0$ of the cloud. Right: Fraction of pictures with vortices as a function of $c_0$. Inset: example picture with at least 4 vortices. Figure adapted from Ref. \cite{Hadzibabic2006}.\label{fig:BKT2006}}
\end{figure}
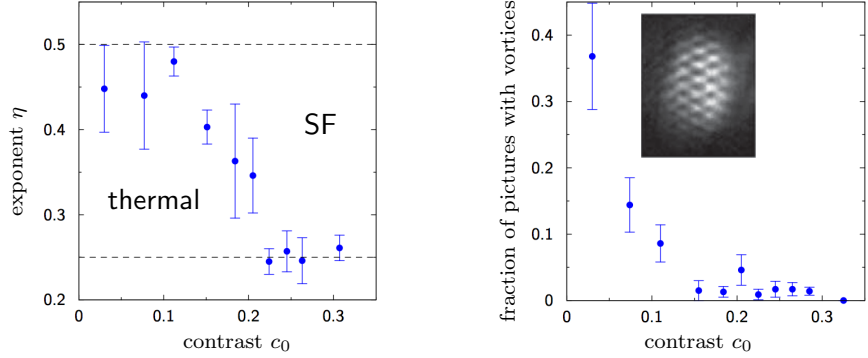

The variation of $\eta$ with the central contrast $c_0$, which is an indication of the cloud's temperature, evidences a transition between $\eta=1/2$ and $\eta=1/4$ for $c_0\approx 0.15$. This first evidence of the BKT transition is confirmed by the more frequent observation of vortices in the region where $c_0<0.15$, see Fig.~\ref{fig:BKT2006}.

Note that vortices can have positive and negative charges, and vortices with opposite charge should pair below the KT critical temperature, with an intervortex distance that is reduced as the temperature is lowered. In 2013, the group of Yong-il Shin in Seoul studied this phenomenon in a 2D harmonic trap \cite{Choi2013a}, and observed a reduced value for the most probable distance between vortices when the temperature was lower. At lower temperature, the vortices where pushed towards the edges of the cloud, where the density ---and thus the superfluid fraction--- is lower.

\subsubsection{Paris experiment revisited}
\begin{figure}[h]
\centering
\includegraphics[height=0.3\linewidth,clip=true, trim=5mm 1.5mm 74mm 1.5mm]{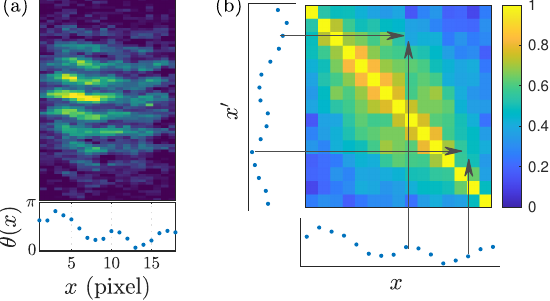}
\includegraphics[height=0.3\linewidth,clip=true, trim=105mm 1.5mm 5mm 1.5mm]{2D_Foot_extract_C}
\caption{Reconstruction of the integrated correlation function $G_1(x,x')$ (c) from the correlations (b) between local phases at positions $x$ and $x'$ of the interference between two expanding 2D gases (a). See text for a detailed description. Figure adapted from \cite{Sunami2022}. \label{fig:Foot_correlation_measure}}
\end{figure}

In 2022, experiments were performed in the group of Chris Foot in Oxford \cite{Sunami2022} in the same spirit, starting with two 2D gases confined in harmonic traps of frequency $\omega_0$, prepared on top of each other with magnetic and radiofrequency fields \cite{Perrin2017,Dubessy2025}. From a similar analysis of the interference fringes after a time-of-flight expansion, the Oxford group studied the decay of the correlation function and the proliferation of vortices in the normal phase. The procedure they used is illustrated in Fig.~\ref{fig:Foot_correlation_measure}:
(i) at a given temperature $T$, prepare two 2D quantum gases parallel to the $xy$ plane, release them and select a slice around $y=0$ for imaging;
(ii) Fig.~\ref{fig:Foot_correlation_measure}(a): recover the local phase $\theta(x)$ at each position $x$ from the fringes observed in time-of-flight;
(iii) Fig.~\ref{fig:Foot_correlation_measure}(b) compute the $(x,x')$ phase correlation $C(x,x')$, and Fig.~\ref{fig:Foot_correlation_measure}(c) represents $C(\bar x)$ as a function of the distance $\bar x=|x-x'|$. The result is averaged over a zone of constant $\bar x$, see the inset of Fig.~\ref{fig:Foot_correlation_measure}(c);
(iv) repeat at various values of $T$.

\begin{figure}[h]
\centering
\begin{tikzpicture}
\node at (-2.2,2.1) {\scriptsize (a) fit $C(\bar x)$ for various $T_0$};
\node at (-2,0) {\includegraphics[height=0.25\linewidth,clip=true,trim=27mm 0mm 0mm 0mm]{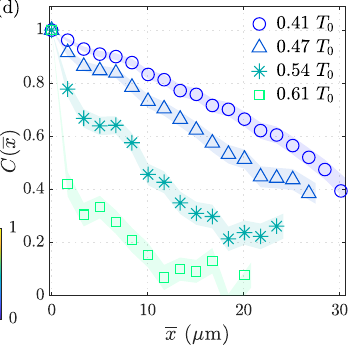}};
\fill[white] (-3.9,-1.7) rectangle (-3.65,-0.4);
\fill[white] (-3.9,2) rectangle (-3.6,1.6);
\node at (2.2,2.1) {\scriptsize (b) algebraic vs exponential decay};
\node at (1.9,0) {\includegraphics[height=0.25\linewidth,clip=true, trim=0mm 0mm 126mm 0mm]{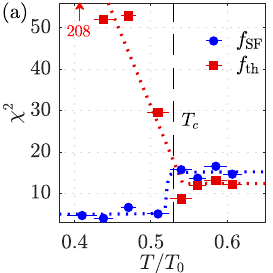}};
\fill[white] (0,1.5) rectangle (0.4,1.9);
\node at (5.35,0) {\includegraphics[height=0.17\linewidth,clip=true, trim=6mm 52.5mm 43mm 0mm]{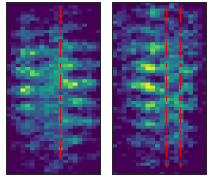}};
\node at (5.4,1.5) {\scriptsize (c) vortex detection};
\node at (9,0) {\includegraphics[height=0.2\linewidth,clip=true, trim=42mm 52.5mm 1mm 0mm]{2D_Foot_vortex}};
\node at (9.2,1.5) {\scriptsize (d) vortex probability $P_v$};
\fill[white] (6.9,1.5) rectangle (7.4,1.1);
\end{tikzpicture}
\caption{Analysis of the behaviour of (a) $C(\bar x)$ at various values of the temperature $T/T_0$ where $k_BT_0=\sqrt{6N}\hbar\omega_0/\pi$ is the critical temperature for BEC in the harmonic trap of frequency $\omega_0$, see Eq.~\eqref{eq:Tc_BEC_2D_harmonic}. (b) Error $\chi^2$ when $C(\bar x)$ is fitted by an algebraic function $f_{\rm SF}(\bar x)\propto {\bar x}^{-2\eta}$ (blue dots) or by an exponential function $f_{\rm th}(\bar x)\propto e^{-\bar x/\ell}$ (red squares). A sudden change of behavior occurs around $T_c=\SI{0.53\pm0.01}{}T_0$, as the algebraic fit is much better in the superfluid phase where $T<T_c$ while the exponential fit is slightly better in the thermal phase. This critical temperature $T_c$ is compared with the one obtained from the probability to detect vortices. (c) Vortices appear as dislocations in the interference pattern (one vortex at the position of the red dashed line in the left picture, two vortices in the right picture). (d) Vortices proliferate above a critical temperature $T_v=\SI{0.52\pm0.01}{}T_0$, equal to $T_c$ within the experimental error. Figure adapted from \cite{Sunami2022}. \label{fig:Foot_BKT}}
\end{figure}

The different curves $C(\bar x)$ obtained at different temperatures are fitted by an algebraic $f_{\rm SF}(\bar x)\propto {\bar x}^{-2\eta}$ and by an exponential function $f_{\rm th}(\bar x)\propto e^{-\bar x/\ell}$, see Fig.~\ref{fig:Foot_BKT}. As the transition from the superfluid to the normal phase is crossed, the best fitting function transitions from an algebraic to an exponential function, as expected from the BKT theory. The exponent at the transition for the algebraic fit is around $\eta=\SI{0.17\pm0.03}{}$, slightly different from $1/4$ due to the inhomogeneity in the harmonic trap, in agreement with Monte-Carlo calculations. The change of behavior of the correlation functions occurs at the same critical temperature than the one where vortices start to proliferate.

\subsubsection{Superfluid jump measured from the second sound}
Finally, let us present a beautiful experiment from the Cambridge group of Zoran Hadzibabic, who observed the superfluid jump $n_s\lambda_{\rm dB}^2=4$ at the transition in a uniform 2D Bose gas \cite{Christodoulou2021}. This is made possible by the trapping of the 2D gas of $^{39}$K atoms in a box, yielding a uniform density, and the tuning of the interaction constant to a relatively large value $\tilde{g}=0.64$ through a Feshbach resonance.

While the total density can be inferred by a measurement of the total atom number, the superfluid density $n_s$ is deduced from a spectroscopic measurement of the first and second speeds of sound in the gas. Indeed, a superfluid at finite temperature not only features the usual, first sound with velocity $c_1$, but also a second sound with a different velocity $c_2$. In incompressible superfluids, the first sound, discussed in Lecture 1, corresponds to density waves, while the second sound can be understood as a heat wave. In a Bose gas near the critical temperature, we cannot make this distinction and the two sounds are combinations of density and heat waves. They are solutions of a quartic equation \cite{Christodoulou2021}
\beq
c^4-(c_{10}^2+c_{20}^2)c^2+\gamma^{-1}c_{10}^2c_{20}^2=0
\label{eq:sound}
\eeq
with solutions $c_1$ and $c_2$, where $c_{20}^2\propto Tn_s/(n-n_s)$ depends explicitly on the superfluid density, and vanishes above the critical temperature, where only the first sound survives and matches the classical expression $c=c_{10} = 1/\sqrt{n M \kappa_s}$ with $\kappa_s$ the isentropic (adiabatic) compressibility.

\begin{figure}[t]
\centering
\begin{tikzpicture}
\node at (-6,-0.5) {\includegraphics[width=0.35\linewidth, clip=true, trim=6mm 0.2mm 137mm 1mm]{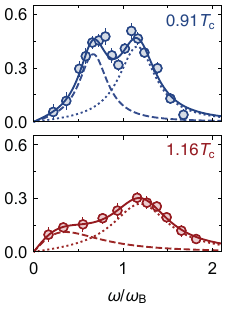}};
\node at (-6,3.2) {\footnotesize (a) Spectroscopy of sound modes};
\node at (-0.05,2) {\includegraphics[width=0.35\linewidth,clip=true,trim=7mm 5mm 100mm 3mm]{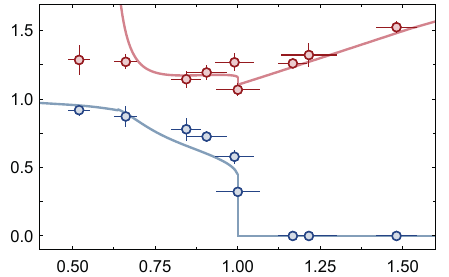}};
\node[red] at (0,3.2) {\sf \scriptsize 1st sound};
\node[blue] at (1.3,1.5) {\sf \scriptsize 2nd sound};
\node at (0,0.2) {\scriptsize $T/T_c$};
\node at (-2.7,2) {\scriptsize $c/c_0$};
\node at (-2.8,3.2) {\footnotesize (b)};
\node at (0,-1.7) {\includegraphics[width=0.34\linewidth,clip=true,trim=98mm 5mm 11mm 3mm]{Cambridge_sound_jump}};
\draw[->,blue] (0.2,-2.2) -- (-0.2,-2.2);
\node[blue] at (1.2,-2.2) {\sf \scriptsize superfluid jump};
\node at (0,-3.5) {\scriptsize $\mathcal{D}/\mathcal{D}_c$};
\node at (-2.7,-1.7) {\scriptsize $\mathcal{D}_s$};
\node at (-2.8,-0.5) {\footnotesize (c)};
\end{tikzpicture}
\caption{Observation of the superfluid jump at the BKT transition in a uniform 2D quantum gas. (a) Amplitude of the center-of-mass motion as a function of the excitation frequency, for a temperature below (upper graph) or above (lower graph) the critical temperature. (b) Measured values of $c_1$ and $c_2$ as a function of temperature. (c) Deduced superfluid phase space density $\mathcal{D}_s$ as a function of the total phase space density, normalized to its value at the transition. $\mathcal{D}_s$ jumps from $0$ to $4$ at the transition. Figure adapted from \cite{Christodoulou2021}. \label{fig:Hadzibabic_BKT}}
\end{figure}

The measurement of $c_1$ and $c_2$ is obtained by driving the center-of-mass motion in the box with an oscillating uniform force, and measuring the amplitude of the response oscillation. Below $T_c$, two peaks are present, see Fig.~\ref{fig:Hadzibabic_BKT}(a), top graph: the lowest peak corresponds to $c_2$ and the highest to $c_1$. Above $T_c$, lower graph, only one peak emerges, corresponding to $c_1$. $c_1$ and $c_2$ as a function of temperature are computed from the resonances, see Fig.~\ref{fig:Hadzibabic_BKT}(b), and the superfluid density $n_s$ is obtained by inverting Eq.~\eqref{eq:sound}, that can also write
\beq
c^4-(c_{1}^2+c_{2}^2)c^2+\gamma c_{1}^2c_{2}^2=0
\eeq
where now the two solutions are $c=c_{10}$ and $c=c_{20}$. Fig.~\ref{fig:Hadzibabic_BKT}(c) shows the superfluid phase space density $\mathcal{D}_s$ as a function of the total phase space density. A jump $\mathcal{D}_s=4$ is observed at the transition, in agreement with BKT theory.

\section{Scaling symmetry in 2D}
In this section, we will discuss yet another specificity of the dimension 2: the fact that the interacting 2D gas obeys a scaling symmetry. The consequence is a universal equation of state, that depends only on the ratio $\mu/k_BT$ of the chemical potential and the temperature, and not on $\mu$ and $T$ independently. Moreover, the 2D gas confined in an harmonic potential of frequency $\omega_0$ presents an undamped monopole mode at $\Omega=2\omega_0$, with a frequency independent of the oscillation amplitude.
\subsection{Scaling symmetry}
A scaling symmetry occurs if, in a dilation of space variables by a factor $\lambda$ and of time variables by $\lambda^2$, the energy scales as $1/\lambda^2$.
\beq
\left.
\begin{array}{c}
\mathbf{r} \longrightarrow \tilde{\mathbf{r}} = \lambda \mathbf{r}\\
t \longrightarrow \tilde{t}=\lambda^2 t
\end{array}
\right\}
\Rightarrow
E \longrightarrow \widetilde{E}= \frac{1}{\lambda^2}E,
\eeq
or more precisely if
\beq
E(\lambda \mathbf{r}, \lambda^2 t) = \frac{1}{\lambda^2}E(\mathbf{r},t).
\eeq

\paragraph{Kinetic energy} This scaling always holds for classical kinetic energy $E_{\rm kin} = Mv^2/2$:
\bea
\mathbf{v} = \frac{d\mathbf{r}}{dt} &\longrightarrow& \frac{1}{\lambda^2}\times \lambda \mathbf{v} = \frac{1}{\lambda}\mathbf{v}\nonumber\\
\Rightarrow E_{\rm kin} &\longrightarrow& \frac{1}{\lambda^2}E_{\rm kin} \nonumber.
\eea
It also holds in quantum mechanics. The transformation of $\psi(\mathbf{r})$ under scaling in dimension $D$ is $\widetilde{\psi}(\tilde{\mathbf{r}}) = \lambda^{-D/2}\psi(\tilde{\mathbf{r}}/\lambda) = \lambda^{-D/2}\psi(\mathbf{r})$ to ensure the normalization of the wave function. Then, the scaled kinetic energy is determined by
\beq
\widetilde{E}_{\rm kin}\propto\int d^D\tilde{\mathbf{r}} \left|\nablagras\widetilde{\psi}\left(\tilde{\mathbf{r}}\right)\right|^2  = \int \frac{d^D\tilde{\mathbf{r}}}{\lambda^D}\left|\frac{1}{\lambda}\nablagras\psi\left(\frac{\tilde{\mathbf{r}}}{\lambda}\right)\right|^2=\frac{1}{\lambda^2}\int d^D\mathbf{r}\left|\nablagras\psi\left(\mathbf{r}\right)\right|^2
\eeq
such that the kinetic energy verifies the scaling symmetry.

\paragraph{Interaction energy} If we now consider contact interactions, the interaction potential in dimension $D$ writes $V_{\rm int}(\mathbf{r})=g\delta^{(D)}(\mathbf{r})$ where $\delta^{(D)}$ is the Dirac-$\delta$ distribution in dimension $D$. The scaled potential writes
\beq
\widetilde{V}_{\rm int}(\tilde{\mathbf{r}}) = g\delta^{(D)}\left(\tilde{\mathbf{r}}\right) = g\delta^{(D)}\left(\lambda\mathbf{r}\right)= \frac{g}{\lambda^D}\delta^{(D)}\left(\mathbf{r}\right) = \frac{1}{\lambda^D}V_{\rm int}\left(\mathbf{r}\right)
\eeq
to ensure that
\beq
\int d^D\tilde{\mathbf{r}}\,\widetilde{V}_{\rm int}(\tilde{\mathbf{r}}) = g.
\eeq
As a result, the interaction energy follows the scaling symmetry only in dimension $D=2$. This is related to the absence of length scale in the contact interaction potential, which is characterized by a dimensionless number $\tilde{g}$, as discussed in Sec.~\ref{sec:2D_GPE} and \ref{sec:3D_2D}.

\subsection{Scale invariant equation of state}
A first consequence of the scaling symmetry is found in the equation of state (EOS) of the 2D gas. The derivation presented here follows the lectures at Collège de France by Jean Dalibard \cite{DalibardCF2017en}. In the grand canonical ensemble, the thermodynamic variables depend a priori on both $\mu$ and $T$, and are derived from the distribution law: the Maxwell-Boltzmann law for a classical gas, or here the Bose-Einstein distribution for our bosonic quantum gas. The probability of a state with $N$ particles at positions and momenta $\{\mathbf{r}_j,\{\mathbf{p}_j\}$ is given in this case by
\beq
\mathcal{P}\left(N,\{\mathbf{r}_j,\mathbf{p}_j\}\right) = \frac{1}{\ds e^{\beta E\left(N,\{\mathbf{r}_j,\mathbf{p}_j\}\right) - \beta\mu}-1}
\eeq
with $\beta^{-1}=k_BT$.

Let's consider a thermodynamic quantity $\mathcal{F}$, typically the phase space density $\mathcal{D}$. Its value at equilibrium is given by a weighted integral of its value $F\left(N,\{\mathbf{r}_j,\mathbf{p}_j\}\right)$ for the various configurations:
\beq
\mathcal{F}(\mu,T) = \sum_N \int \mathcal{P}\left(N,\{\mathbf{r}_j,\mathbf{p}_j\}\right) F\left(N,\{\mathbf{r}_j,\mathbf{p}_j\}\right) \prod_{j=1}^N \dd^D\mathbf{r}_j \, \dd^D\mathbf{p}_j.
\eeq
The first sum scans all possible particle number, and at fixed $N$ the integral counts the contribution of all possible configurations $\left(N,\{\mathbf{r}_j,\mathbf{p}_j\}\right)$.

Assume now that $F$ obeys a scaling law with exponent $2\nu$, such that
\beq
F\left(N,\{\lambda\mathbf{r}_j,\frac{1}{\lambda}\mathbf{p}_j\}\right) = \frac{1}{\lambda^{2\nu}} F\left(N,\{\mathbf{r}_j,\mathbf{p}_j\}\right).
\eeq
The scaled thermodynamic quantity is $\mathcal{F}\left(\mu/\lambda^2,T/\lambda^2\right)$, such that $\beta\mu$ in the distribution is unchanged, while $\beta E$ now becomes $\lambda^2\beta E$:
\beq
\mathcal{F}\left(\frac{1}{\lambda^2}\mu,\frac{1}{\lambda^2}T\right) = \sum_N \int \frac{1}{\ds e^{\lambda^2\beta E\left(N,\{\mathbf{r}_j,\mathbf{p}_j\}\right) - \beta\mu}-1} F\left(N,\{\mathbf{r}_j,\mathbf{p}_j\}\right) \prod_{j=1}^N \dd^D\mathbf{r}_j \, \dd^D\mathbf{p}_j.
\eeq
We now introduce a change in the integration variables
\beq
\mathbf{r}'_j = \frac{1}{\lambda}\mathbf{r}_j \quad \mathbf{p}'_j = \lambda\mathbf{p}_j, \quad \mbox{or equivalently} \quad \mathbf{r}_j = \lambda\mathbf{r}'_j \quad \mathbf{p}_j = \frac{1}{\lambda}\mathbf{p}'_j,
\eeq
and we use the scaling symmetry
\bea
\lambda^2 E\left(N,\{\mathbf{r}_j,\mathbf{p}_j\}\right) = \lambda^2 E\left(N,\left\{\lambda\mathbf{r}'_j,\frac{1}{\lambda}\mathbf{p}'_j\right\}\right) = \lambda^2\times \frac{1}{\lambda^2} E\left(N,\{\mathbf{r}'_j,\mathbf{p}'_j\}\right) = E\left(N,\{\mathbf{r}'_j,\mathbf{p}'_j\}\right)
\eea
as well as the scaling of the $F$ function:
\beq
F\left(N,\{\mathbf{r}_j,\mathbf{p}_j\}\right) = F\left(N,\left\{\lambda\mathbf{r}'_j,\frac{1}{\lambda}\mathbf{p}'_j\right\}\right) = \frac{1}{\lambda^{2\nu}}F\left(N,\{\mathbf{r}'_j,\mathbf{p}'_j\}\right).
\eeq
Finally, we remark that $\dd^D\mathbf{r}_j \, \dd^D\mathbf{p}_j = \dd^D\mathbf{r}'_j \, \dd^D\mathbf{p}'_j$. We finally get for the scaled $\mathcal{F}$
\beq
\mathcal{F}\left(\frac{1}{\lambda^2}\mu,\frac{1}{\lambda^2}T\right) = \frac{1}{\lambda^{2\nu}} \mathcal{F}(\mu,T).
\eeq
We chose $\lambda^2=T/T_0$ with some arbitrary reference temperature $T_0$ and we find
\beq
\mathcal{F}\left(\frac{\mu}{k_BT}\, k_BT_0,T_0\right) = \left(\frac{T_0}{T}\right)^{2\nu} \mathcal{F}(\mu,T).
\eeq
The left-hand side is a function $f(\mu/k_BT)=f(\alpha)$ of $\alpha=\beta\mu=\mu/k_BT$ only, such that we can finally write
\beq
\mathcal{F}(\mu,T) = (k_BT)^\nu \,f\left(\frac{\mu}{k_BT}\right).
\label{eq:scaling}
\eeq

Let us examine the case of the phase space density $\mathcal{D}=n(\mathbf{r})\lambda_{\rm dB}^D$, where the density is given by
\beq
n\left(N,\{\mathbf{r}_j,\mathbf{p}_j\}\right) = \sum_{j=1}^N \delta^{(D)}(\mathbf{r}-\mathbf{r}_j).
\eeq
In the scaling transformation, the density scales as $1/\lambda^D$, which yields $\nu=D/2$. From Eq.~\eqref{eq:scaling}, we get for the density
\beq
n(\mu,T) = (k_BT)^{D/2} \,f_n\left(\frac{\mu}{k_BT}\right).
\eeq
Note that the function $f_n$ depends on the strength $\tilde{g}$ of the interactions. We can thus write the phase space density as
\beq
\mathcal{D} = \lambda_{\rm dB}^D n(\mu,T) = \left(\frac{2\pi\hbar^2}{Mk_BT}\right)^{D/2}(k_BT)^{D/2}\,f_n\left(\frac{\mu}{k_BT}\right) = f_\mathcal{D}\left(\frac{\mu}{k_BT}\right)
\eeq
with $f_\mathcal{D}(u) = (2\pi\hbar^2/M)^{D/2}f_n(u)$. The terms in $(k_BT)^{D/2}$ simplify and we find that the phase space density depends only on the ratio $\mu/T$ (and on $\tilde{g}$), but not independently on $\mu$ and $T$. It yields a \textbf{universal law} for the equation of state, valid if the scaling symmetry holds:
\beq
\boxed{
\mathcal{D} = f(\alpha,\tilde{g})
}
\eeq
where we have made explicit the dependence on $\tilde{g}$.

\begin{figure}[t]
\centering
\begin{tikzpicture}
\node at (0,0) {\includegraphics[width=0.7\linewidth,clip=true,trim=7mm 7.5mm 0mm 5mm]{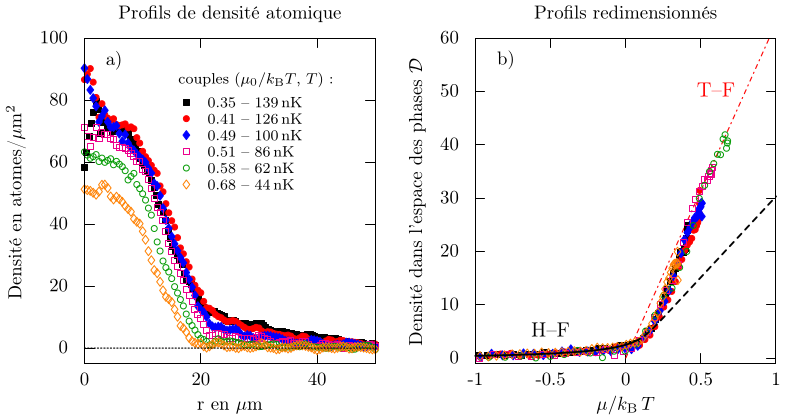}};
\node[rotate=90] at (-5.5,0) {\footnotesize angular-averaged density $n(r)$ in $\mu$m$^{-2}$};
\node at (-2.5,-2.8) {\footnotesize $r$ in $\mu$m};
\node at (3,-2.8) {\footnotesize $\alpha=\mu/k_BT$};
\fill[white] (0,-2) rectangle (0.3,2.1);
\node[rotate=90] at (0.1,0) {\footnotesize phase space density $\mathcal{D}$};
\end{tikzpicture}
\caption{(a): Angular-averaged density profiles $n(r)$ of a harmonically trapped 2D Bose gas, for different temperatures. (b): Same data in a plot of the local phase space density as a function of the local scaled chemical potential $\alpha=\mu_{\rm loc}/k_BT$. All curves collapse on a single curve giving the EOS $\mathcal{D}=f(\alpha,\tilde{g})$. Here, $\tilde{g}=0.11$ is fixed. Figure adapted from Tarik Yefsah thesis \cite{YefsahThese}. See also \cite{Yefsah2011}. H-F: Hartree-Fock regime (degenerate phase). T-F: Thomas-Fermi regime (superfluid phase).
\label{fig:EOS_Yefsah}}
\end{figure}
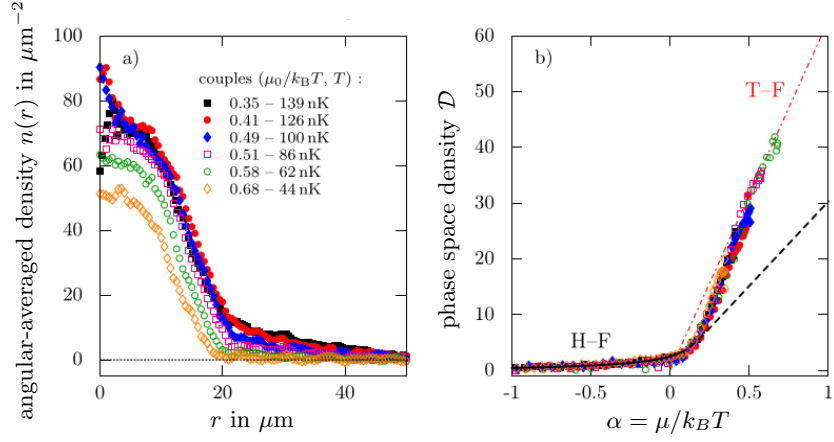

\paragraph{Measurement of the EOS} This prediction has been verified experimentally for 2D quantum gases confined in harmonic traps, taking advantage of the built-in density distribution in the trap to measure simultaneously the equation of state at different values of the density \cite{Yefsah2011,Hung2011,Ha2013}. The idea is to assume that, at equilibrium, the equation of state holds locally, in a volume large enough to average over fluctuations but small enough to neglect the variation of the trapping potential on that scale. The density profile $n(\mathbf{r})$ gives access to all values of the phase space density between $0$ and $\mathcal{D}_{\rm max} = n_{\rm max}\lambda_{\rm dB}^2$ as a function of the local chemical potential $\mu_{\rm loc}(\mathbf{r})=\mu - V(\mathbf{r})$, which belongs to the interval $\mu_{\rm loc}\in(0,V_{\rm max})$ where $V_{\rm max}$ is the trapping potential at the edge of the atomic cloud, and we have taken the potential minimum $V_{\rm min}=0$ as the origin of energy. Fig.~\ref{fig:EOS_Yefsah} presents the result of this procedure applied on a 2D Bose gas of rubidium 87 atoms in a harmonic trap, at fixed value of the interaction constant $\tilde{g}=0.11$ \cite{Yefsah2011,YefsahThese}.

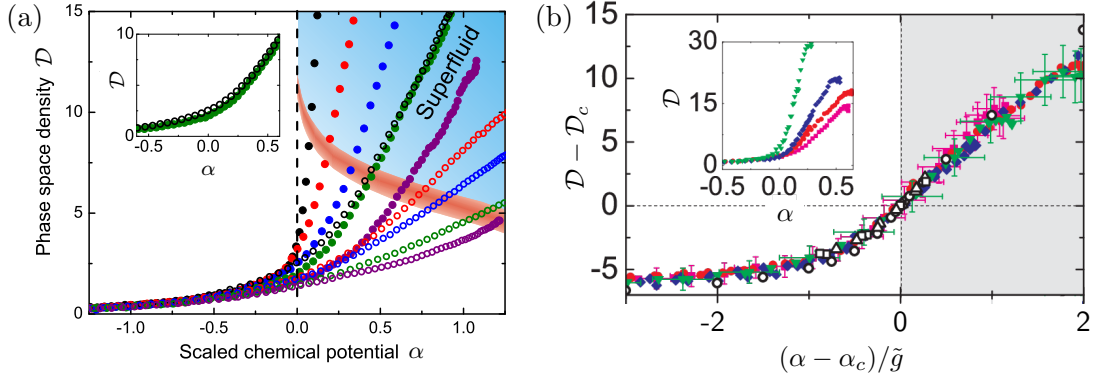
\begin{figure}
\centering
\begin{tikzpicture}
\node at (-7.5,-0.5) {\includegraphics[width=70mm]{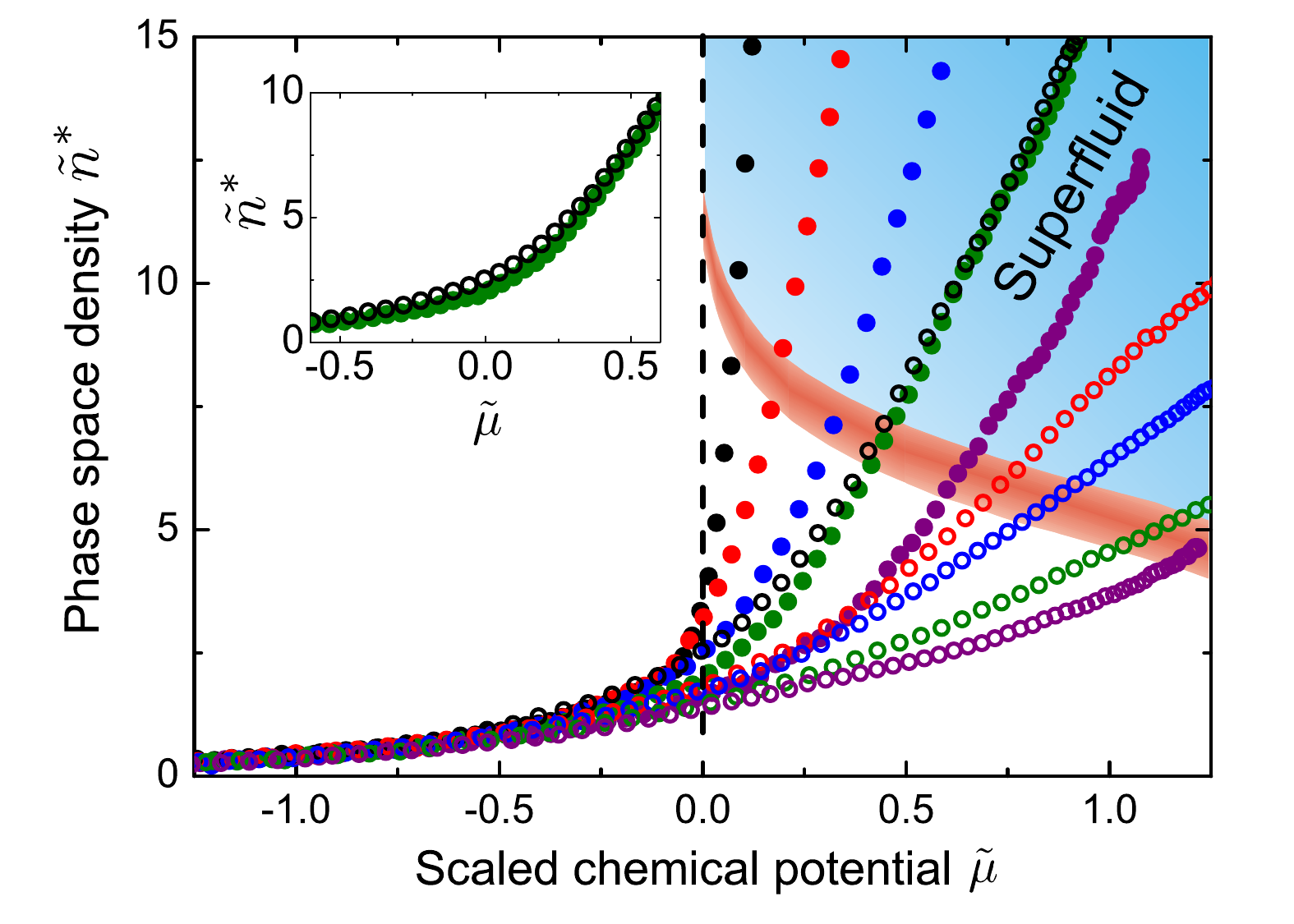}};
\node[rotate=90,fill=white] at (-10.55,1.2) {\footnotesize $\mathcal{D}$};
\node[rotate=90,fill=white] at (-9.58,0.9) {\scriptsize $\mathcal{D}$};
\node[fill=white] at (-8.37,-0.3) {\scriptsize $\alpha$};
\node[fill=white] at (-5.62,-2.72) {\footnotesize $\alpha$};
\node at (0,0) {\includegraphics[width=70mm,clip=true,trim=30mm 134mm 0mm 0mm]{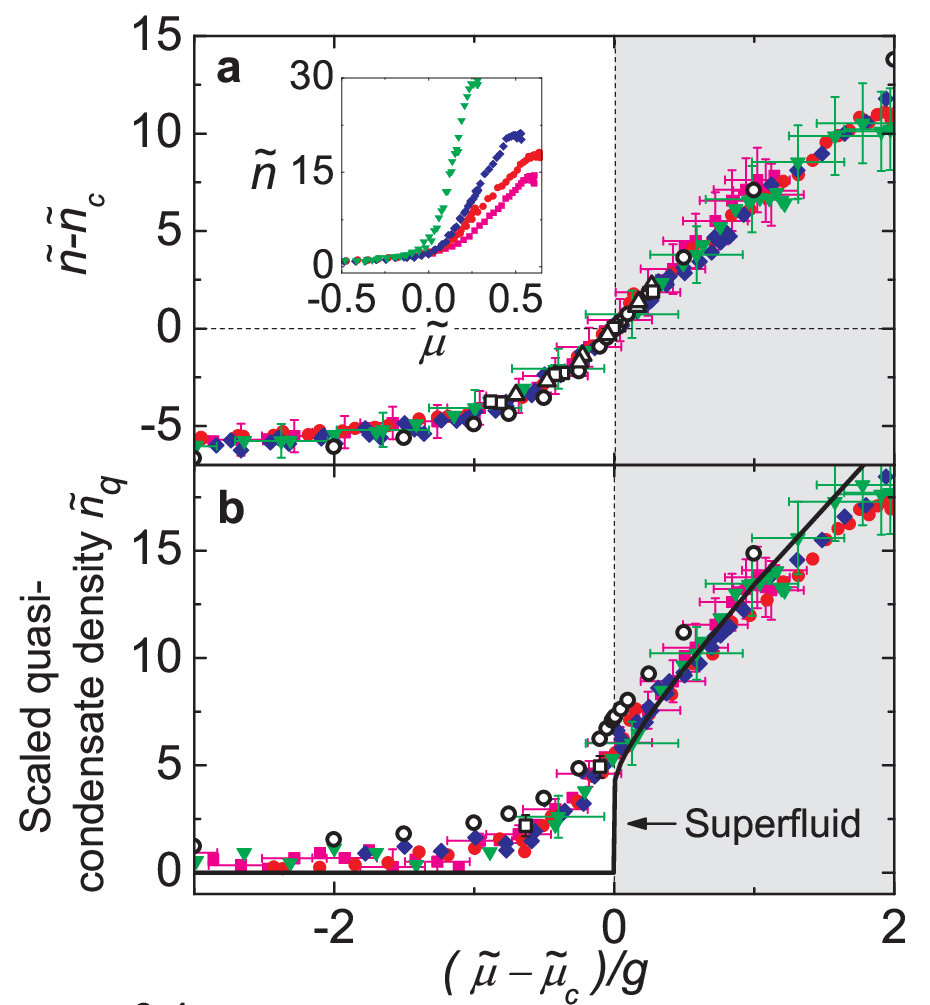}};
\node[rotate=90,fill=white] at (-3.5,0) {\small $\mathcal{D}-\mathcal{D}_c$};
\node[rotate=90,fill=white] at (-2.25,0.65) {\footnotesize $\mathcal{D}$};
\node[fill=white] at (-0.74,-0.9) {\small $\alpha$};
\node[fill=white] at (0,-2.8) {\small $(\alpha-\alpha_c)/\tilde{g}$};
\draw[white,fill=white] (-2.4,1.2) rectangle (-2.7,1.6);
\draw[white,fill=white] (0.805,-1.973) rectangle (1.97,-2.1);
\draw[white,fill=white] (2.01,-1.973) rectangle (3.175,-2.1);
\node at (-1.7,-2.3) {\sf -2};
\node at (0.78,-2.3) {\sf 0};
\node at (3.2,-2.3) {\sf 2};
\node at (-10.8,1.7) {(a)};
\node at (-3.8,1.7) {(b)};
\end{tikzpicture}
\caption{(a) Phase space density as a function of the scaled chemical potential $\alpha$, for various values of $\tilde{g}$. Close circles: $\tilde{g}=0.05$ (black), 0.15 (red), 0.24 (blue), 0.41 (green), and 0.66 (purple). Open circles: $\tilde{g}=0.45$ (black), 0.85 (red), 1.2 (blue),
1.9 (green), and 2.8 (purple). For these data (open circles), a lattice was added to enhance the value of $\tilde{g}$ with an increased effective mass ($M^*>M$). Inset: $\mathcal{D}(\alpha)$ for $\tilde{g}=0.4$, obtained without (green dots) and with (open black circles) the lattice. Figure adapted from \cite{Ha2013}. (b) Universal behavior around the critical point: all data overlap when $\mathcal{D}-\mathcal{D}_c$ is plotted against $(\alpha-\alpha_c)/\tilde{g}$. Inset: $\mathcal{D}(\alpha)$ for $\tilde{g}=0.05$ (green), 0.13 (blue), 0.19 (red) and 0.26 (magenta). Figure adapted from \cite{Hung2011}.
\label{fig:EOS_Chin}}
\end{figure}

This procedure was repeated in the group of Cheng Chin using cesium atoms, which gives access to a tuning of the interaction constant $\tilde{g}$ \cite{Hung2011,Ha2013}. A series of EOS for each value of $\tilde{g}$ between 0.05 and 2.8 was recorded, see Fig.~\ref{fig:EOS_Chin}. To reach the largest values of $\tilde{g}$, an square optical lattice was added in the plane: this results in an increased effective mass $M^*>M$, which in turn increases $\tilde{g}\propto \sqrt{M^*}$, see Eq.~\eqref{eq:gtilde}.

In addition, close to the critical point $\alpha=\alpha_c$, the equation of state presents a universal behavior for all values of $\tilde{g}$ \cite{Hung2011}:
\beq
\mathcal{D}-\mathcal{D}_c = f(\alpha,\tilde{g}) - f(\alpha_c,\tilde{g}) = \mathcal{Y}\left(\frac{\alpha-\alpha_c}{\tilde{g}}\right) \quad \mbox{near } \alpha\simeq\alpha_c
\eeq
where $\mathcal{Y}$ is a universal function. All curves collapse on the same curve around the critical point, yielding a universal phase diagram, a signature of the scaling symmetry.

\subsection{Monopole mode}
Another signature of the scaling symmetry is given by the frequency of the monopole (or breathing) mode in a harmonically trapped 2D quantum gas. In a harmonic trap, the potential energy, quadratic in the spatial coordinate, is modified through the scaling transformation as $E_{\rm pot} \longrightarrow \lambda^2 E_{\rm pot}$. Together with the scaling invariance for kinetic and interaction energy, it brings the $\lambda^2 + 1/\lambda^2$ isochronous potential\footnote{The harmonic potential $V(r)\propto r^2$ is isochronous, in the sense that the oscillation frequency doesn't depend on the amplitude of motion. This is also true for a potential of the form $V(r)\propto 1/r^2$, or any combination of the two.} into the equation of motion for the radius of the trapped quantum gas.

\subsubsection{Monopole mode in the classical-field regime}
Let us examine the hydrodynamic equations in the case of the breathing mode excitation of a 2D Bose gas in an isotropic harmonic trap of frequency $\omega_0$, which corresponds to an oscillation of the radius of the cloud. This mode affects the cloud at large scales, such that the cloud will keep its Thomas-Fermi profile. We can neglect the quantum pressure in the Euler-like hydrodynamic equation and write, see Lecture 1, Sec. 3.5, Eq.~(79) and (80):
\bea
\partial_tn &=& -\nablagras\cdot(n\mathbf{v})\label{eq:continuity}\\
\partial_t\mathbf{v} &=& -\frac{1}{M}\nablagras\left(\frac{1}{2}Mv^2+\frac{1}{2}M\omega_0^2r^2+gn\right) = -\nablagras\left(\frac{1}{2}v^2+\frac{1}{2}\omega_0^2r^2+\frac{gn}{M}\right).\label{eq:Euler}
\eea
The equilibrium density distribution in the Thomas-Fermi regime reads
\beq
n_0(r)=n_0\left(1-\frac{r^2}{R^2}\right) 
\eeq
where $R=\omega_0^{-1}\sqrt{2\mu/M}=\omega_0^{-1}\sqrt{2gn_0/M}$ is determined from
\beq
gn_0\left(1-\frac{r^2}{R^2}\right)+\frac{1}{2}M\omega_0^2r^2 = \mu.
\eeq
Note that, in contrast to the approach we followed in Lecture 1, we do not assume small deviations $n(\mathbf{r},t)-n_0(\mathbf{r})$ here.
As the breathing mode corresponds to a dilation of the cloud, we look for a solution at time $t$ that is scaled with respect to the initial solution, i.e. that transforms the initial position as
\beq
\mathbf{r} \longrightarrow \mathbf{r}'(t) = \lambda(t)\mathbf{r}.
\eeq
We deduce the velocity and the density at time $t$ using the scaling transformation in dimension $D$:
\bea
\mathbf{v}(\mathbf{r},t) & = & \frac{1}{\lambda}\frac{\dd\mathbf{r}'}{\dd t}=\frac{\dot\lambda}{\lambda}\mathbf{r}\label{eq:scaled_velocity}\\
\partial_t\mathbf{v}(\mathbf{r},t) + \frac{1}{2}\nablagras v^2 & = & \left(\frac{\ddot\lambda}{\lambda}-\frac{\dot{\lambda}^2}{\lambda^2}\right)\mathbf{r}+\frac{\dot{\lambda}^2}{\lambda^2}\mathbf{r} = \frac{\ddot\lambda}{\lambda}\mathbf{r}\\
n(r,t) &=& \frac{n_0}{\lambda^D}\left(1-\frac{r^2}{\lambda^2R^2}\right).\label{eq:scaled_density}
\eea

We can easily check that this form for $n$ and $\mathbf{v}$ already satisfies the continuity equation \eqref{eq:continuity} using Eqs.~\eqref{eq:scaled_velocity} and \eqref{eq:scaled_density}, and the identities $\nablagras\cdot \mathbf{r}=D$ and $\nablagras \cdot(r^2\mathbf{r}) = (D+2)r^2$. Let us now plug them into Euler's equation \eqref{eq:Euler}. Using $v^2=r^2\dot\lambda^2/\lambda^2$ and $\nablagras r^2 = 2\mathbf{r}$, we get:
\beq
\frac{\ddot\lambda}{\lambda}\mathbf{r} = -\omega_0^2\mathbf{r} + \frac{gn_0}{M\lambda^D}\frac{2}{\lambda^2R^2}\mathbf{r}.
\eeq
Using $R^2=2gn_0/(M\omega_0^2)$, we obtain a simple differential equation for $\lambda$:
\beq
\ddot\lambda+\omega_0^2\left(\lambda-\frac{1}{\lambda^{D+1}}\right)=0.
\label{eq:diff_lambda}
\eeq
For the initial conditions, we consider a superfluid prepared initially at rest with a radius $\lambda_{\rm max}R$ larger than its equilibrium radius $R$ by a factor $\lambda_{\rm max}$, that starts a breathing oscillation at $t=0$. This gives $\lambda(0)=\lambda_{\rm max}$ and $\dot\lambda(0)=0$. We can multiply the differential equation by $\dot\lambda$ and integrate using the initial condition. We get:
\beq
\dot{\lambda}^2+\omega_0^2\left(\lambda^2+\frac{2}{D}\frac{1}{\lambda^D}\right) = \omega_0^2\left(\lambda_{\rm max}^2+\frac{2}{D}\frac{1}{\lambda_{\rm max}^D}\right).
\label{eq:breathing_MF}
\eeq
In dimension $D=2$, it is the same equation as a particle in a potential of the form $r^2+1/r^2$, which yields an isochronous motion, with an amplitude-independent period. Namely, the solution is periodic at frequency $\Omega_m=2\omega_0$, whatever the breathing amplitude $\lambda_{\rm max}$, with a radius oscillating between $\lambda_{\rm max}R$ and $R/\lambda_{\rm max}$:
\beq
\lambda(t) = \sqrt{\frac{1}{2}\left(\lambda_{\rm max}^2+\frac{1}{\lambda_{\rm max}^2}\right)+\frac{1}{2}\left(\lambda_{\rm max}^2-\frac{1}{\lambda_{\rm max}^2}\right)\cos 2\omega_0 t}.
\eeq
Note that in general the oscillation is not harmonic, although for small amplitudes $\lambda_{\rm max}=1+\varepsilon$ with $\varepsilon\ll 1$ we recover an harmonic oscillation $\lambda(t)=1+\varepsilon\cos 2\omega_0 t$.

\begin{figure}[h]
\centering
\begin{tikzpicture}
\node at (-3.6,0) {\includegraphics[width=0.65\linewidth]{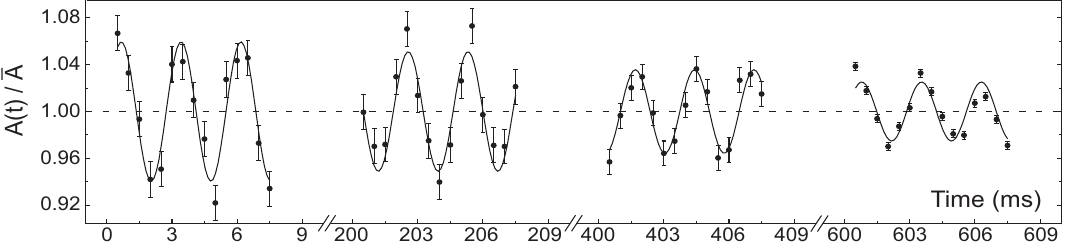}};
\node at (-3.6,-1.45) {\scriptsize Holding time (ms)};
\node at (-8,1.5) {\footnotesize (a)};
\node at (4,0) {\includegraphics[width=0.3\linewidth]{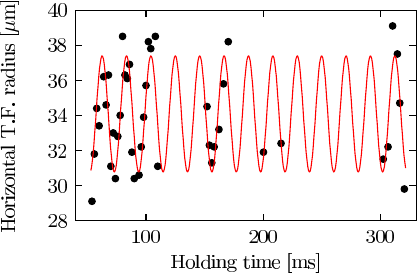}};
\node[fill=white] at (4.4,-1.45) {\scriptsize Holding time (ms)};
\node at (1.5,1.5) {\footnotesize (b)};
\end{tikzpicture}
\caption{(a) Observation of the transverse monopole oscillation in an elongated 3D gas. The damping is very small, 200 periods are visible. Figure adapted from \cite{Chevy2002}. (b) Observation of the monopole mode in a 2D Bose gas. No damping is observable on the timescale of the experiment (13 periods). Figure adapted from \cite{Merloti2013a}. The data are fitted with sine functions because the oscillation amplitude is small.\label{fig:monopole}}
\end{figure}
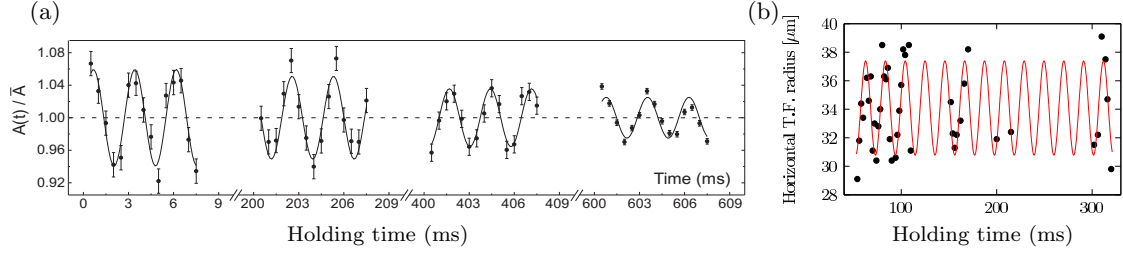

The amplitude-independent frequency is a signature of the scaling symmetry, as predicted by Pitaevskii and Rosch \cite{Pitaevskii1997}. This mode has been observed in a magnetically trapped 2D quantum Bose gas \cite{Merloti2013a}, see Fig.~\ref{fig:monopole}(b). It has also been observed earlier in very elongated cigar-shaped 3D condensates for the transverse breathing mode, see Fig.~\ref{fig:monopole}(a), as the longitudinal degree of freedom doesn't play any role there \cite{Chevy2002}.

\subsubsection{Monopole mode as a probe of the EOS}
In fact, the frequency of the monopole mode can probe the zero temperature equation of state of the gas $\mu(n)$. To derive the frequency $2\omega_0$ above, we have assumed an equation of state of the form $\mu=gn$ with an interaction constant $g=\tilde{g}\hbar^2/M$ which doesn't depend on $n$, which is the result for the classical field approach \cite{Olshanii2010}. If instead the EOS for $\mu(n)$ departs from the linear case $\mu(n)=gn$, we should replace $gn$ by $\mu(n)$ in the Thomas-Fermi version of the Gross-Pitaevskii equation and in the Euler equation that results from it \cite{Olshanii2010}.

Let us first check the expression of the density at equilibrium in the Thomas-Fermi regime. We write at equilibrium in the local density approximation
\beq
\mu[n_{\rm eq}(r)] + \frac{1}{2}M\omega_0^2r^2 = \mu[n_{\rm eq}(0)]=\mu[n_0]=\mu_0.
\eeq
The density profile at equilibrium is thus related to the inverse of the EOS $\mu(n)$:
\beq
n_{\rm eq}(r) = \mu^{-1}\left(\mu_0-\frac{1}{2}M\omega_0^2r^2\right).
\eeq
As before, we will look for scaling solutions $\lambda(t)$ of the form
\beq
n(r,t) = \frac{1}{\lambda(t)^2}\,n_{\rm eq}\left(\frac{r}{\lambda(t)}\right).
\eeq

Now we replace $gn$ by $\mu(n)$ in Eq.~\eqref{eq:Euler} and it will modify the last term of Eq.~\eqref{eq:diff_lambda} because of the term
\beq
-\frac{1}{M}\nablagras \mu[n(r,t)] = -\frac{1}{M} \mu'[n(r,t)]\nablagras n \qquad \mbox{with} \quad \mu'(n) = \frac{\dd\mu}{\dd n}(n).
\eeq
$\nablagras n$ is linked to $\nablagras n_{\rm eq}$ though:
\beq
-\frac{1}{M}\nablagras n(r,t) =-\frac{1}{M}\frac{1}{\lambda^2}\frac{1}{\lambda}\nablagras n_{\rm eq}\left(\frac{r}{\lambda}\right).
\eeq
We compute $\nablagras n_{\rm eq}$ and get:
\beq
\nablagras n_{\rm eq}(r) = \frac{-M\omega_0^2\mathbf{r}}{\mu'\left[n_{\rm eq}(r)\right]},
\quad \mbox{hence} \quad
-\frac{1}{M}\nablagras n_{\rm eq}\left(\frac{r}{\lambda}\right) = \frac{\omega_0^2}{\lambda}\frac{1}{\mu'\left[n_{\rm eq}\left(r/\lambda\right)\right]}\mathbf{r}.
\eeq
We finally have
\beq
-\frac{1}{M}\nablagras \mu[n(r,t)] = \frac{\omega_0^2}{\lambda^4} \frac{\mu'\left[\frac{1}{\lambda^2}n_{\rm eq}\left(r/\lambda\right)\right]}{\mu'\left[n_{\rm eq}\left(r/\lambda\right)\right]}\mathbf{r}
\eeq
and for the time evolution of $\lambda$
\beq
\ddot{\lambda}+\omega_0^2\lambda - \frac{\omega_0^2}{\lambda^3} \frac{\mu'\left(n_{\rm eq}/\lambda^2\right)}{\mu'\left(n_{\rm eq}\right)} = 0
\eeq
where $n_{\rm eq}$ stands for $n_{\rm eq}(r/\lambda)$.
The fact that $\mu(n)$ is non linear in $n$ breaks the scaling symmetry, and we must restrict ourselves to the small amplitude regime to get the monopole frequency, writing $\lambda=1+\varepsilon$. Then
\beq
\mu'\left(\frac{n_{\rm eq}}{\lambda^2}\right)\simeq \mu'\left[(1-2\varepsilon)n_{\rm eq}\right]\simeq \mu'(n_{\rm eq}) -2\varepsilon n_{\rm eq} \mu''(n_{\rm eq}).
\eeq
Eq.~\eqref{eq:diff_lambda} now reads to first order in $\varepsilon$ \cite{Olshanii2010}
\beq
\ddot\varepsilon+\omega_0^2\left[1+\varepsilon-(1-3\varepsilon)\left(1-2\frac{n_{\rm eq} \mu''(n_{\rm eq})}{\mu'(n_{\rm eq})}\varepsilon\right)\right]\simeq\ddot{\varepsilon}+\omega_0^2\left[4+2\frac{n_0 \mu''(n_0)}{\mu'(n_0)}\right]\varepsilon=0.
\label{eq:diff_lambda_shift}
\eeq
In the last equality, we have replaced $n_{\rm eq}$ by $n_0$, assuming $n\mu''(n)/\mu'(n)$ doesn't depend much on $n$ (we will see a natural case below where it doesn't depend at all on $n$). The frequency of the monopole mode is shifted from its classical field value $2\omega_0$ and now reads
\beq
\boxed{
\Omega_m=\omega_0\sqrt{4+2\frac{n_0 \mu''(n_0)}{\mu'(n_0)}}.
}
\eeq
It can be used to \textbf{probe the equation of state} $\mu(n)$. Consider in particular a polytropic equation of state $\mu(n)\propto n^\gamma$. In this case, $n\mu''(n)/\mu'(n)=\gamma-1$, independent of the density, and the monopole frequency reads
\beq
\Omega_m = \omega_0\sqrt{2(1+\gamma)}.
\label{eq:monopole_polytropic}
\eeq

\begin{itemize}
\item In the classical field limit where $\mu(n)=gn$ with $g$ independent of $n$, $\mu''(n)=0$ and we recover the undamped mode of the scaling symmetry with $\Omega_m=2\omega_0$ \cite{Pitaevskii1997}.
\item In the case of a very oblate 3D condensate, which still has a transverse Thomas-Fermi distribution but with a size $R_z\ll R_r$, we can write a 2D equation of state by integrating out the transverse direction. The chemical potential is related to the 3D density and the 2D integrated density through $\mu=g_{3D}n_{3D}\propto n_{2D}/R_z$ with $R_z\propto\sqrt{\mu}$, which yields $\mu\propto n_{2D}^{2/3}$. The equation of state is again polytropic with an exponent $\gamma=2/3$, and the monopole frequency is $\Omega_m=\omega_0\sqrt{10/3}$, as also derived by Stringari by the more usual approach \cite{Stringari1998}. The cross-over between the 2D regime and the 3D regime as the chemical potential approaches $\hbar\omega_z$ has been studied in a quasi-2D Bose gas in a tunable oblate magnetic trap \cite{Merloti2013b}.
\item For very tight transverse confinement, the EOS again departs from the classical field case: if the length of the transverse harmonic oscillator $a_z=\sqrt{\hbar/M\omega_z}$ is not very large with respect to the scattering length $a$, collisions should be treated in 2D and a new length scale $a_{2D}$ appears. The dimensionless character of the interaction constant $\tilde{g}$ is lost due to a \textit{quantum anomaly} \cite{Olshanii2010}, which yields a monopole frequency exceeding $2\omega_0$ by a small amount. To first order in $a/a_z$, we expect \cite{Olshanii2010}
\beq
\Omega_m\simeq 2\omega_0\left(1+\frac{1}{4\sqrt{2\pi}}\frac{a}{a_z}\right).
\eeq
This effect is in general very small and below the experimental resolution for usual values of the scattering length \cite{Merloti2013b}. However, using a Feshbach resonance, a similar anomaly \cite{Hofmann2012} has been observed in Fermi gases, together with the crossover to the 3D monopole frequency \cite{Holten2018,Peppler2018}.
\end{itemize}

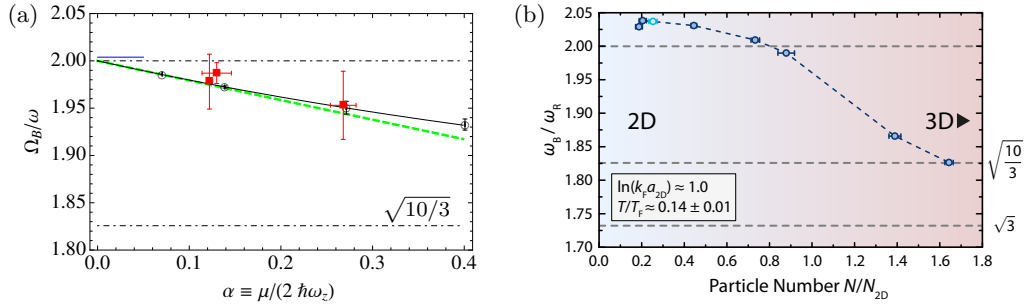
\begin{figure}
\centering
\begin{tikzpicture}
\node at (-3,0) {\includegraphics[width=0.4\linewidth]{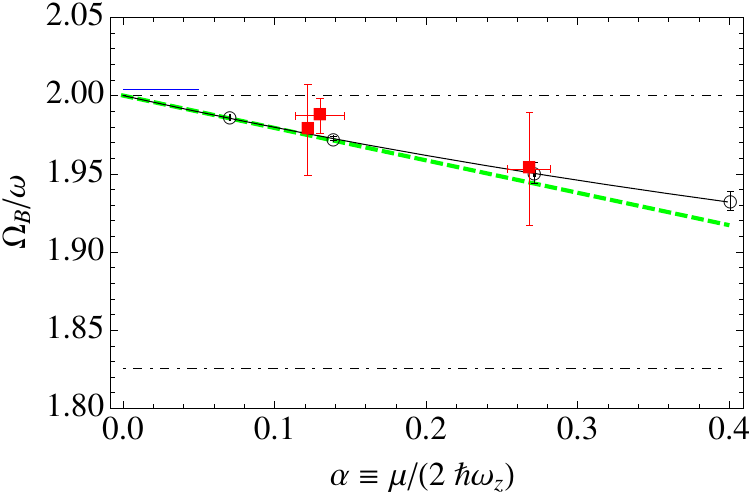}};
\node at (-0.8,-0.75) {\scriptsize $\sqrt{10/3}$};
\node at (-6,1.8) {\footnotesize (a)};
\node at (4,0) {\includegraphics[width=0.44\linewidth]{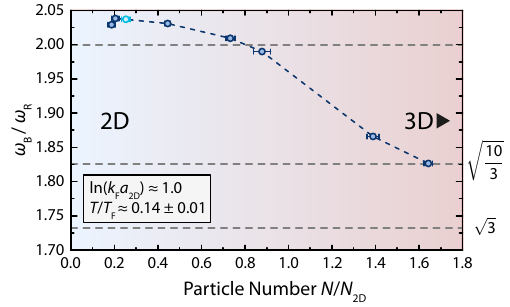}};
\node at (0.7,1.8) {\footnotesize (b)};
\end{tikzpicture}
\caption{(a) Monopole frequency for a rubidium quantum gas in a very oblate trap, as a function of the ratio $\mu_{2D}/(2\hbar\omega_z) = (\mu_{3D}-\hbar\omega_z/2)/(2\hbar\omega_z)$ which quantifies the 2D vs 3D character of the gas. The EOS transitions slowly from a pure 2D quantum gas, with $\Omega_m=2\omega_0$, to a 3D integrating quantum gas, near $\Omega_m=\omega_0\sqrt{10/3}$, evidencing the role of the third dimension that breaks the scaling symmetry. The uncertainty in the data do not allow for the search of the quantum anomaly effect, represented with a blue line. Figure from \cite{Merloti2013a}. (b) Monopole frequency, presented as a function of the particle number, in a quantum Fermi gas of lithium 6, with a the scattering length enhanced with a Feshbach resonance. The quantum anomaly is visible at low particle number, and the system eventually transitions to the $\Omega_m=\omega_0\sqrt{10/3}$ frequency expected in the weakly interacting case. Figure from \cite{Holten2018}. Similar data are reported in \cite{Peppler2018}.
\label{fig:anomaly}}
\end{figure}

\clearpage

\section{Appendix}
In this Appendix, I give explicit derivation for a number of results presented in the main text.
\subsection{Average of $e^{iX}$}
\label{sec:proof}
In this section I give a proof of the result Eq.~\eqref{eq:average_phase} we used in Sec.~\ref{sec:QLRO}: given a random variable $X$ obeying a Gaussian distribution $p(X) = \exp[-(X-X_0)^2/2\sigma_X^2]/(\sqrt{2\pi}\sigma_X)$ with average value $\langle X\rangle=X_0$ and variance $\langle(X-\langle X\rangle)^2\rangle=\sigma_X^2$, we have
$$
\langle e^{iX} \rangle = e^{i\langle X\rangle}e^{-\langle(X-\langle X\rangle)^2\rangle/2}=e^{iX_0}e^{-\sigma_X^2/2}.
$$
\textbf{Proof}:
\bea
\langle e^{iX}\rangle &=& \langle e^{iX_0}e^{i(X-X_0)}\rangle\nonumber\\
&=& e^{iX_0}\sum_{n=0}^\infty \frac{i^n}{n!}\langle (X-X_0)^n\rangle\nonumber\\
&=& e^{iX_0}\sum_{n=0}^\infty \frac{(-1)^n}{(2n)!}\langle (X-X_0)^{2n}\rangle\nonumber
\eea
where we have used the fact that a Gaussian distribution is symmetric around its average value $X_0$, such that $\langle (X-X_0)^n\rangle=0$ for $n$ odd. Moreover, the even moments are linked to the second order moment (the variance $\sigma_X^2$). We can write for $n\geq 1$:
\bea
\langle (X-X_0)^{2n}\rangle &=& \ds\frac{1}{\sqrt{2\pi}\sigma_X}\int_{-\infty}^{+\infty}dX\, (X-X_0)^{2n} e^{\frac{-(X-X_0)^2}{2\sigma_X^2}} = \ds\frac{1}{\sqrt{2\pi}\sigma_X}\int_{-\infty}^{+\infty}dX\, u^{2n} e^{\frac{-u^2}{2\sigma_X^2}}\nonumber\\
&=&\ds\frac{1}{\sqrt{2\pi}\sigma_X}\left\{\left[-\sigma_X^2 u^{2n-1}e^{\frac{-u^2}{2\sigma_X^2}}\right]_{-\infty}^{+\infty} + (2n-1)\sigma_X^2\int_{-\infty}^{+\infty}dX\, u^{2n-2} e^{\frac{-u^2}{2\sigma_X^2}}\right\}\nonumber\\
&=&(2n-1)\sigma_X^2\langle (X-X_0)^{2n-2}\rangle.\nonumber
\eea
We then have
\bea
\langle (X-X_0)^{2n}\rangle &=& (2n-1)\sigma_X^2\langle (X-X_0)^{2n-2}\rangle =(2n-1)(2n-3)\sigma_X^4\langle (X-X_0)^{2n-4}\rangle\nonumber\\ 
&=&\dots = (2n-1)(2n-3)\dots 3\times 1\times \sigma_X^{2n}\nonumber\\
&=& \frac{(2n)!}{n!2^n}\sigma_X^{2n}.\nonumber
\eea
We finally get
\bea
\langle e^{iX} \rangle &=& e^{iX_0}\sum_{n=0}^\infty \frac{(-1)^n}{(2n)!}\langle (X-X_0)^{2n}\rangle\nonumber\\
& = &e^{iX_0}\sum_{n=0}^\infty \frac{(-1)^n}{(2n)!}\frac{(2n)!}{n!2^n}\sigma_X^{2n} = e^{iX_0}\sum_{n=0}^\infty \frac{1}{n!}\left(\frac{-\sigma_X^2}{2}\right)^n\nonumber\\
&=&e^{iX_0}e^{-\sigma_X^2/2}.
\eea

\subsection{Relation between $G_1(r)$ and $\mathcal{N}(\mathbf{p})$}
\label{sec:G1_TF}
We show here that the one-body correlation function $G_1$ is the Fourier transform of the momentum distribution. We follow the approach of Jean Dalibard's lectures at Collège de France \cite{DalibardCF2017en} (see also Volume 3 of \cite{CohenQuantiqueAnglais}). We consider particles in a cubic box of volume $L^D$ in dimension $D$. The eigenstates constitute the plane wave basis $\left\{\ket{\mathbf{p}}\right\}$, with wave functions
\beq
\psi_{\mathbf{p}}(\mathbf{r}) = \braket{\mathbf{r}}{\mathbf{p}} = \frac{1}{L^{D/2}}e^{i\mathbf{p}\cdot\mathbf{r}/\hbar}.
\eeq
Here, $\mathbf{p}$ is quantized and can take any value $\mathbf{p}=\frac{2\pi\hbar}{L}\sum_i n_i \mathbf{e}_i$ with $n_i\in\mathbb{Z}$.

The occupation of each state \ket{\mathbf{p}} is denoted as $n_\mathbf{p}$, such that
\beq
\sum_\mathbf{p} n_\mathbf{p} = N
\eeq
with $N$ the total number of particles. In the grand canonical ensemble, $N$ is the average total number of particles and $n_\mathbf{p}$ is given by the Bose-Einstein distribution:
\beq
n_\mathbf{p} = \ds\frac{1}{e^{\beta\left(\mathbf{p}^2/2M-\mu\right)}-1}.
\eeq
We can also write $n_\mathbf{p}= \left\langle\mathbf{p}|\rho_1|\mathbf{p}\right\rangle$ with $\rho_1$ the one-body density matrix.

In the thermodynamic limit where $N,L\to\infty$ with $N/L^D=n_0=\mbox{cst}$, we can replace sums over $\mathbf{p}$ by integrals:
\beq
\sum_\mathbf{p} \longrightarrow \frac{L^D}{(2\pi\hbar)^D}\int d\mathbf{p}.
\eeq
$n_\mathbf{p}$ is related to the momentum distribution $\mathcal{N}(\mathbf{p})$ (normalized to unity) through
\beq
\mathcal{N}(\mathbf{p}) = \frac{1}{N}\frac{L^D}{(2\pi\hbar)^D} n_\mathbf{p} = \frac{n_0^{-1}}{(2\pi\hbar)^D} n_\mathbf{p}.
\eeq

$n_\mathbf{p}$ is related to $G_1(r)$ by Fourier transform. We can check this by inserting closure relations in momentum space:
\bea
\left\langle\mathbf{p}|\rho_1|\mathbf{p}\right\rangle &=& \int_{\rm box} \!\!\!\! d\mathbf{r}\int_{\rm box} \!\!\!\! d\mathbf{r}'\braket{\mathbf{p}}{\mathbf{r}}\left\langle\mathbf{r}|\rho_1|\mathbf{r}'\right\rangle\braket{\mathbf{r}'}{\mathbf{p}}\nonumber\\
&=&\frac{1}{L^D}\int_{\rm box} \!\!\!\! d\mathbf{r}\int_{\rm box} \!\!\!\! d\mathbf{r}'\,e^{-i\mathbf{r}\cdot\mathbf{p}/\hbar}G_1(\mathbf{r},\mathbf{r}') \, e^{i\mathbf{r}'\cdot\mathbf{p}/\hbar}\nonumber\\
&=&\frac{1}{L^D}\int_{\rm box} \!\!\!\! d\mathbf{r}\int_{\rm box} \!\!\!\! d\mathbf{r}'\,G_1(\mathbf{r}-\mathbf{r}',0)\,e^{-i(\mathbf{r}-\mathbf{r}')\cdot\mathbf{p}/\hbar} \nonumber\\
&=&\frac{1}{L^D}\int_{\rm box} \!\!\!\! d\mathbf{u}\,G_1(\mathbf{u},0)\,e^{-i\mathbf{u}\cdot\mathbf{p}/\hbar}\int_{\rm box} \!\!\!\! d\mathbf{r}'\, 1\nonumber\\
&=&\int_{\rm box} \!\!\!\! d\mathbf{u}\,G_1(\mathbf{u},0)\,e^{-i\mathbf{u}\cdot\mathbf{p}/\hbar}\nonumber
\eea
and thus
\beq
\mathcal{N}(\mathbf{p}) = \frac{n_0^{-1}}{(2\pi\hbar)^D}\int d\mathbf{r}\,G_1(\mathbf{r})\,e^{-i\mathbf{r}\cdot\mathbf{p}/\hbar}
\eeq
where we have written $G_1(\mathbf{r})\equiv G_1(\mathbf{r},0)$ in the last equation and let $L\to\infty$ for the integration domain. We can thus deduce $G_1$ from the inverse Fourier transform of the momentum distribution when it is easy to estimate, which is the case for an ideal gas:
\beq
\boxed{
G_1(\mathbf{r}) = n_0\int d\mathbf{p}\,\mathcal{N}(\mathbf{p})\,e^{i\mathbf{r}\cdot\mathbf{p}/\hbar}.
}
\eeq
\subsection{Calculation of $G_1(r)$ for a thermal gas}
\label{sec:G1_thermal}
We start from an ideal, uniform thermal gas. The momentum distribution is simply a Gaussian
\beq
\mathcal{N}(\mathbf{p}) = \frac{1}{(2\pi Mk_BT)^{D/2}}e^{-p^2/2Mk_BT} = \left(\frac{\lambda_{\rm dB}}{2\pi\hbar}\right)^De^{-p^2\lambda_{\rm dB}^2/4\pi\hbar^2}.
\eeq
Its Fourier transform is also a Gaussian:
\bea
G_1(\mathbf{r}) &=& \frac{n_0\lambda_{\rm dB}^D}{(2\pi\hbar)^D}\int d\mathbf{p}\,e^{i\mathbf{r}\cdot\mathbf{p}/\hbar}e^{-p^2\lambda_{\rm dB}^2/4\pi\hbar^2}\nonumber\\
&=&\frac{n_0}{(\sqrt{2\pi})^D}\left(\frac{\lambda_{\rm dB}}{\sqrt{2\pi}}\right)^D\int d\mathbf{q}\,e^{i\mathbf{q}\cdot\mathbf{r}}e^{-q^2\lambda_{\rm dB}^2/4\pi}\nonumber\\
G_1(\mathbf{r})&=&n_0\,e^{-\pi r^2/\lambda_{\rm dB}^2}.
\eea
We find that the size over which the gas is coherent is typically the de Broglie wave length. The correlation function decays rather fast, with a Gaussian shape, over a length that increases as the temperature is reduced. This result is valid in any dimension $D$.

\subsection{Another derivation of $G_1(r)$ for the thermal gas}
\label{sec:G1_other}
In a uniform thermal gas, the system is described by plane waves $e^{i\mathbf{p}\cdot\mathbf{r}/\hbar}$ and the distribution of $p_i$ in each spatial direction obeys a Gaussian law with width $\Delta p_i^2=Mk_BT$, independent in each direction. In dimension $D$ we then have
\bea
G_1(\mathbf{r},\mathbf{0})&=&n_0\left\langle \ds e^{i\mathbf{p}\cdot\mathbf{r}/\hbar}\right\rangle  = n_0\left\langle e^{i\sum_{i=1}^D p_i x_i/\hbar}\right\rangle\nonumber\\
&=& n_0\left\langle \prod_i \ds e^{ip_i x_i/\hbar}\right\rangle = n_0\prod_i \ds \left\langle e^{ip_i x_i/\hbar}\right\rangle\nonumber\\
&=&n_0\prod_i \ds  e^{-\left\langle p_i^2\right\rangle x_i^2/2\hbar^2} = n_0\prod_i \ds  e^{-Mk_BT x_i^2/2\hbar^2}\nonumber\\
&=&n_0 \, \exp\left({\ds -\frac{\pi}{\lambda_{\rm dB}^2}\sum_i x_i^2}\right)\nonumber
\eea
where we have introduced the thermal de Broglie wave length $\lambda_{\rm dB}=2\pi\hbar/\sqrt{Mk_BT}$. The first order correlation function for a thermal gas is Gaussian and decays with a characteristic length $\sim\lambda_{\rm dB}$, the size of thermal wave packets:
\beq
\boxed{
G_1(r) = n_0\, e^{\ds -\pi r^2/\lambda_{\rm dB}^2}.
}
\eeq
We recover the result of Sec.~\ref{sec:QLRO}.

\subsection{Calculation of $G_1(r)$ for an ideal degenerate Bose gas \cite{DalibardCF2017en}}
\label{sec:G1_ideal}
More generally, the occupation number of state \ket{\mathbf{p}} is given by the Bose-Einstein distribution
\beq
n_\mathbf{p} = \frac{1}{e^{\beta\left(p^2/2M-\mu\right)}-1}
\eeq
or in the thermodynamic limit (see Appendix~\ref{sec:G1_TF})
\beq
\mathcal{N}(\mathbf{p}) = \frac{n_0^{-1}}{(2\pi\hbar)^2}\frac{1}{e^{\beta\left(p^2/2M-\mu\right)}-1}.
\label{eq:momentum_Bose}
\eeq
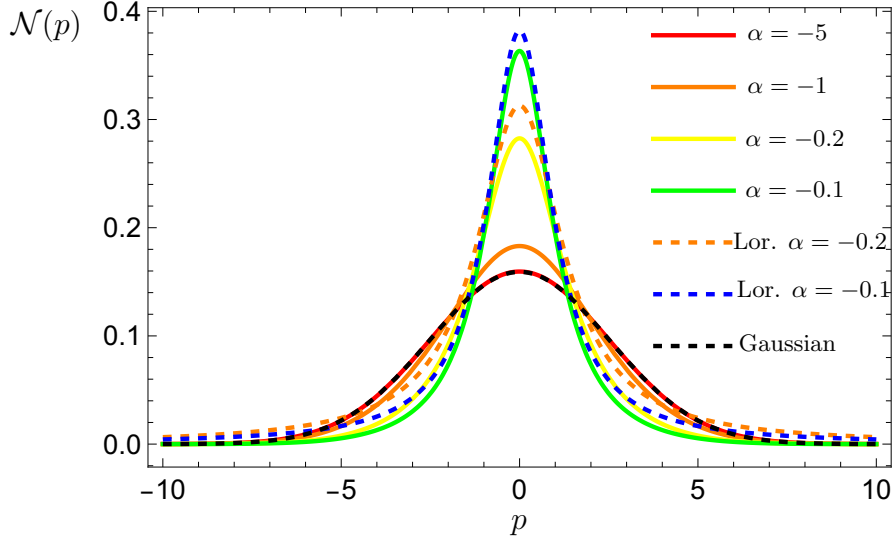
\begin{figure}
\centering
\begin{tikzpicture}
\node at (0,0) {\includegraphics[width=0.7\linewidth]{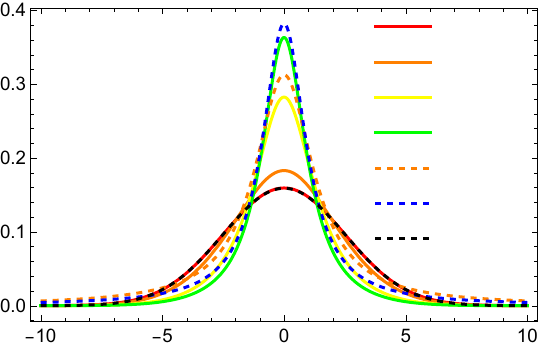}};
\node at (3.8,2.9) {\footnotesize $\alpha=-5$};
\node at (3.8,2.2) {\footnotesize $\alpha=-1$};
\node at (3.9,1.5) {\footnotesize $\alpha=-0.2$};
\node at (3.9,0.82) {\footnotesize $\alpha=-0.1$};
\node at (4.1,0.14) {\footnotesize Lor. $\alpha=-0.2$};
\node at (4.15,-0.5) {\footnotesize Lor. $\alpha=-0.1$};
\node at (3.8,-1.2) {\footnotesize Gaussian};
\node at (0.25,-3.6) {\large $p$};
\node at (-6,3) {\large $\mathcal{N}(p)$};
\end{tikzpicture}
\caption{Momentum distribution $\mathcal{N}(p)$ as given by the Bose-Einstein distribution Eq.~\eqref{eq:momentum_Bose} for different values of $\alpha=\mu/k_BT$. When $\alpha$ approaches zero, the approximation by a Lorentz distribution, Eq.~\eqref{eq:momentum_Lorentz}, gives an accurate description of the central peak. In the other limit where $\alpha$ is large and negative, the Gaussian function of the Maxwell-Boltzmann distribution $\mathcal{N}(p) \propto \exp(p^2/2Mk_BT)$ gives a good approximation.
\label{fig:momentum_distribution}
}
\end{figure}
The limit of the thermal gas corresponds to $\mu<0$ and $|\mu|\gg k_BT$, where the exponential is dominant in the denominator and $\mathcal{N}(\mathbf{p})$ is well approximated by a Gaussian, see Fig.~\ref{fig:momentum_distribution}. The degenerate gas corresponds to the other limit where $\beta|\mu|\ll 1$ or $|\mu|\ll k_BT$. In this limit, $e^{\beta\mu}\simeq 1+\beta\mu$ and using Eq.~\eqref{eq:psd_exc} we can write $|\mu|/k_BT\simeq e^{-\mathcal{D}}$. For $p$ sufficiently large (much larger than $p_T=\sqrt{2Mk_BT}$) the tails of the momentum distribution are still Gaussian. In the central region where $p\ll p_T$, the distribution can be approximated by a Lorentzian, see Fig.~\ref{fig:momentum_distribution}, as is clear from an expansion for small values of the exponent $\beta\left(p^2/2M-\mu\right)$:
\beq
\mathcal{N}(\mathbf{p}) \simeq \frac{n_0^{-1}}{(2\pi\hbar)^2}\frac{k_BT}{p^2/2M + |\mu|} = \frac{1}{\pi n_0\lambda_{\rm dB}^2}\frac{1}{p^2+p_\mu^2}\qquad\mbox{for }p\ll p_T.
\label{eq:momentum_Lorentz}
\eeq
The half width at half maximum of this Lorentzian is $p_\mu = \sqrt{2M|\mu|}\simeq \sqrt{2Mk_BT}e^{-\mathcal{D}/2}$, and we have $p_\mu\ll p_T$ in the degenerate regime $|\mu|\ll k_BT$. As a result, most of the particles have a momentum in the central region and we can ignore the Gaussian tails.\footnote{The tails are important to ensure that the sum over $p$ converges in the limit $p\to+\infty$.} We then take the Fourier transform to compute the first order correlation function:
\beq
G_1(\mathbf{r}) = n_0\int d\mathbf{p}\,\mathcal{N}(\mathbf{p})\,e^{i\mathbf{r}\cdot\mathbf{p}/\hbar}=\frac{1}{\pi\lambda_{\rm dB}^2}\int d\mathbf{p}\,\frac{e^{i\mathbf{r}\cdot\mathbf{p}/\hbar}}{p^2+p_\mu^2}=\frac{1}{\pi\lambda_{\rm dB}^2}\int_0^{+\infty}\!\!\!\!\!\! q\,dq\int_0^{2\pi}\!\!\!\! d\varphi\,\frac{e^{iqr\cos\varphi}}{q^2+q_\mu^2}
\eeq
with $q_\mu=p_\mu/\hbar$. We use the decomposition of the exponential in terms of Bessel functions
\beq
e^{iqr\cos\varphi} = J_0(qr) + 2\sum_{n=0}^\infty i^n J_n(qr)\cos(n\varphi).
\eeq
All contributions with $n\neq 0$ vanish due to the integration of the cosine term, and we get
\beq
G_1(r) = \frac{2}{\lambda_{\rm dB}^2}\int_0^{+\infty}\!\!\!\!\!\! dq\,\frac{qJ_0(qr)}{q^2+q_\mu^2} = \frac{2}{\lambda_{\rm dB}^2}\int_0^{+\infty}\!\!\!\!\!\! du\,\frac{uJ_0(u)}{u^2+q_\mu^2r^2}=\frac{2}{\lambda_{\rm dB}^2}\left|K_0(q_\mu r)\right|
\eeq
where $K_0$ is the modified Bessel function of the second kind. Its behavior at large argument is
\beq
K_0(q_\mu r)\simeq \sqrt{\frac{\pi}{2q_\mu r}}e^{-q_\mu r},\qquad \mbox{for }q_\mu r\gg 1.
\eeq
We get finally
\beq
G_1(r)\simeq \sqrt{\frac{2\pi}{q_\mu r}}\frac{1}{\lambda_{\rm dB}^2}e^{-q_\mu r}\propto \sqrt{\frac{\ell}{r}}e^{-r/\ell}\qquad \mbox{for }q_\mu r\gg 1.
\eeq
The first order correlation function decays on a typical size $\ell=q_\mu^{-1}$. Replacing $q_\mu$ by its value $q_\mu=p_\mu/\hbar\simeq e^{-\mathcal{D}/2}\sqrt{2Mk_BT}/\hbar = \sqrt{4\pi}e^{-\mathcal{D}/2}/\lambda_{\rm dB}$, we get for the decay length
\beq
\ell = \frac{\lambda_{\rm dB}}{\sqrt{4\pi}}e^{\mathcal{D}/2}.
\eeq

\subsection{Calculation of $G_1(r)$ for a weakly interacting degenerate 2D gas}
\label{sec:G1_degenerate}
This section is based on Refs.~\cite{Petrov2004,DalibardCF2017en}. We will see that in this case $G_1$ decays more slowly, giving rise to a quasi long-range order. In order to derive the expression for $G_1(r)$, we will describe the 2D weakly interacting quantum gas with a classical field
\beq
\psi(\mathbf{r}) = \sqrt{n(\mathbf{r})} e^{i\theta(\mathbf{r})},
\eeq
such that
\beq
G_1(\mathbf{r}) = \left\langle\psi(\mathbf{r})\psi^*(\mathbf{0})\right\rangle = \left\langle\sqrt{n(\mathbf{r})n(\mathbf{0})}e^{i\left[\theta(\mathbf{r}) - \theta(\mathbf{0})\right]}\right\rangle.
\eeq
The use of a classical field will be justified by the fact that the 2D gas is at least partially superfluid below the critical temperature for BKT, see Sec.~\ref{sec:BKT}, such that we can attribute a density and phase with small fluctuations to this superfluid fraction. Our aim is thus to estimate these fluctuations.

We recall the hydrodynamic equations (67) and (68) of Lecture 1 that are derived from the Gross-Pitaevskii equation for the classical field, but keeping the phase $\theta$ instead of the velocity and considering the uniform case:
\bea
\partial_t n &=& -\frac{\hbar}{M}\nablagras(n\nablagras\theta),\\
-\hbar\partial_t\theta &=& -\frac{\hbar^2}{2M}\frac{\nablagras^2\sqrt{n}}{\sqrt{n}}+\frac{\hbar^2}{2M}(\nablagras\theta)^2 + \frac{\hbar^2}{M}\tilde{g}n.
\eea

The fields $n$ and $\theta$ fluctuate at non zero temperature around their average value $n_0=N/L^2$ and $\theta_0(t) = -\hbar\tilde{g}n_0t/M$ due to the population in low-energy modes. We will follow the Bogolubov approach of Sec.~3.3 of Lecture 1: We write $n(\mathbf{r},t) = n_0+\delta n(\mathbf{r},t)$ and $\theta(\mathbf{r},t) = \theta_0(t) + \delta\theta(\mathbf{r},t)$ and expand the hydrodynamic equations to first order in $\delta n$ and $\delta\theta$, using the expansion $\sqrt{n_0+\delta n}=\sqrt{n_0} + \delta n/(2\sqrt{n_0})$ to first order:
\bea
\partial_t \delta n &=& -\frac{\hbar}{M} n_0\nablagras^2\delta\theta,\label{eq:dt_delta_n}\\
-\partial_t\delta\theta &=& -\frac{\hbar}{2M}\frac{\nablagras^2\delta n}{2n_0} + \frac{\hbar}{M}\tilde{g}\delta n.\label{eq:dt_delta_theta}
\eea

We now decompose $\delta n$ and $\delta\theta$ in Fourier series in a box of size $L\times L$:
\bea
\delta n &=& n_0\sum_{\mathbf{q}} \alpha_\mathbf{q}(t) e^{i\mathbf{q}\cdot\mathbf{r}}\qquad\mbox{with }\mathbf{q} = (n_x,n_y)\frac{2\pi}{L}, \quad n_i\in\mathbb{Z},\\
\delta\theta &=& \sum_{\mathbf{q}} \beta_\mathbf{q}(t) e^{i\mathbf{q}\cdot\mathbf{r}}.
\eea
We note that $\alpha_{-\mathbf{q}}=\alpha_\mathbf{q}^*$ and $\beta_{-\mathbf{q}}=\beta_\mathbf{q}^*$ to ensure that $\delta n$ and $\delta\theta$ are real. We now inject this decomposition into the expansion Eqs.~\eqref{eq:dt_delta_n} and \eqref{eq:dt_delta_theta}:
\bea
\dot\alpha_\mathbf{q} &=& \frac{\hbar q^2}{M}\beta_\mathbf{q},\\
-\dot{\beta}_\mathbf{q} &=& \frac{\hbar}{4M}\left(q^2 +4\tilde{g}n_0\right)\alpha_\mathbf{q}.
\eea
If we differentiate once again, we get two similar second order equations $\ddot\alpha_\mathbf{q}+\omega_\mathbf{q}^2\alpha_\mathbf{q}=0$ and $\ddot{\beta}_\mathbf{q} + \omega_\mathbf{q}^2\beta_\mathbf{q}=0$, with:
\beq
\omega_\mathbf{q} = \frac{\hbar q}{2M}\sqrt{q^2+4\tilde{g}n_0}.
\eeq
We recover the Bogolubov spectrum of Lecture 1, with its two regimes separated by the healing length $\xi=1/\sqrt{2\tilde{g}n_0}$: sound waves $\omega_\mathbf{q}=cq$ for $q\ll \xi^{-1}$, with a speed of sound $c=\hbar\sqrt{\tilde{g}n_0}/M$, and free particles on top of the condensate $\hbar\omega_\mathbf{q} = \mu+\hbar^2q^2/2M$ for $q\gg \xi^{-1}$, with a chemical potential $\mu=\tilde{g}n_0\hbar^2/M$. 

If we write $\alpha_\mathbf{q}(t)=\alpha_\mathbf{q}(0)e^{-i\omega_\mathbf{q} t}$ and $\beta_\mathbf{q}(t)=\beta_\mathbf{q}(0)e^{-i\omega_\mathbf{q} t}$, we can derive a relation between the amplitudes $\alpha_\mathbf{q}$ and $\beta_\mathbf{q}$:
\beq
\alpha_\mathbf{q} = i\frac{\hbar q^2}{M\omega_\mathbf{q}}\beta_\mathbf{q} = i\frac{2q}{\sqrt{q^2+4\tilde{g}n_0}}\beta_\mathbf{q}.
\eeq
In the sound wave regime where $q\ll \xi^{-1}$, $\alpha_\mathbf{q}$ is much smaller than $\beta_\mathbf{q}$, and the main source of fluctuations are phase fluctuations (see also below for a more detailed discussion).

We assume that the modes are weakly populated up to some cut-off energy $\hbar\omega_c$. This yields a cut-off $q_c$ for the momentum $q$:
\beq
\frac{\hbar^2 q_c}{2M}\sqrt{q_c^2+4\tilde{g}n_0}=\hbar\omega_c \quad\Rightarrow\quad q_c = \xi^{-1}\left(\sqrt{1+\left(\frac{\hbar\omega_c}{\mu}\right)^2}-1\right)^{1/2}.
\label{eq:qc_app}
\eeq
At very low temperature, with $k_BT<\mu$, quantum fluctuations dominate and we can set $\hbar\omega_c\sim\mu$ and thus $q_c\sim\xi$. In the other regime, thermal excitations are dominant and we set the cut-off energy to $\hbar\omega_q\simeq k_BT$. The condition $k_BT>\mu$ is equivalent to $\tilde{g}\mathcal{D}<2\pi$. $\tilde{g}$ is small to ensure that the gas is in the weakly interacting regime, its value is typically around $0.1$ in many experiments, which allows us to have both $\mathcal{D}>1$ and $\tilde{g}\mathcal{D}<2\pi$. In this regime $\hbar\omega_c/\mu = k_BT/\mu = 2\pi/\tilde{g}\mathcal{D}$ and we get for the expression of $q_c$:
\beq
q_c = q_T = \xi^{-1}\left(\sqrt{1+\left(\frac{2\pi}{\tilde{g}\mathcal{D}}\right)^2}-1\right)^{1/2}.
\label{eq:qT_app}
\eeq
Note that in the weakly degenerate case with $\tilde{g}\mathcal{D}\ll 2\pi$, $q_T$ is approximately given by
\beq
q_T\simeq \sqrt{\frac{4\pi\tilde{g}n_0}{\tilde{g}\mathcal{D}}}=2\sqrt{\pi}\lambda_{\rm dB}^{-1}.
\eeq

\paragraph{Phase fluctuations} We will assume in the following that the phase space density is large enough to neglect the density fluctuations in the calculation of $G_1$, i.e. $\mathcal{D}\gg 1$ (see below). We can then write
\beq
G_1(\mathbf{r})=n_0\left\langle e^{i\left[\delta\theta(\mathbf{r})-\delta\theta(\mathbf{0})\right]}\right\rangle.
\eeq
With weak population in the normal modes of the phase fluctuations, we have Gaussian fluctuations for $\delta\theta$ set by the Maxwell-Boltzmann law, and we will make use of a mathematical result derived in Appendix~\ref{sec:proof}. Given a random variable $X$ obeying a Gaussian distribution $p(X) $ with average value $\langle X\rangle=X_0$ and variance $\langle(X-\langle X\rangle)^2\rangle=\sigma_X^2$, we can show that
\beq
\langle e^{iX} \rangle = e^{i\langle X\rangle}e^{-\langle(X-\langle X\rangle)^2\rangle/2}=e^{iX_0}e^{-\sigma_X^2/2}.
\label{eq:average_phase}
\eeq
We get for $G_1$:
\beq
G_1(\mathbf{r})=n_0 e^{-\frac{1}{2}\left\langle\left[\delta\theta(\mathbf{r})-\delta\theta(\mathbf{0})\right]^2\right\rangle}.
\label{eq:G1Gaussian}
\eeq

We need to compute the variance of the phase fluctuations
\beq
\Delta\theta^2=\left\langle\left[\delta\theta(\mathbf{r})-\delta\theta(\mathbf{0})\right]^2\right\rangle = \left\langle\left[\sum_{\mathbf{q}} \beta_\mathbf{q} (e^{i\mathbf{q}\cdot\mathbf{r}}-1)\right]^2\right\rangle.
\eeq
If we decompose $\beta_\mathbf{q}$ in real and imaginary parts $\beta_\mathbf{q} = \beta'_\mathbf{q} + i\beta''_\mathbf{q}$, the conditions for the phase to be real read
\beq
\beta'_{-\mathbf{q}} = \beta'_\mathbf{q} \quad \mbox{and} \quad \beta''_{-\mathbf{q}} = -\beta''_\mathbf{q}.
\eeq
$\beta'_\mathbf{q}$ and $\beta''_\mathbf{q}$ are independent, centered random Gaussian variables, and we limit the sum to the half-plane $q_x>0$ to account for the conditions:
\bea
\Delta\theta^2 &=& \left\langle\left|\sum_{q_x>0,q_y} \beta_\mathbf{q} (e^{i\mathbf{q}\cdot\mathbf{r}}-1)+\beta^*_\mathbf{q} (e^{-i\mathbf{q}\cdot\mathbf{r}}-1)\right|^2\right\rangle\nonumber\\
&=& \left\langle\left|\sum_{q_x>0,q_y} 2\beta'_\mathbf{q} \left[\cos(\mathbf{q}\cdot\mathbf{r})-1\right]-2\beta''_\mathbf{q} \sin(\mathbf{q}\cdot\mathbf{r})\right|^2\right\rangle\nonumber\\
&=&4\sum_{q_x>0,q_y}\sum_{q'_x>0,q'_y}\langle \beta'_\mathbf{q}\beta'_\mathbf{q'}\rangle\left[\cos(\mathbf{q}\cdot\mathbf{r})-1\right]\left[\cos(\mathbf{q'}\cdot\mathbf{r})-1\right]+\langle \beta''_\mathbf{q}\beta''_\mathbf{q'}\rangle\sin(\mathbf{q}\cdot\mathbf{r})\sin(\mathbf{q'}\cdot\mathbf{r})\nonumber\\
&&-\langle \beta'_\mathbf{q}\beta''_\mathbf{q'}\rangle\left[\cos(\mathbf{q}\cdot\mathbf{r})-1\right]\sin(\mathbf{q'}\cdot\mathbf{r})-\langle \beta''_\mathbf{q}\beta'_\mathbf{q'}\rangle\sin(\mathbf{q}\cdot\mathbf{r})\left[\cos(\mathbf{q'}\cdot\mathbf{r})-1\right]\nonumber\\
&=&4\sum_{q_x>0,q_y}\langle \beta'^2_\mathbf{q}\rangle\left[\cos(\mathbf{q}\cdot\mathbf{r})-1\right]^2+\langle \beta''^2_\mathbf{q}\rangle\sin^2(\mathbf{q}\cdot\mathbf{r})\label{eq:variance1}
\eea
because the modes $\beta'_\mathbf{q}$ and $\beta''_\mathbf{q}$ are independent.

We now need to estimate $\langle \beta'^2_\mathbf{q}\rangle$ and $\langle \beta''^2_\mathbf{q}\rangle$. We can assume than we have an energy $\frac{1}{2}k_BT$ per degree of freedom (width of the Boltzmann law). The energy functional reads:
\beq
E[\psi]=\frac{\hbar^2}{2M}\int d\mathbf{r}\, \left(\left|\nablagras\psi\right|^2+\tilde{g}\left|\psi\right|^4\right)=E_{\rm kin} + E_{\rm int}.
\label{eq:energy_functional}
\eeq
The phase variations contribute to the first term corresponding to the kinetic energy:
\bea
E_{{\rm kin},\theta}&=&\frac{\hbar^2}{2M}\int d\mathbf{r} \,n\left|\nablagras\theta\right|^2 \simeq \frac{\hbar^2}{2M}n_0\int d\mathbf{r} \,\left|\sum_{\mathbf{q}} \mathbf{q}\beta_\mathbf{q} e^{i\mathbf{q}\cdot\mathbf{r}}\right|^2\nonumber\\
&=&N\frac{\hbar^2}{2M}\sum_{\mathbf{q}} q^2\left|\beta_\mathbf{q}\right|^2 = N\frac{\hbar^2}{M}\sum_{q_x>0,q_y} q^2\left(\beta'^2_\mathbf{q}+\beta''^2_\mathbf{q}\right).
\eea
To get on average $\frac{1}{2}k_BT$ per degree of freedom, we must thus have
\beq
\langle \beta'^2_\mathbf{q}\rangle = \langle \beta''^2_\mathbf{q}\rangle = \frac{Mk_BT}{2N\hbar^2q^2}=\frac{\pi}{N\lambda_{\rm dB}^2q^2}. 
\eeq
We inject this result into Eq.~\eqref{eq:variance1}:
\bea
\Delta\theta^2&=&\frac{4\pi}{N\lambda_{\rm dB}^2}\sum_{q_x>0,q_y}\frac{\left[\cos(\mathbf{q}\cdot\mathbf{r})-1\right]^2+\sin^2(\mathbf{q}\cdot\mathbf{r})}{q^2}\nonumber\\
&=&\frac{4\pi}{N\lambda_{\rm dB}^2}\sum_\mathbf{q}\frac{1-\cos(\mathbf{q}\cdot\mathbf{r})}{q^2}\qquad\mbox{with a sum over all }\mathbf{q}\nonumber\\
&=&\frac{4\pi}{N\lambda_{\rm dB}^2}\frac{L^2}{4\pi^2}\int d\mathbf{q}\frac{1-\cos(\mathbf{q}\cdot\mathbf{r})}{q^2}\nonumber\\
&=&\frac{1}{\pi n_0\lambda_{\rm dB}^2}\int_{0}^{q_T}\frac{dq}{q}\int_0^{2\pi}\!\!d\varphi\,\left[1-\cos(qr\cos\varphi)\right]\nonumber\\
&=&\frac{2}{n_0\lambda_{\rm dB}^2}\int_{0}^{q_T}dq\,\frac{1-J_0(qr)}{q}=\frac{2}{\mathcal{D}}\int_0^{rq_T}du\,\frac{1-J_0(u)}{u}
\eea
where $J_0$ is the zero order Bessel function. The integral converges\footnote{Anyway, a physically relevant low momentum cut-off at $1/r$ may be introduced.} at small $u$ where $J_0(u)\simeq 1-u^2/4$. At large $u$, the term with the Bessel function also converges because $J_0(u)\sim 1/\sqrt{u}$. However the term in $1/u$ diverges logarithmically at large $u$ if $r\to\infty$, it is the dominant term. The large $r$ behavior is thus dominated by
\beq
\Delta\theta^2 \simeq \frac{2}{\mathcal{D}}\log(rq_T).
\eeq
Reporting this behavior into Eq.~\eqref{eq:G1Gaussian}, we finally get
\beq
G_1(r) \simeq n_0\,e^{\ds-\frac{1}{\mathcal{D}}\log(rq_T)} = n_0\left(\frac{\ell_T}{r}\right)^{1/\mathcal{D}}\quad\mbox{for }r\to\infty
\eeq
with $\ell_T=1/q_T$ and $q_T$ is given at Eq.~\eqref{eq:qT_app}.

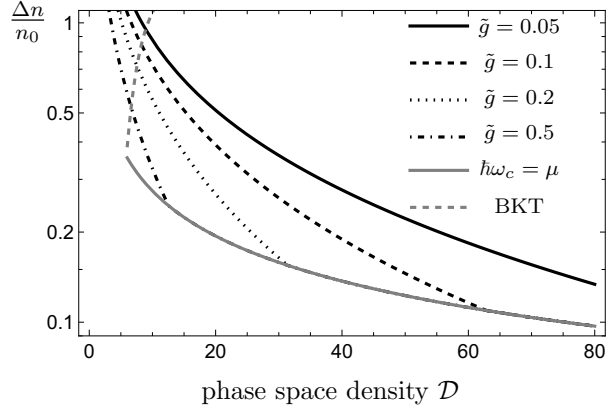
\begin{figure}
\centering
\begin{tikzpicture}
\node at (0,0) {\includegraphics[width=0.5\linewidth]{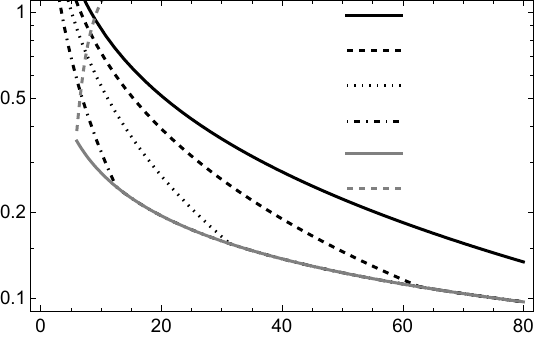}};
\node at (2.5,2.12) {\scriptsize $\tilde{g}=0.05$};
\node at (2.5,1.65) {\scriptsize $\tilde{g}=0.1$};
\node at (2.5,1.18) {\scriptsize $\tilde{g}=0.2$};
\node at (2.5,0.71) {\scriptsize $\tilde{g}=0.5$};
\node at (2.5,0.24) {\scriptsize $\hbar\omega_c=\mu$};
\node at (2.5,-0.23) {\scriptsize BKT};
\node at (0,-2.7) {\small phase space density $\mathcal{D}$};
\node at (-4,2.2) {\large $\frac{\Delta n}{n_0}$};
\end{tikzpicture}
\caption{Relative density fluctuations $\Delta n/n_0$ obtained from Eq.~\eqref{eq:Dnovern0}, for different values of $\tilde{g}$ (various black lines). Also shown is the limit value we get taking $\hbar\omega_c=\mu$, relevant for phase space densities above $2\pi/\tilde{g}$ (gray line). The dashed gray line is the BKT limit beyond which the quantum gas is superfluid.
\label{fig:Dnovern0}
}
\end{figure}

\paragraph{Density fluctuations} Finally, let us double check the condition for neglecting the density fluctuations. Coming back to Eq.~\eqref{eq:energy_functional}, we can write for the term for which density fluctuations contribute:
\bea
E_{{\rm kin},n} + E_{\rm int} &=& 
\frac{\hbar^2}{2M}\int d\mathbf{r} \,\left(\frac{\left|\nablagras n\right|^2}{4n}+\tilde{g}n^2\right)\nonumber\\
&\simeq & \frac{\hbar^2}{2M}\int d\mathbf{r} \,\left(\frac{n_0}{4}\left|\sum_{\mathbf{q}} \mathbf{q}\alpha_\mathbf{q} e^{i\mathbf{q}\cdot\mathbf{r}}\right|^2+\tilde{g}n_0^2+2\tilde{g}n_0\delta n+\tilde{g}n_0^2\left|\sum_{\mathbf{q}} \alpha_\mathbf{q} e^{i\mathbf{q}\cdot\mathbf{r}}\right|^2\right)\nonumber\\
&=&N\frac{\hbar^2}{2M}\tilde{g}n_0 + N\frac{\hbar^2}{2M}\sum_{\mathbf{q}} \frac{q^2+4\tilde{g}n_0}{4}\left|\alpha_\mathbf{q}\right|^2\qquad\mbox{as }\int d\mathbf{r} \,\delta n = 0\nonumber\\
&=&N\frac{\hbar^2}{2M}\tilde{g}n_0 + N\frac{\hbar^2}{M}\sum_{q_x>0,q_y} \frac{q^2+4\tilde{g}n_0}{4}\left(\alpha'^2_\mathbf{q}+\alpha''^2_\mathbf{q}\right).
\eea
We have $\frac{1}{2}k_BT$ per degree of freedom, which means
\beq
\langle\alpha'^2_\mathbf{q}\rangle = \langle \alpha''^2_\mathbf{q}\rangle = \frac{2Mk_BT}{N\hbar^2(q^2+4\tilde{g}n_0)}=\frac{4\pi}{N\lambda_{\rm dB}^2(q^2+4\tilde{g}n_0)}.
\eeq
From this expression we can compute the variance of relative density fluctuations
\bea
\frac{\Delta n^2}{n_0^2} &=& \frac{1}{n_0^2} \left\langle\left[n(\mathbf{r}) - n(\mathbf{0})\right]^2\right\rangle = \left\langle\left[\sum_\mathbf{q}\alpha_\mathbf{q} \left(e^{i\mathbf{q}\cdot\mathbf{r}}-1\right)\right]^2\right\rangle\nonumber\\
&=&4\sum_{q_x>0,q_y}\langle \alpha'^2_\mathbf{q}\rangle\left[\cos(\mathbf{q}\cdot\mathbf{r})-1\right]^2+\langle \alpha''^2_\mathbf{q}\rangle\sin^2(\mathbf{q}\cdot\mathbf{r})\nonumber\\
&=&
4\sum_\mathbf{q}\langle\alpha'^2_\mathbf{q}\rangle\left[1-\cos(\mathbf{q}\cdot\mathbf{r})\right]\qquad\mbox{with a sum over all }\mathbf{q}\nonumber\\
&=&\frac{16\pi}{N\lambda_{\rm dB}^2}\frac{L^2}{4\pi^2}\int_0^{q_T} d\mathbf{q}\frac{1-\cos(\mathbf{q}\cdot\mathbf{r})}{q^2+4\tilde{g}n_0}
\nonumber\\
&=&
\frac{4}{n_0\lambda_{\rm dB}^2}\int_0^{q_T}\frac{2q[1-J_0(qr)]\,dq}{q^2+2/\xi^2} = \frac{4}{\mathcal{D}}\int_0^{rq_T}\frac{2u[1-J_0(u)]\,dq}{u^2+2r^2/\xi^2}.
\eea
The integral converges for $u\to 0$, and diverges for $u\to\infty$ at large $r$. In this limit, it is dominated by the term $1/(u^2+2r^2/\xi^2)$, which can be integrated exactly. We finally have
\beq
\frac{\Delta n^2}{n_0^2} \simeq \frac{4}{\mathcal{D}}\log\left(1+\frac{q_T^2\xi^2}{2}\right)=\frac{4}{\mathcal{D}}\log\left[\frac{1}{2}+\frac{1}{2}\sqrt{1+\left(\frac{2\pi}{\tilde{g}\mathcal{D}}\right)^2}\right].
\label{eq:Dnovern0}
\eeq
The log term is typically a few units even in the regime where $\tilde{g}\mathcal{D}<2\pi$, such that density fluctuations are small as soon as $\mathcal{D}\gg 1$, which is the definition of the degenerate regime. The numerical result is plotted in Fig.~\ref{fig:Dnovern0} for various values of $\tilde{g}$, showing that $\Delta n/n_0$ is indeed very small for $\mathcal{D}$ exceeding a few tens.


\addcontentsline{toc}{section}{\protect
\end{document}